\documentclass[12pt]{article}
\usepackage[utf8]{inputenc}
\usepackage{amsmath, amsthm, amsxtra}
\usepackage{scalerel,stackengine}
\usepackage{geometry}
\usepackage{comment}
\usepackage{longtable}
\usepackage{booktabs}
\usepackage{bm}
\usepackage{multirow}
\usepackage{CJKutf8}
\usepackage{scalefnt}
\usepackage[table]{xcolor}
\usepackage{amsfonts}
\usepackage{threeparttable}
\usepackage{amssymb}
\usepackage{relsize}
\newcommand{\sym}[1]{\textsuperscript{#1}}
\usepackage{graphicx}
\usepackage[hidelinks]{hyperref}
\usepackage[utf8]{inputenc}
\usepackage{xcolor}
\usepackage{booktabs}
\usepackage{tabularx}
\usepackage{colortbl}
\usepackage{float}
\usepackage{setspace}
\usepackage{xparse}
\usepackage{microtype}
\usepackage{xurl}
\usepackage{mathtools}
\usepackage[skip=0pt]{caption}
\usepackage{subcaption}
\usepackage{etoolbox}
\usepackage{authblk}
\usepackage[bottom]{footmisc}
\usepackage{float}
\usepackage[backend=biber, style=authoryear, sorting=nyt, uniquename = false]{biblatex}
\stackMath
\newcommand\reallywidehat[1]{%
\savestack{\tmpbox}{\stretchto{%
  \scaleto{%
    \scalerel*[\widthof{\ensuremath{#1}}]{\kern-.6pt\bigwedge\kern-.6pt}%
    {\rule[-\textheight/2]{1ex}{\textheight}}
  }{\textheight}%
}{0.5ex}}%
\stackon[1pt]{#1}{\tmpbox}%
}
\makeatletter
\newcommand*\bigcdot{\mathpalette\bigcdot@{.5}}
\newcommand*\bigcdot@[2]{\mathbin{\vcenter{\hbox{\scalebox{#2}{$\m@th#1\bullet$}}}}}
\makeatother

\newenvironment{assumption}[2]{%
    \par\noindent\textbf{Assumption #1: #2} 
    \def\currentassum{#1}
    \ignorespaces
}{%
    \par%
}

\makeatletter
\newcommand{\assumlabel}[1]{%
    \edef\@currentlabel{\currentassum}
    \phantomsection 
    \label{#1}
}
\makeatother

\title{Anticorruption Enforcement and Sale Mechanism Choice in China’s Land Market}

\author{Julia Manso\thanks{Correspondence: Julia Manso, Department of Statistics, 24-29 St Giles', Oxford OX1 3LB, United Kingdom. \newline Email: jumanso@stats.ox.ac.uk. I thank Kosuke Imai, Steve Bond, Frank Windmeijer, and David Steinsaltz for their advice and feedback on this paper. }}
\affil{\small{\textit{Department of Statistics and Nuffield College, University of Oxford, Oxford, U.K.}}}
\date{}

\begin{document}

\maketitle
\singlespacing
\vspace{-10mm}
\begin{abstract}
    Upon taking office in late 2012, Chinese President Xi Jinping launched one of the most intensive anticorruption campaigns in the history of the People's Republic of China. Prior to the campaign, China's land market suffered from corruption, particularly surrounding sale method selection (auction versus listing). Listing is a two-stage sale mechanism that prior research has identified as more susceptible to corruption, leading to lower prices. This paper examines the campaign's impact on land allocation, focusing on whether corruption influences the choice of sale method and, in turn, land sale prices. This paper is the first to utilize Blackwell and Yamauchi (2021, 2024)'s marginal structural model with fixed effects in the inverse probability of treatment weighting model; absorbing time-invariant unobserved confounding and utilizing a set of time-varying covariates as controls, this model can estimate causal effects in the land sale case. I find that indictments in a prefecture cause a statistically significant drop in the probability that land is sold via listing\textemdash an effect that is further compounded when indictments occur in consecutive months. Sensitivity analyses indicate that any violations of the identification assumptions would bias estimates towards zero, confirming the negative effect. A second marginal structural model shows that both mean and median land sale prices increase in the presence of indictments. Together, these results suggest that the anticorruption campaign not only deterred actual corrupt allocation practices, but also impacted the discretionary use of listings.
\end{abstract}
\noindent \textbf{Keywords\textemdash} Chinese real estate, causal inference, marginal structural model, land allocation, auction design

\vspace{-7mm}

\onehalfspacing

\section{Introduction}

In the years leading up to the liquidation of Evergrande\textemdash formerly China's largest real estate developer and one of the largest companies in the world\textemdash the Chinese real estate market was growing rapidly, with housing prices increasing exponentially since 2000 (\cite{zhao_playing_2017}). Underlying this growth lay a distinctive incentive structure: local governments relied heavily on land sale revenue to fill fiscal deficits and were thus incentivized to raise the standing and attractiveness of their locality. This often involved strategic management of the land supply. For instance, local governments would sometimes sell industrial land for low prices in order to attract desirable industries to the area, but residential land was more tightly controlled to drive up the land's selling price and value. In this market characterized by such rapid price appreciation, any discount could yield a significant profit for developers\textemdash and in these circumstances, corruption naturally found a foothold.

Ultimately, by 2012, corruption had become endemic to the Chinese land sale market, effectively shaping every aspect of land development\textemdash how much land was sold, in what way, to whom, when, and at what price. In this environment, shortly after taking control of the Party in late 2012, President Xi launched a broad-sweeping, top-down anticorruption campaign that investigated officials at every level of government. The campaign took the form of anticorruption waves: teams of investigators were dispatched to a set of provinces and conducted investigations into corruption for a few months, returning to Beijing afterwards to report findings. Under political pressure from the central government, local governments were then empowered to prosecute offenders and deter further corruption. This centralized and extensive anticorruption campaign provides an interesting setting to directly investigate corruption's impacts in the land market, and this paper focuses on one pathway in particular\textemdash corruption's impact on the sale method used by a local government when selling a property, and subsequently, how corruption influences the sale price faced by developers.  In essence, one sale method (the listing) is believed to be associated with corruption in the literature.

To investigate this relationship, I utilize a dataset on corruption indictments reported at the prefecture level, as well as comprehensive land transaction records scraped from the Chinese government's land transaction database. After performing a two-way fixed effects specification and finding the assumptions are not met, I use a marginal structural model (MSM) with unit-level fixed effects in the inverse probability of treatment weighting (IPTW) model, which is used to create the weights for each unit. MSMs are an underutilized inference tool in the social sciences, particularly given the shortcomings of traditional two-way fixed effects specifications, as described in \textcite{imai_use_2021}. Developed by Blackwell and Yamauchi (2021, 2024), the MSM with fixed effects in the IPTW is a novel tool to bypass the traditional shortcomings of MSMs, and this paper is the first to apply the Blackwell and Yamauchi methodology. Further, the question at the heart of this paper\textemdash corruption's impact on the sale method and sale price of land, as reflected by the anticorruption campaign\textemdash has not yet been examined in the literature. This paper also makes important contributions on the data side with its new, accurate data set\textemdash an important development given that a dominant data set in the Chinese real estate space (\cite{chen_busting_2019}) was revealed to have significant accuracy issues.\footnote{See \textcite{manso_are_2026} for a discussion of the data problems of \textcite{chen_busting_2019}.}

In this analysis, corruption investigations serve as the treatment, and the sale method is the outcome. I ultimately find a statistically significant negative effect when prefecture-level fixed effects are included in the IPTW model: using marginal effects, a prefecture having corruption indictments causes a 1.16 percentage point decrease in the probability of having any listings. However, this effect is further compounded when sequential periods are treated, and additionally, uncertainty in the timing of the indictment means that the effect on the outcome may also spill into neighboring time periods. Investigating the cumulative effect for the five-month period around having a corruption indictment reveals a 7.78 percentage point decline in the probability of having any listings. Sensitivity checks are conducted on these results and evaluate whether the assumptions are satisfied, finding that any violation of the assumptions would likely cause the coefficients to be underestimates. 

This paper then examines the relationship between sale method and price via an additional MSM and finds that the mean and median land sale price rise in the presence of corruption indictments. Specifically, a prefecture having corruption indictments in a given month/year causes, on average, a 6.78\% increase in the mean price per square meter, and the median increases similarly. Given that the mean is 2,200 yuan per square meter (US\$350) and that properties tend to be tens of thousands of square meters, if not more, a nearly 7\% increase represents a significant amount of cash. Interpreting these results together, I posit that the anticorruption campaign does deter actual corrupt behavior (i.e., having listings on favorable properties), but that it also impacts behavior for properties on the margin (i.e., those that could be sold as either auctions or listings). 

This paper is structured as follows. After offering background on the land sale process and President Xi's anticorruption campaign (Section \ref{sec: background}), Section \ref{sec: literature review} offers a literature review. Section \ref{sec: data} gives an overview of the data and its collection/cleaning process. Section \ref{sec: methodology} briefly discusses the two-way fixed effects specification and why it falls short in this context before outlining the MSM with fixed effects. Section \ref{sec: results} details the results, and Section \ref{sec: Regressing corruption indictments on price} examines the relationship between corruption indictments and price to offer insight into how market behavior changes in the presence of the anticorruption campaign. Section \ref{sec: conclusion} concludes.

\section{Background} \label{sec: background}

First, to understand how the Chinese land finance system operates, I provide background on China's land use system and land auction system, discussing why corruption is so attractive and prevalent in the land use environment. I then introduce Xi's anticorruption campaign and its effects. 

\subsection{Land ownership and the path to sale} \label{land ownership and path to sale}

Under the Chinese system, the government is the ultimate owner of land, and funding from land use right sales (dubbed ``land finance") is a critical part of local governments' funding. The system is incredibly complex and multifaceted, but in essence, the local government creates new urban land use rights (LURs) via land acquisition and subsequently sells these LURs to private and public land users (\cite{gyourko_land_2022}).\footnote{In China, all urban land is owned by the State while rural and suburban land can be owned by rural collectives (``collective land") (\cite{huang_land_2018}). Thus, when a local government purchases new land, it is either purchasing rural land from one of these collectives or buying back the land it previously sold\textemdash that is, reacquiring LURs for land already developed (\cite{gyourko_land_2022}; \cite{zhang_china_2015}). Because the government is the sole party able to purchase land (having eliminated the secondary market for LURs in 2004) and has strict rules about how land prices can be calculated, the State is able to purchase land from collectives and from current LUR holders at prices advantageous to itself (\cite{li_urban_2019}; \cite{gyourko_land_2022}). Thus, when a local government sells a LUR, it is selling the right to use the land for a fixed period\textemdash for instance, 70 years for residential, 40 years for commercial, and 50 years for industrial, although these may be shorter depending on the municipality\textemdash but it is still the ultimate land owner (\cite{su_visible_2012}).} When a local government sells a LUR, it receives an upfront, lump sum payment from the developer, and the difference between the purchase and sale price is the local government's profit, which it then uses to finance other aspects of its budget (\cite{wang_political_2016}). 

Every 15 years, the State releases the national-level ``Land Use Master Plan," which effectively caps the amount of rural-to-urban land conversions during the specified 15-year period in each province. The provincial government then issues quotas to each city, which are subsequently passed down to lower levels of government (\cite{gyourko_land_2022}). Every year, each local government (specifically the local land resource bureau) creates a ``Land Use Annual Plan," setting its expected rural-to-urban land conversion and urban land supply for the year. It is this plan that includes the list of the land parcels that the State will auction that year (\cite{wu_primary_2020}).\footnote{There are also pathways for local governments to raise the quota, for instance, by trading quotas with other local governments or negotiating with higher-level (provincial) governments. Local governments could also disregard the quotas and accept punishments, which are relatively minor if certain requirements are also met (\cite{gyourko_land_2022}). } Then, an independent committee of the city's political leaders and key figures from government agencies, such as the land resource bureau and the urban planning bureau, decide key constraints for each property to be sold, and when the land comes up for sale later during the year, they determine sale price as well (\cite{cai_build_2017}; \cite{cai_chinas_2013}). Responsibility for the land is finally passed to the local land bureau, which executes the land sale and decides the auction style (\cite{cai_chinas_2013}). 

\subsection{Sale method} \label{auction style}
The local government subsequently decides the timeline of LUR sales for the year and each parcel's auction method. After land sale reforms implemented in September 2004, ``leaseholds are, in principle, all sold at public auction," and there are three main types of auctions used in China: an English auction ($Paimai$, in Chinese), which is a standard ascending auction; a listing/two-stage auction ($Guapai$), an ascending open-bid auction in two related stages that last for a period of time; and a sealed-bid auction/tendering ($Zhaobiao$), where bidding is granted through invitation (\cite{cai_chinas_2013}, p. 2; \cite{liao_boundedly_2023}; \cite{zhu_shadow_2012}). About 97\% of LUR sales in major cities are comprised of English and listing auctions, with tenders (sealed-bid auctions) occurring only in Beijing and Shanghai for selected transactions. 

A standard auction wherein the bidder who offers the highest price wins, the English auction is usually announced 20 working days in advance, at which time basic details like a property's use restrictions, reserve price, and location are released (\cite{cai_chinas_2013}). Interested parties can also obtain additional information for a small fee and/or inspect the site itself. Participation in the auction then requires a cash deposit of around 10\% of the reserve price, and the auction itself is often public, videotaped with press in attendance (\cite{cai_chinas_2013}). Announced 20 working days in advance, listings are also known as ``two-stage" auctions because of their format: the first stage typically lasts 10 working days after the auction begins, although it can last much longer in certain cases, and during this time, qualified bidders submit a bid (at or above the reserve price) in person or online. Bids are posted as they are made on the ``trading board of the land bureau, as well as typically on the internet, although the identity of bidders is not posted" (\cite{cai_chinas_2013}, p. 6; \cite{wu_primary_2020}). Bidders can bid incrementally as many times as they want, and if there is only one active bidder at the end of the first stage, the land is sold to that bidder at their final bidding price (\cite{wu_primary_2020}). If, however, more than one bidder was active at the end of the first stage, all active bidders enter the second stage\textemdash an English auction. 

Finally, tendering/sealed-bid auction involves more complex criteria: the bid evaluation committee rates each bid on several factors, including not only the price of the bid itself, but also the credibility of the bidder, how much ``social responsibility" the bidder is willing to take on, and the proposed development plan (\cite{cai_chinas_2013}, p. 5; \cite{wu_primary_2020}).\footnote{As described by \textcite{cai_chinas_2013}, credibility focuses on ``the quality and reputation of the projects the bidder has developed in the past," in addition to the bidder's financial capacity (p. 5). ``Social responsibility" is rooted in attempts to curb housing prices, for instance: developers who commit to upper bounds on housing prices receive greater scores.} As highlighted, these auctions are relatively rare, comprising only 2.69\% of LUR sales between 2000-2015, according to data from the China Real Estate Index System (\cite{wu_primary_2020}). 

In addition to deciding which auction type to use, the local government determines all details about which way each parcel of land will be auctioned off, selecting the auction date and time, auction format, bidder qualification, and reserve price\textemdash and it is this latitude that allows corruption and under-the-table deals to occur.

\subsection{Pathway of corruption in land sale} \label{Pathway of Corruption in Land Sale} 

Corruption in the land market is vast and complex, rising from a multitude of incentive structures and pathways. I detail only the theoretical pathway of corruption's influence on sale method here; others are explored in Appendix \ref{appendix: pathway of corruption in land sale}. In effect, auctions are not fully ``open to competition" because local governments decide all facets of the auction (\cite{zhu_shadow_2012}). The key manipulation tools include raising the bar of entry into an auction, for instance, by requiring a minimum level of capital or certain ratings; merging small land parcels into one large parcel so that only the wealthiest firm can bid, and offering deals to provide infrastructure for new properties. The local government may also instruct firms to make deals amongst themselves before bidding for a plot opens (\cite{zhu_shadow_2012}). 
Listings (``two-stage" auctions) are particularly susceptible to corruption. For instance,  although the auction is announced roughly 20 working days in advance, the exact start date of the first stage is often announced at a much later date. Additionally, any firm wanting to bid in the first stage has to submit a bidding application and other basic materials, and local officials subsequently review the bidder's development qualifications and ``integrity records"  
\begin{CJK*}{UTF8}{gbsn} 
(\cite{noauthor__2013-1}, trans.).
\end{CJK*} In cases of corruption, those who ``fail to meet the criteria" can be denied outright, or approvals to bid can be ``delayed" until the first stage is already underway, excluding would-be bidders (\cite{cai_chinas_2013}). 

\begin{figure}[H]
    \centering
    \caption{Sale timeline diagram}
    \includegraphics[width=1.02\linewidth]{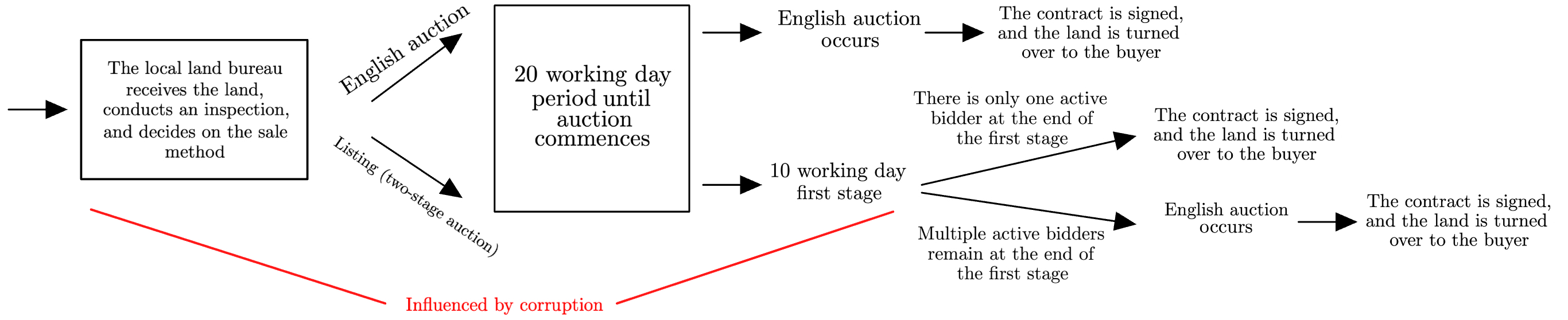}
    \label{fig:Timeline diagram}
\end{figure}

Indeed, \textcite{cai_chinas_2013} investigated data from land sales in 15 cities across China from 2003-2007 and found that officials divert ``hot" properties to the more corruptible ``two-stage" auction, with many of these auctions often having only one bidder (and thus no competition). They further find that English auctions have competition and tend to result in higher prices, controlling for differential property characteristics such as distance from the business district. This result indeed makes sense given that English auctions are required to have a minimum of three ``qualified bidders" (\cite{wang_are_2017}, p. 199).\footnote{While some papers such as \textcite{wang_are_2017} use the three ``qualified bidders" requirement as reflecting a completely fair auction, I expect this requirement is not enough to wholly signal competition because developers can enter the auction with their different subsidiaries. That is, if China Vanke is the listed developer, two if its subsidiaries, Nanjing Yuyue Real Estate and Nanjing Yuxiao Real Estate, could enter the auction, comprising two of the three bidders, but they fundamentally represent the same developer (\cite{noauthor_2023_2024}). This behavior became particularly rife in later periods, such as when the central government mandated mass land auctions in 2021: some of these auctions had over 300 entities participating\textemdash which actually represented only 30 developers (\cite{yu_chinese_2021}). Nevertheless, \textcite{cai_chinas_2013} find evidence that sale prices tend to be higher for English auctions than two-stage ones, suggesting higher levels of competition.} Figure \ref{fig:Timeline diagram} below highlights this timeline and where corruption can interfere. 

\medskip

Note, however, as detailed fully in Appendix \ref{appendix: pathway of corruption in land sale}, ``even though [local governments] had been directed by the central government to increase land supply and cap home prices to keep a lid on social discontent" (\cite{shao_beijing_2013}), they also had an incentive to raise revenue as much as possible, and LUR sales are an important source of this revenue. Additionally, prior to the anticorruption campaign, leading local officials' promotions could even be tied to Gross Regional Product (GRP) increases.\footnote{To curb this behavior, the government announced in late June 2013 that it will stop evaluating party officials solely based on their contributions to growing GRP, aiming to diminish the role of land sales funding local governments' budgets and reduce overinvestment in the sector (\cite{zhu_president_2013}; \cite{guilford_china_2013}).} Intuitively, this could lend itself to higher land prices, as officials would want to sell fewer properties for very high prices to maximize revenue (\cite{li_political_2005}). However, viewing this from a utility maximization perspective provides insight: firms are simple profit maximizers, wanting to acquire land at the lowest cost, build as much as possible and as quickly as possible on that land, and sell units for as much as possible. Along the way, they are happy to spend money to ``grease the wheels" (i.e., bribe officials) as long as it results in a higher revenue for the project. Officials are utility maximizers: those who get more utility out of bribes accept them, while those who do not abstain. The profits from corrupt sales, even if they are lower than they could be under greater levels of competition, often still allow the government to profit from the LUR sale\textemdash and this money is then funneled into infrastructure investment, which creates further demand for property in the region, allows governments to set higher reserve prices, and increases access to borrowing. This cycle continually repeats such that both officials and developers get what they want at every level. Officials can curry both political favor and profit at the same time by behaving as clever utility maximizers. 

Interesting questions result when these motivations clash: for instance, when an official's boss seeks to follow the central government directive to limit land sale price growth and instructs the official to ensure no discounted land sales occur, but the official is offered a tempting bribe. Alternatively, when fear of detection rises, as in an anticorruption drive, the official may be tempted to bring job security into the calculation. This is the central question of this paper\textemdash is there evidence of the theoretical pathway between corruption and sale method, which in turn impacts sale price? 

\subsection{Anticorruption campaign overview} \label{Anticorruption campaign overview}

Prior to President Xi's top-down anticorruption campaign, anticorruption campaigns in China had been largely performative and had little impact on underlying corruption levels.\footnote{See Appendix \ref{sec: President Xi's Anticorruption campaign background} for further details.} Xi's campaign, beginning in November 2012 and gaining steam in early 2013, launched what many consider to be the most intensive and protracted anticorruption campaign in the history of the People's Republic of China, catalyzing a period of reckoning and corruption indictments\textemdash and then, towards the end of 2016, the beginning of the transformation of the anticorruption infrastructure itself (\cite{wedeman_four_2016}; \cite{deng_national_2018}). 

In the first months of 2013, Xi began preparing the anticorruption machinery before overseeing the creation of online reporting platforms wherein citizens could participate in supervision and reporting \begin{CJK*}{UTF8}{gbsn} 
(\cite{noauthor__2013}).
\end{CJK*}

Then, in mid-May, Xi dispatched 10 Central Inspection teams to a set of 6 provinces (Inner Mongolia, Chongqing, Guizhou, Hubei, Jiangxi, and Beijing) and 4 organizations (China Publishing Group, the Ministry of Grain Reserves and Water, the Export-Import Bank of China, and Renmin University of China, located in Beijing). These Central Inspection teams had until the end of July or early August to investigate their region/organization before reporting their findings to the inspected areas and the units of inspection (\cite{wang_76_2013}). The Central Inspection teams then returned to Beijing to report their findings to party leadership\textemdash which includes all the parties and government members of the inspected areas (\cite{wang_76_2013}). Findings, as presented in these meetings, held nothing back, often offering up scathing critiques of provincial problems, reporting everything from ``abuse [of] power for personal gain," to corrupt hiring practices, to ``inadequate management and supervision of top leaders, $\dots$ shaken beliefs, ideological decline, and moral deviance" (\cite{wang_76_2013}, trans.). 

As they progressed with further rounds of inspections, the Central Inspection teams sought to be nimble and unpredictable\textemdash one deputy director focusing on anticorruption remarked that if there was any kind of pattern or cycle to the anticorruption campaign, the officials under inspection would detect it and change their behavior, instigating ```falsification and cover-up'" (qtd. in \cite{wang_76_2013}, trans.). The Central Inspection teams thus had revolving leaders and targets on their quest as they were dispatched to different provinces in several waves. Further, the Central Inspection team assigned to a province would not notify the region/units to be inspected until 10 days in advance, in writing, of their arrival, among several other measures to involve the public in reporting and enforcement. This nuanced and carefully planned workflow, designed to ascertain the true state of corruption in each area/unit as much as possible, was the blueprint for subsequent waves of the anticorruption campaign. 

In this second wave, 10 Central Inspection teams were sent to 6 provinces (Shanxi, Jilin, Yunnan, Anhui, Hunan, Guangdong) and 4 organizations (Xinhua News Agency, Ministry of Land and Resources, Ministry of Commerce, and the State-owned company overseeing the construction of Hubei province's Three Gorges Dam). \begin{CJK*}{UTF8}{gbsn} (\cite{li__2013}; \cite{tian__2014}). \end{CJK*} The third round of central inspections began in late February/early March 2014 with officials being dispatched to 10 provinces (Beijing, Tianjin, Liaoning, Fujian, Shandong, Henan, Hainan, Gansu, Ningxia, and Xinjiang/Xinjiang Production Corps) and 3 organizations (the Ministry of Science and Technology, State-owned food processing company COFCO, and Shanghai's Fudan University) \begin{CJK*}{UTF8}{gbsn}(\cite{noauthor__2014}). \end{CJK*} 

In total, there were 10 anticorruption waves before August 2016, but for brevity, only those most relevant are discussed above. 

\vspace{-3mm}
\section{Literature review} \label{sec: literature review}

This section offers a broader overview of the literature on corruption's impacts before moving into a review of studies on the impact of President Xi's anticorruption campaign. 

Broadly defined as dishonest or fraudulent conduct by those in power, corruption has been extensively studied in the economic and statistical literature, particularly with investigations into its impact and mechanisms. Early studies such as \textcite{becker_law_1974} and \textcite{rose-ackerman_corruption_1978} focus chiefly on the relationship between the top level of government (principals) and officials (agents) who accept bribes. \textcite{shleifer_corruption_1993} develop a theoretical framework to understand how corruption distorts government policies and outcomes, distinguishing two types of corruption: with and without theft. They describe the low-competition, low-detection settings that breed corruption, positing that a weak, decentralized central government allows multiple officials across government agencies to simultaneously collect bribes from the same private agent. They argue that corruption will not be deterred until its incentive structure and/or operational mechanism are overhauled, for instance, by implementing intragovernmental competition (i.e., making several agencies compete over the provision of a good). 

On the heels of these early analyses, several studies began to quantify the empirical impacts of corruption, particularly focusing on the relationship between corruption, investment, and government income. Many of these studies have yielded diverging results based on the data, design, and statistical methods used: in one of the first and most prominent studies, \textcite{mauro_corruption_1995}, for instance, examines the impact of corruption on private investment, using an index of ethnolinguistic fractionalization as an instrument and finding that corruption lowers economic growth.\footnote{As described by \textcite{mauro_corruption_1995}, ethnolinguistic fractionalization (ELF) ``measures the probability that two persons drawn at random from a country's population will not belong to the same ethnolinguistic group" (p. 682-3). Subsequently, many have critiqued this instrument for failing to meet exclusion restrictions: \textcite{bentzen_how_2012} argues that based on the findings of \textcite{acemoglu_unbundling_2005}, ELF is likely correlated with other determinants of growth excluded from Mauro's analysis. Bentzen introduces a modified statistical approach and finds a similar negative effect of corruption on investment.} On the other hand, \textcite{meon_is_2010} find that corruption can actually increase efficiency in highly inefficient regimes, and \textcite{pradhan_impact_1999} emphasize that high levels of predictable corruption do not necessarily hinder economic growth. 

Similarly, \textcite{tanzi_corruption_1997} find that higher corruption is associated with higher public investment and lower government expenditure while \textcite{poveda_relations_2019} find that higher corruption is associated with lower public investment and government expenditure. These results converge towards two alternate ideas/hypotheses: that corruption ``greases" the wheels of bureaucracy, by allowing inefficiencies to be overcome with a private exchange, or ``sands" the wheels, being harmful to economic growth and investment (\cite{meon_is_2010}).

The overall consensus, though, is that corruption is generally associated with lower levels of economic growth and that whether or not net public investment and government spending increase with higher levels of corruption, these expenditures are less efficient and may be misdirected, tending to reduce productivity (\cite{mauro_corruption_1995}; \cite{bentzen_how_2012}; \cite{tanzi_corruption_1997}; \cite{poveda_relations_2019}; \cite{pradhan_impact_1999}; \cite{meon_is_2010}; \cite{meon_does_2005}; and \cite{aldieri_corruption_2023}, among others). 

There is markedly less literature that analyzes the impacts of anticorruption campaigns, partly because national, top-down, sustained anticorruption investigations are quite rare. While there have been prominent anticorruption campaigns in South Korea (\cite{min_impact_2023}), Brazil (\cite{castro_contextual_2017}), Indonesia (\cite{widojoko_indonesias_2017}), and India (\cite{riley_corruption_2016}), among other countries, in recent years, the degree of efficacy has widely varied, and for the most part, detailed information on corruption indictments is not widely available. Yet, as explored in Section \ref{china anticorruption background}, the case of China is an exception; a number of studies have emerged to investigate the extent of corruption and the impacts of Xi's anticorruption campaign. 

\subsection{Investigating the impacts of China's anticorruption campaign} \label{china anticorruption background}

The Chinese corruption literature has converged into a few strands. First, that highlighting the positive impacts of anticorruption: areas with stronger anticorruption efforts saw improved attitudes towards government credibility (\cite{zhang_anti-corruption_2019}), improved acquisition of research and development (R\&D) funding (\cite{xu_how_2017}), reductions in stock price crashes (\cite{chen_does_2018}), and increased entry of new firms (\cite{ding_equilibrium_2020}). Other studies find the effects are more moderate and depend heavily on the level and number of officials arrested: \textcite{kim_value_2018}, for instance, analyze stock market price responses of all firms listed on the Shenzhen and Shanghai stock exchanges, finding that the market reaction to the campaign is ``more positive for provinces where more department-level officials are arrested" but is negligible for the indictments of lower-ranking officials (p. 116).

Another strand focuses on how benefits of the anticorruption campaign diverge by firm ownership status, with \textcite{kong_effects_2017} detecting differing firm performance trends based on whether a firm is a state-owned enterprise (SOE). Specifically, they find that the anticorruption campaign greatly improves performance for SOEs but significantly reduces it for non-SOEs. \textcite{tian_how_2018} investigate how anticorruption measures affect corporate governance, finding that the reduction in executive incentives accompanying the anticorruption campaign was stronger for SOEs than non-SOEs. This evidence, they argue, suggests SOEs have been larger targets of Xi's campaign than non-SOEs. On the other hand, \textcite{zhang_public_2018} finds that the anticorruption campaign and its new mechanisms have a greater impact on listed non-SOEs than SOEs, implying private firms are more sensitive to the campaign. \textcite{zhang_public_2018} further notes that the campaign has a larger impact on firms in poor areas and areas with weak legal environments.  \textcite{cao_anti-corruption_2018} examine how corruption investigations shape firm information release. They note that regions with higher scrutiny experienced significantly lower negative information release in comparison to firms in other regions, finding a more pronounced effect for SOEs\textemdash who presumably have stronger ties to local authorities\textemdash compared to non-SOEs. Finally, \textcite{alonso_value_2022} find that political connections have increased in importance for non-SOEs in the wake of the anticorruption campaign, as they help these firms receive more governmental subsidies. Comparatively, politically-connected SOEs still experienced ``access to lower cost of debt, [but] at a lower magnitude than before" the anticorruption campaign (p. 785).\footnote{A smaller subset of analyses distinguish among the impacts of SOE status for ``event firms"\textemdash those implicated in corruption investigations/scandals\textemdash and ``non-event firms" who were not implicated (e.g., \textcite{pan_political_2020}, \textcite{griffin_is_2022}, and \textcite{he_political_2017}). The exact impacts of Xi's anticorruption campaign on SOEs vs. non-SOEs are complex, involve many different pathways, and are difficult to quantify, but the consensus is that the effect differs across both firm types and their event status.}

Other relevant papers investigating anticorruption in China include \textcite{zhao_impact_2020}, which uses a difference-in-differences design to identify the effects of the anticorruption campaign on land supply in China. They specifically focus on corruption of top leaders (mayors and Party secretaries) in prefecture-level cities, examining a date range of 2006-2016. They find that following the indictment of a major official for corruption, not only would the total amount of land supply drop sharply, but the proportion of profitable commercial and residential use land would decrease while that of public use land increased\textemdash in essence, the type of land supplied changed.\footnote{One recent paper by \textcite{arslan_auctions_2025} compares listings and tenders in the Chinese market in the context of corruption. While the paper seems quite relevant, it utilizes \textcite{chen_busting_2019}'s data\textemdash which suffers from a significant number of erroneous duplicates and a mistransformed measure of area, as highlighted in \textcite{manso_are_2026}. I thus do not give weight to its conclusions.}

Therefore, while these papers shed light on the impacts of corruption and the anticorruption campaign itself, my paper moves beyond the focus and methods of previous studies, not only investigating how the anticorruption campaign shapes the sale method, but making important contributions on the data side\textemdash a particularly relevant development given the prevalence of \textcite{chen_busting_2019}'s faulty data in the literature.

\section{Data} \label{sec: data}

I amalgamate data from multiple sources to investigate these questions. First, the data for corruption investigations was extracted from Tencent, the largest internet/multimedia company in China, by \textcite{wang_how_2020}. During Xi's anticorruption campaign, Tencent ``launched a searchable online database of all corruption investigations across China since 2011" (p. 13).  Combining and synthesizing information from ``Party disciplinary committees, courts, and procuratorates from the central to local levels" (p. 13), this Tencent database is the most comprehensive publicly accessible repository of China's corruption investigations. Further, this database was provided for users to explore the extent of corruption in their town and province; as highlighted by Wang and Dickson, it is the ``only place where Chinese citizens can find out this information with a single click" (p. 14) and has been widely circulated via Tencent's messaging app WeChat.\footnote{WeChat is China's most popular social media/messaging app, surpassing 1 billion users in 2018 (\cite{noauthor_one_2018}).} 

The data was scraped by Wang and Dickson in August 2016 via Python and contains detailed information on official indictments, including each official's name, position, locality, province, and the reason for investigation, as well as the official's rank of importance on a scale from 1-10. This scale defines 1 as state-level officials while 10 is deputy directors and below. On this scale, rank 3 is provincial governors and top officials in the party committee of each province. Rank 4 is mayors of major cities, vice governors, and lesser officials in the province-level party committee, among others of similar positions. A full description of the remaining ranks 5-10 is listed in Table \ref{tab:breakdown of officials' rank}. 

Entries also contain information on when the investigation into each official began, although approximately 1,600 out of 19,000 entries lacked the specific date of investigation (for instance, missing a month and/or year); the vast majority of these were officials of the lowest importance (levels 9 and 10 on the 1-10 scale). Where possible, I conducted additional research to fill in missing date entries, utilizing party newspapers and media to discern the dates that investigation into each official began. With this cleaning, I was able to complete over 1/3 of the incomplete entries, bringing the total number of entries missing the year and month of investigation to 933, and those missing only the month to 1,099. The relative counts of each rank of official, as well as their description, for the completed entries are included in Table \ref{tab:breakdown of officials' rank}. 

\vspace{1mm}
\begin{table}[H]
\centering

\caption{Ranking Scale of Officials with Count and Share}
\label{tab:breakdown of officials' rank}
\begin{tabularx}{\textwidth}{cccX}
\toprule
\textbf{Rank} & \textbf{Count} & \textbf{Share (\%)} & \textbf{Description} \\ 
\midrule
1 & 0 & 0\%   & Most critical State-level officials (National Politburo Standing Committee members) \\
2 & 0 & 0\%   & State-level officials (National Politburo Committee members) \\ 
3 & 8 & 0.04\% & Provincial governors and top officials in the party committee of each province \\ 
4  & 71 & 0.40\%  & Mayors of major cities, vice governors, and lesser officials in the province-level party committee (such as Provincial-level Standing Committee members), SOE leaders\\ 
5 & 400 & 2.24\%  & Mayors of minor cities and districts, deputy mayors of major cities, directors of provincial administrative departments, chairs of provincial initiatives, SOE executives \\ 
6 & 858 & 4.81\%  & Mid-level provincial administrative roles, deputy mayors of minor cities and districts, leaders of county or district party committees, general managers of major SOEs, deputy directors of provincial departments\\ 
7 & 2,158 & 12.10\%  & Directors of minor local bureaus, top officials of county or district party committees, general managers of minor SOEs, deputy general managers of major SOEs, principals of major educational institutions \\ 
8 & 1,931 & 10.82\%  & Deputy county magistrates, vice chairs of county or district party committee, deputy directors of local bureaus and committees, general managers of minor SOEs, deputy general managers of major SOEs, principals of mid-level educational institutions \\ 
9 & 8,255& 46.27\%  &  Directors and managers of small, specialized bureaus and agencies at the county or district level; directors of technical and administrative departments within public service institutions; principals of minor educational institutions; leaders (directors, chairmen, captains) of localized programs, services, or projects  \\ 
10 & 4,160 & 23.32\%  & Deputy directors and below \\ 
\midrule
\textbf{Total} & \textbf{17,841} & \textbf{100\,\%} & \\
\bottomrule
\end{tabularx}
\end{table}

I obtain transaction-level data from the Ministry of Land and Resources' Land Transaction Monitoring System (\href{http://www.landchina.com/}{\texttt{www.landchina.com}}). As each municipality's ``bureau of land and resources is required to report each land transaction in their jurisdiction electronically on this website," per the Law of Land Management (\cite{chen_busting_2019}, p. 199), the dataset captures granular sale details for all land transactions. I focus on residential real estate land transactions for the years 2010-2017 that were sold via auction and listing, obtaining 209,706 total transactions.\footnote{A description of data cleaning procedures for the scraped data can be found in Appendix \ref{data cleaning appendix}.}

Control data comes from several sources: demographic and geographic coordinate information is derived from \textcite{dong_census_2022}, who utilize a more robust version of China's Sixth Census Yearbook. Prefecture-level GDP is obtained from the yearly volumes of the China Statistical Yearbook for Regional Economy, compiled by the Department of Comprehensive Statistics and the Department of Rural Survey of the National Bureau of Statistics for 2011-2013 (\cite{ma_china_2013}). Data for 2014-2016 combines annual provincial prefecture-level GDP reports to form a comprehensive dataset, with provincial websites and other annual reports used to fill in any missing data. When determining the number of corruption indictments in a prefecture in each month, for provincial level officials, I distribute the effect across all prefectures in a province; however, only binary measures of corruption indictment are used in computing the below results.

\section{Methodology} \label{sec: methodology}

To examine the relationship between corruption indictments and the share of listings in a prefecture, I first estimated a set of two-way fixed effects regressions but ultimately found that the assumptions are not met. Appendix \ref{sec: two-way FE} provides full details and regression results, but in essence, linear fixed effects models require two causal identification assumptions: first, that past outcomes do not impact the current treatment, and second, that past treatments do not directly influence the current outcome (\cite{imai_when_2019}). 

In the case of the Chinese anticorruption campaign, both of these assumptions are violated as there is likely strong feedback between the treatment (the number of corruption indictments) and the outcome (the share of listings). Further, I expect that there are some time-varying confounders like gross regional product (GRP), which may make a unit more likely to be treated\textemdash for instance, if richer areas are treated sooner. At the same time, being treated likely affects the area's GRP starting from the time of treatment. Such variables are both simultaneous confounders and intermediate variables, and including time fixed effects therefore blocks part of the causal pathway, biasing estimates. 

Given the causal identification assumptions for two-way linear fixed effects are not met, I implement a marginal structural model (MSM) to investigate the causal relationship between corruption indictments and sale method. Developed by Robins (1986, 1998ab, 1999ab), \nocite{robins_association_1999} \nocite{robins_marginal_1999} \nocite{robins_correction_1998} \nocite{robins_marginal_1998} the MSM is a multi-step estimation tool that aims to estimate the causal effect of a treatment on an outcome in the presence of time-dependent covariates that may be both simultaneously confounders and intermediate variables. As implemented by Robins, the MSM relies on inverse probability of treatment weighting (IPTW): IPTW in effect creates a pseudo-population by weighting each unit by the inverse of the conditional probability of receiving the treatment that it received (\cite{hernan_causal_2020}). Then, passing these weights into the final MSM regression is  ``conceptually identical to running an unweighted, regular regression model in the pseudopopulation in which confounders and treatments are independent of each other" (\cite{thoemmes_primer_2016}, p. 42), allowing a causal effect to be estimated. When the MSM's assumptions are met, IPTW can consistently estimate the model parameters, allowing for identification of the marginal mean of potential outcomes under any treatment sequence.\footnote{In this sense, the model is ``marginal" because it utilizes the marginal distribution of the treatment and ``structural" because it models the probabilities of counterfactual variables, which is referred to as ``structural" in the econometrics/social sciences literature (\cite{williamson_marginal_2017}; \cite{robins_marginal_2000}).} 

In the Chinese case, as described below, the basic MSM is a poor fit, as one key assumption (sequential ignorability) is not met, and another assumption (consistency) is doubtful, given many confounders that influence treatment are difficult to capture in conventional variables. I expect that many of these are time-invarying prefecture-level confounders, and this paper's main MSM specification of focus is Blackwell and Yamauchi (2021, 2024)'s MSM with fixed effects in the IPTW model.\footnote{Note that ``fixed effects" is used in the econometric sense, controlling for unobserved heterogeneity\textemdash in this case\textemdash at the unit level. Note also that details on the basic specification of the MSM can be found in Appendix \ref{appendix: msm more math}.} This model is ideal to apply when time-constant unmeasured confounding is likely present\textemdash if units have differing baseline probabilities of treatment due to difficult-to-measure traits/features, sequential ignorability may not be met, as is likely the case here (\cite{blackwell_adjusting_2021}). As Blackwell and Yamauchi highlight, the MSM with fixed effects requires restrictions beyond the typical MSM case. They concentrate on truncated MSMs, which focus on a treatment history of fixed length rather than the entire treatment history. 

\subsection{MSM with fixed effects in the IPTW} \label{sec: MSM with fixed effects in the IPTW}

In implementing the MSM, I fit the following binary pooled logistic outcome model with weights calculated via IPTW with fixed effects:
\begin{align}
   \text{logit Pr(}AnyListings_{ij} = 1) =  \psi_0 + \psi_1ACI_{ij} 
   \label{eq: binary outcome no time}
\end{align}
using weights $w_{ij}^*$ generated from IPTW, where logit is the natural logarithm of the odds, $log\left( \frac{p}{1-p} \right)$. Here, $ACI_{ij}$ is a binary variable that captures whether any corruption indictments (hence ``ACI") were made in month $j$ for prefecture $i$. $AnyListings_{ij}$ is a binary variable representing whether the prefecture had any properties sold as listings in a given month/year $j$.\footnote{In the two-way fixed effects estimation mentioned, the continuous versions of these variables were used. Note also that tenders are excluded from this analysis, as they represent a very small proportion of the data and are only used in limited, more tightly regulated settings (\cite{cai_chinas_2013}; \cite{wu_primary_2020}).}

Following the literature (e.g., \textcite{robins_marginal_2000}, \textcite{imai_robust_2015}, and \textcite{blackwell_framework_2013}), the setup of the MSM is as follows: suppose we observe $N$ units indexed by $i = 1, 2, \dots , N$ at each of $J$ time periods. At each time period $j = 1, 2,\dots, J$, we observe the time-dependent treatment variable $T_{ij}$ and the time-varying covariates $X_{ij}$ that could be impacted by past treatments. Define treatment $T_{ij}$ to be a binary treatment variable where $T_{ij} =1 $ implies that unit $i$ is treated in period $j$. Conversely, $T_{ij} =0$ suggests that unit $i$ is not treated in period $j$. Assume that $X_{ij}$ is already realized before the treatment at time $j$ and is therefore not impacted by the treatment in period $j$, $T_{ij}$.

The observed treatment history for each unit $i$ up to time $j$ is represented by $\overline{T} _{ij} = \{T_{i1}, T_{i2}, \dots, T_{ij}\}$ while the observed time-invarying covariate history for unit $i$ up to time $j$ is captured as $\overline{X} _{ij} = \{X_{i1}, X_{i2}, \dots, X_{ij}\}$. The set of possible treatment and covariate values at time $j$ is $\overline{\mathcal{T}} _j$ and $\overline{\mathcal{X}}_j$, respectively. The outcome of interest $Y_{ij}$ is observed at the time period $j$. In the basic MSM, this outcome is impacted by the entire treatment history up until $j$, and $Y_{ij}(\overline{t}_j)$ is thus used to denote the potential value of the outcome variable for unit $i$ at the time period $j$ under the treatment history $\overline{T}_{ij} = \overline{t}_j$, where $\overline{t}_j \in \mathcal{\overline{T}}_j$. 

Yet, as Blackwell and Yamauchi highlight, the MSM with fixed effects requires restrictions beyond the typical MSM case. They concentrate on truncated MSMs, which focus on a treatment history of fixed length rather than the entire treatment history. This truncated MSM consistently estimates $\mathbb{E}\{Y_{ij}(\overline{t}_{[j-k, j]})\}$ where $\overline{t}_{[j-k, j]} = \{t_{j-k},{t}_{j-(k-1)}, \dots, t_j \}$ and $k$ is a fixed number of the last $k$ periods.

For any unit $i$ and time $j$, only one of these potential outcomes can be observed, as a unit cannot follow multiple treatment paths over the same time window. As described further below, the consistency assumption is therefore used to connect the potential outcome to the observed outcome; it states, in effect, that the observed outcome and the potential outcome are the same for the observed history. In this framework, $X_{i,[j-k,j]}(\overline{t}_{[j-k-1,j-1]})$ reflects the potential values of covariates for unit $i$ at each time period $j$ given the relevant treatment history up to time $j-1$. The assumptions of the MSM with fixed effects in the IPTW are as follows.

\medskip

\begin{assumption}{1}{Consistency}
\assumlabel{MSM Assump 1}    
Consistency states that $Y_{ij} = Y_{ij}(\overline{T}_{i,[j-k,j]})$. These ``shorter" potential outcomes can be defined as $Y_{ij}(\overline{t}_{[j-k, j]}) \equiv Y_{ij}(\overline{T}_{i, j-k-1}, \overline{t}_{[j-k, j]})$ so that the treatment history before $k$ lags acts ``more like a baseline confounder" (\cite{blackwell_effect_2024}, p. 10). Implicit in this definition of consistency is that the treatment history can impact the outcome via the history of the time-varying covariates. 
\end{assumption}

\medskip

\begin{assumption}{2}{Positivity}
\assumlabel{MSM Assump 2}
This assumption states that the conditional probability of treatment assignment is between zero and one, exclusive, for each time period. 
\end{assumption}

\medskip

\begin{assumption}{3}{Sequential ignorability}
\assumlabel{MSM Assump 3}
Let $\alpha_i$ be an unmeasured, time-constant random variable. For all $i$, $j$, and $\overline{T}_{ij}$, $Y_{ij}(\overline{t} _j) \perp \!\!\! \perp T_{ij} \ | \ \overline{T}_{i,j-1} = \overline{t} _{j-1}, \overline{X}_{ij} = \overline{x} _j, \alpha_i.$
This version of sequential ignorability effectively states that conditional on unit-specific effects, treatment history, and covariate history, treatment is randomized with respect to covariates and the outcome. This assumption implicitly allows for both time-varying confounding by measured covariates $\overline{X}_{ij}$ and time-invariant confounding by measured and unmeasured covariates, as is captured by $\alpha_i$. Further, as highlighted by Blackwell and Yamauchi, the requirements of sequential ignorability extend not only to the treatments of interest in the MSM, but to the potential outcomes for the entire treatment history, applying to the MSM with fixed effects in the IPTW model for truncated treatment histories, $\mathbb{E}(Y_{i}(\overline{t}_{[J-k, J]}))$.\footnote{Blackwell and Yamauchi then posit sampling assumptions regarding asymptotics and across/within-unit dependence to nonparametrically identify the mean of the potential outcomes in this structure. As I am focused only on parametric identification, I do not delve into their full theoretical framework but note that because I have a sufficiently large $J$ ($J = 65$), mean potential outcomes can be consistently estimated under fixed $J$ in this case.}
\end{assumption}

Weights for unit $i$ under hypothetical treatment history $\overline{t}_{[j-k, j]}$ are estimated as
\begin{align}
    w_{ij}^*(\overline{t}_{[j-k, j]}, \overline{X}_{i,[j-k, j]}(\overline{t} _{[j-k-1, j-1]}), \alpha_i) & = \frac{Pr(\overline{T}_{i,[j-k, j]} = \overline{t}_{[j-k, j]} \ | \ \overline{T}_{i,[j-k-1, j-1]})}{Pr(\overline{T}_{i,[j-k, j]} = \overline{t}_{[j-k, j]} \ | \ \overline{X}_{i,[j-k, j]}(\overline{t} _{[j-k-1, j-1]}), \alpha_i)} \nonumber \\ 
    & = \prod_{s =j-k}^j  \frac{Pr(T_{is} = {t}_{is}\ | \ \overline{T}_{i,s-1} = \overline{t} _{s-1})} {Pr(T_{is} = {t}_{is} \ | \ \overline{T}_{i,s-1} = \overline{t} _{s-1},  \overline{X}_{is}(\overline{t} _{s-1}), \alpha_i)}.
    \label{stabilized_weights FE}
\end{align}
Note that weights are stabilized, with the numerator representing the baseline probability of receiving the treatment history, as estimated by a model with no covariates.\footnote{Note that unstabilized weights have 1 in the numerator.}

As these weights $w^*$ are unknown in an observational study, they must be estimated; following the MSM literature, this usually entails specifying a parametric model to estimate the propensity score, ``the conditional probability of assignment to a particular treatment given a vector of observed covariates" (\cite{rosenbaum_central_1983}, p. 41). Several types of MLE estimators can be used, and I focus on the propensity score behavior when using a pooled logistic sigmoid regression, as is standard in the literature for binary treatments (\cite{imai_robust_2015}; \cite{thoemmes_primer_2016}; \cite{hernan_causal_2020}). I omit further technical details of this specification here but refer the reader to Appendix \ref{msm with FE}. 

In my estimation, the treatment variable $T_{ij}$ is $ACI_{ij}$, a binary indicator of whether there were any corruption indictments in the given prefecture $i$ at time (month, year) $j$. To estimate the probability of unit $i$ receiving a given treatment in a given period $j$, I estimate numerator and denominator models, per (\ref{stabilized_weights FE}). The numerator and denominator models are each calculated using a pooled logistic model that treats each prefecture-month as an observation, generating the weight for each unit.

\section{Results} \label{sec: results}

Once the weights are calculated as detailed above through IPTW, treatment effects can now be estimated using an outcome model incorporating the weights. The basic outcome model detailed in equation (\ref{eq: binary outcome no time}) is used, estimating the impact of having any corruption indictments in a prefecture in month/year $j$ on whether there are any listings in the given prefecture. In line with \textcite{robins_marginal_2000}, covariates are not included in this outcome model to capture the causal effect.

I highlight a few final technical details. First, the most extreme weights are removed via truncation, as is standard in the literature (\cite{cole_constructing_2008}; \cite{thoemmes_primer_2016}; \cite{chesnaye_introduction_2022}); truncation at the $1^{\text{st}}$/$99^{\text{th}}$ percentiles is used in this analysis.\footnote{Results with aggressive truncation are largely similar and are included in Appendix \ref{appendix binary aggressive truncation}.} As bootstrap standard errors are generally recommended in the literature (see, for instance, \cite{thoemmes_primer_2016}; \cite{cole_constructing_2008}; \cite{yiu_joint_2022}; \cite{robins_new_1986}; \cite{robins_marginal_2000}; and \cite{blackwell_framework_2013}), the standard errors in this paper are bootstrapped with clustering over 500 replications (a ``pairs clustered bootstrap"). 

\begin{figure}[H]
    \centering
    \caption{Covariate Balance: MSM with Fixed Effects, binary treatment}
    \includegraphics[width=1\linewidth]{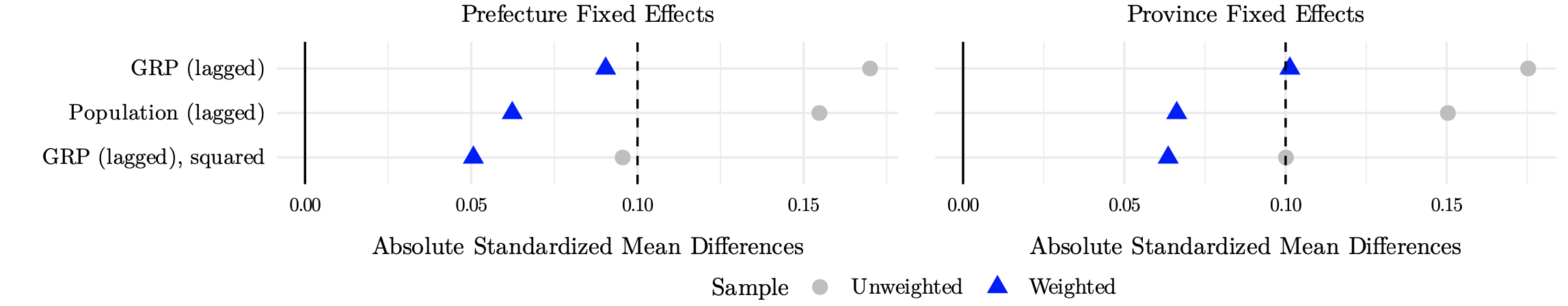}
    \label{fig:MSM Covar Balance FE binary}
    \caption*{\raggedright \footnotesize Note: The 0.1 threshold is marked with the dotted black line.}
\end{figure}

Specifications with prefecture-level fixed effects (column (3)) and province-level fixed effects (column (4)) are conducted in Table \ref{tab: binary MSM outcome model no time outcome}. In addition to these unit-level fixed effects, several time-varying covariates were experimented with, and the selected model has the best covariate balance. As such, the final model has lagged GRP and its square, as well as the lag of population. The standard mean differences before and after weighting are plotted, as shown in Figure \ref{fig:MSM Covar Balance FE binary}.\footnote{As highlighted in the literature (for instance, \cite{chesnaye_introduction_2022} and \cite{zhu_boosting_2015}), the standardized differences should be less than 0.10 for the weighted sample for all characteristics/covariates, although 0 is the ultimate target.}

\vspace{3mm}

Here, weighting significantly improves the standardized mean differences, particularly in the presence of prefecture fixed effects (Panel A). Now, all covariates have mean differences below 0.1, confirming that the model is balanced with respect to observed confounders. In Panel B, the standardized mean difference is slightly above the threshold for the lag of GRP, measuring 0.1013. Further checks of covariate balance and evaluation of the role of outliers are included in Appendix \ref{appendix: further details on covar balance}. The results table is below.

\vspace{3mm}
\begin{table}[H] 
\centering
\caption{Presence of Corruption Impacting $AnyListings$, Regular Truncation}
\label{tab: binary MSM outcome model no time outcome}
\small
\begin{threeparttable}
\setlength{\tabcolsep}{10pt} 
\renewcommand{\arraystretch}{1.3} 
\begin{tabular}{l*{4}{c}}
\hline \hline
& \multicolumn{4}{c}{AnyListings} \\
\cmidrule(lr){2-5}
                    & (1) & (2) & (3) & (4) \\
\midrule
ACI & 0.0759 & -0.0069 & -0.1135* & -0.0588  \\
& (0.0792) & (0.0820) & (0.0490) & (0.0536) \\
\midrule
Weights & No & Yes & Yes & Yes \\
P-value & 0.623 & 0.933 & 0.021 & 0.276 \\

Fixed Effects & None & None & Prefecture-level & Province-level \\
Number of Prefectures & 343 & 343 & 346 & 346 \\
Number of Provinces & 31 & 31 & 31 & 31 \\
Number of Observations & 17,852 & 17,449 & 16,463 & 16,455 \\
Effective Sample Size & 100\% & 98.31\% & 95.03\% & 96.78\% \\
\hline \hline
\end{tabular}
\end{threeparttable}
\caption*{\footnotesize Note: Standard errors are in parentheses (\sym{*} \(p<0.05\), \sym{**} \(p<0.01\), \sym{***} \(p<0.001\)); they are estimated via a pairs clustered bootstrap, clustered at the prefecture-level, over 500 replications. As is highlighted in the table, columns differ by fixed effects included. Unweighted estimates (column (1)) do not have a causal interpretation and are shown for comparison purposes only. Column (2) illustrates the basic MSM with no fixed effects, although it does include several other time-invariant covariates not reported in the table above; it is offered for comparison and should not be interpreted causally. The intercept is also not reported for all columns. The above results are for weights truncated at the $1^{\text{st}}$ and $99^{\text{th}}$ percentiles. Recall that these results are in log-odds as the logistic regression is used.}
\end{table}

Column (3), with its balanced covariates, can be interpreted causally, suggesting that a prefecture having corruption indictments causes a 1.16 percentage point decrease in the probability of having any listings (95\% CI: (-0.0223, -.000909)), calculated via incremental effects. Column (4)'s coefficient is statistically insignificant. If interpreted causally using incremental effects, it would imply that a prefecture having any corruption indictments leads to a 0.584 percentage point decline in the probability of having any listings, but this effect is not statistically different from zero. Province-level fixed effects in the IPTW model absorb province-level time-invariant variation to test whether prefecture-level effects persist in the presence of these additional controls. This statistically insignificant result suggests that treatment $T_{ij}$, which remains at the prefecture level, indeed has a different correlation with the province-level fixed effect than the prefecture fixed effects\textemdash that is, that $T_{ij}$ is more a correlate of the prefecture-level unobservables than the province-level unobservables.  

\subsection{Contextualizing column (3) of Table \ref{tab: binary MSM outcome model no time outcome}} \label{MSM contextualizing results}

While the coefficient of column (3) is statistically significant, a natural question is whether it is likewise economically/practically significant, and the sale timeline becomes important here. As highlighted in Section \ref{Anticorruption campaign overview}, areas to be inspected by the Central Inspection team were only notified 10 days before their arrival in the province, and officials therefore (in theory) had little opportunity to change their behavior in advance. Further, the decision of whether a property is diverted to auction or listing is made quite close to the sale date: after the date for the property sale nears and the price has been calculated, the property is then passed to the local land bureau, which inspects the property, decides the auction style, and conducts the sale (\cite{cai_build_2017}; \cite{cai_chinas_2013}; see Sections  \ref{auction style} and \ref{Pathway of Corruption in Land Sale} for further details). 

As described in Section \ref{sec: background}, most corruption waves are 2 full months, with indictments often following in the month or two after. Under this timescale, if officials do indeed divert properties towards auctions as a result of the anticorruption campaign, such behavior may occur not only in the month of corruption indictments but in those just before and after. In essence, I expect that the true trigger of their behavioral change is the investigations themselves. I anticipate that officials divert properties to auctions to either a) avoid being detected for corruption or b) avoid the appearance of such corruption. The anticorruption campaign's true causal power lies in its investigations, and indictments are only a symptom/measure of the severity of these anticorruption investigations in each prefecture. Thus, given the amount of time between deciding what sale method to use for each property and the signing date, I expect that if there is a drop in the probability of having any listings when a prefecture is being investigated, it will appear sometime around the time of corruption indictments, not necessarily $only$ in the month(s) of the indictments. 

This hypothesis was tested with several versions of the standard outcome model (equation (\ref{eq: binary outcome no time})) wherein $AnyListings_{ij}$ is converted to leads to capture future months ($AnyListings_{i,j+1}$, $AnyListings_{i,j+2}$, $\dots$) and lags to reflect past months ($AnyListings_{i,j-1}$, $AnyListings_{i,j-2}$, $\dots$), all while $ACI_{ij}$ remains as is.\footnote{Note that changing the outcome variable like this is possible in an MSM because the weights are only calibrated with the specific treatment; any binary outcome can be estimated with this logistic specification, as long as the treatment used in IPTW is used as the treatment in the outcome model, such that the weights correctly balance covariates.} The coefficients on leads 1-3 were highly statistically significant and of a similar magnitude/direction as that in column (3)\textemdash in probability terms, no lead exceeded a decrease of 1.87 percentage points on the probability of having any listings\textemdash while lags were significant and of similar magnitude only for a one-period lag.\footnote{Given the possibility that the standard weights for $ACI_{ij}$ do not fully adjust for post-treatment confounders, I examined whether these results persist (and covariate balance is maintained) when the model is re-estimated using weights from period $j+1$ while maintaining treatment timing at $j$; this specification allows me to test whether the confounder adjustment needs to be ``future-aligned" for the leads\textemdash that is, whether adjustment for confounders up to $j+1$ (rather than $j$) meaningfully alters the effect of $ACI_{ij}$ on the outcomes. I find that the result is stable, suggesting that the original weights adjusted to time $j$ already account for relevant confounders whose effects may persist into $j+1$ and that the treatment effect is robust to any minor temporal misalignment in confounder adjustment timing. } These coefficient estimates are included in Table \ref{tab: binary MSM outcome model no time, varied ACI outcome} below.

\vspace{3mm}
\begin{table}[H] 
\centering
\caption{Presence of corruption impacting $AnyListings_{i,[j-1, j+3]}$, regular truncation}
\label{tab: binary MSM outcome model no time, varied ACI outcome}
\small
\begin{threeparttable}
\setlength{\tabcolsep}{0pt} 
\renewcommand{\arraystretch}{1.3} 
\begin{tabular}{l*{5}{c}}
\hline \hline
                    & AnyListings$_{ij}$ & AnyListings$_{i,j-1}$ & AnyListings$_{i,j+1}$ & AnyListings$_{i,j+2}$ & AnyListings$_{i,j+3}$ \\
                    & (1) & (2) & (3) & (4) & (5) \\
\midrule
ACI & -0.1135* & -0.1495* & -0.1782*** & -0.1688** & -0.1402* \\
& (0.0490) & (0.0605) & (0.0500) & (0.0514) & (0.0555) \\

\midrule
Weights & Yes & Yes & Yes & Yes & Yes \\
P-value & 0.0206 & 0.0134 & 0.00036 & 0.00102 & 0.0114\\
Percentage Point Change & -1.160 pp & -1.539 pp & -1.868 pp & -1.758 pp & -1.454 pp \\
Number of Prefectures & 346 & 346 & 346 & 346 & 346 \\
Number of Provinces & 31 & 31 & 31 & 31 & 31 \\
Number of Observations & 16,463 & 16,502 & 16,180 & 15,890 & 15,611 \\
Effective Sample Size & 95.03\% & 95.03\%& 95.03\% & 95.03\% & 95.03\% \\
\hline \hline
\end{tabular}
\end{threeparttable}
\caption*{\footnotesize Note: Standard errors are in parentheses (\sym{*} \(p<0.05\), \sym{**} \(p<0.01\), \sym{***} \(p<0.001\)); they are estimated via a pairs clustered bootstrap, clustered at the prefecture-level, over 500 replications. All columns have fixed effects at the prefecture level. The intercept is not reported for all columns. The above results are for weights truncated at the $1^{\text{st}}$ and $99^{\text{th}}$ percentiles. Percentage point changes are calculated via marginal effects.}
\end{table}

Contextually, 89.31\% of prefectures in any given month have at least one listing transaction. The mean number of properties sold (of both auction and listing types) in each prefecture per month is 7, while the maximum is 240 properties. For listings alone, the mean is 5 properties per prefecture per month, and the maximum is 240. Thus, a prefecture actually moving from having any listings to no listings (as a change in $AnyListings$ from 1 to 0 would indicate) would mean that the prefecture changes the sale type of the majority of its properties\textemdash and considering the time scale, this would be quite a strong effect: for there to be a statistically significant drop in the likelihood of having any listings by the time corruption indictments officially occur, prefectures must seriously shift their behavior very rapidly right at/after the start of the local Central Inspection team investigations. Thus, not only is the overall decrease accompanying corruption indictments magnified, as there are effectively 5 months ($j-1$, $j$, $j+1, j+2,$ and $j+3$) with a statistically significant negative coefficient on $ACI$\textemdash each with approximately a 1.56 percentage point decline in the probability of having any listings, on average\textemdash but this spillover behavior means that the original coefficient estimated in equation (\ref{eq: binary outcome no time}) may be underestimated.  

Therefore, ultimately, while the 1.16 percentage point decline found in column (3) seems small, when compounded across the 5 months, this amounts to 7.78 percentage points (95\% CI: (-0.1113, -0.0443)). Indeed, the cumulative effect is still relatively small, but if a prefecture does substitute all of its would-be listings for auctions, this action could have a critical impact on the distribution of developers, allowing other players to enter previously closed markets and fostering a more competitive market environment long-term. 

\subsection{Sensitivity checks}
I next discuss whether the assumptions are met and briefly highlight some further sensitivity checks conducted and their (high-level) results; full details and any relevant graphs can be found in Appendix \ref{appendix: technical details on sensitivity checks}.

First, for consistency to hold, the value of $Y_{ij}$ when exposed to treatment $T_{ij}$ will be the same 1) no matter what mechanism is used to assign treatment $T_{ij}$ to unit $i$ and 2) no matter what treatments the other units receive. Both of these statements should hold for all units $i$ (\cite{rubin_comment_1986}). I expect there are likely some anticipation effects wherein unit $r$ sees that unit $i$ is being treated in time $j$ and adapts its behavior, effectively anticipating, and I discuss this further in light of the sensitivity test findings highlighted below. 

Positivity is met in this case because in any given period $j$, there is a nonzero probability of treatment (\cite{petersen_diagnosing_2012}). In the data, all prefectures experience treatment at least once, if not more. Further, some prefectures experience treatment in multiple sequential periods, so even after being treated in $j$, there is still the possibility for treatment in period $j+1$. Conversely, no prefecture is certain to experience treatment in any period. 

For sequential ignorability to hold under the MSM with fixed effects in the IPTW model, treatment should be effectively randomized with respect to future covariates and the outcome, conditional on the past and time-invariant features of unit $i$. In effect, we must examine if there is still ``unmeasured confounding" after controlling for time-invariant confounding via the unit fixed effects. I expect that these fixed effects capture much (if not all) of the confounding that influences treatment\textemdash for instance, the two biggest threats to sequential ignorability, suspected initial corruption level and ties of provincial leaders to Politburo Standing Committee members\textemdash because these factors are time-invariant.

I conduct several sensitivity analyses and robustness checks. I first perform \textcite{robins_association_1999}'s sensitivity check, as implemented by \textcite{ko_estimating_2003}, to investigate the sensitivity of estimates to the presence of unmeasured confounders. In essence, if sequential ignorability holds per (\ref{sequential ignorability}), we have that $\mathbb{E}\{Y_{ij}(\overline{t}_j) \ | \ \overline{X}_{ij}, \overline{T}_{i, t-1} = \overline{t}_{i,j-1}, T_{ij} \}$ = $\mathbb{E}\{Y_{ij}(\overline{t}_j) \ | \ \overline{X}_{ij}, \overline{T}_{i, t-1} = \overline{t}_{i,j-1}\}$ since sequential ignorability implies that $Y_{ij}(\overline{t}_j)$ is mean-independent of $T_{ij}$ given the past $\overline{X}_{ij}, \overline{T}_{i, t-1}$. 
However, if unmeasured confounders are present, this equality no longer holds. The function 
\begin{align}
    q_{ij}(t_j, t_j') = \mathbb{E}\{Y_{ij}(\overline{t}_j) \ | \ \overline{X}_{ij}, \overline{T}_{i, t-1} = \overline{t}_{j-1}, T_{ij} = t_j\} - \mathbb{E}\{Y_{ij}(\overline{t}_j) \ | \ \overline{X}_{ij}, \overline{T}_{i, t-1} = \overline{t}_{j-1}, T_{ij} = t_j'\}
\end{align}
will be nonzero when $t_j \neq t_j'$, where $t_j'$ is an alternate treatment (\cite{ko_estimating_2003}); $q_{ij}(t_j, t_j')$ is the natural measure of the ``magnitude of noncomparability with respect to the mean of" $Y_{ij}(\overline{t}_j)$ of the two groups due to unmeasured confounding (\cite{robins_association_1999}, p. 168).
\textcite{robins_association_1999} proposes a sensitivity analysis based on interpretable parameterizations of $q$. This is often selected as $q_{ij}(t_j, t_j') = \varphi\times(t_j - t_j')$ where $t_j' = 1- t_j$, and $\varphi$ is the expected difference between $Y_{ij}(\overline{t}_j)$ given $T_{ij} = t_j$ versus  $T_{ij} = t_j'$ (and conditional on $\overline{X}_{ij}, \overline{T}_{i, t-1}$). Under this definition, $\varphi = 0 $ corresponds to the assumption of no unmeasured confounders. Unmeasured confounders are marked by nonzero $\varphi$ values: if $\varphi > 0$, then on average, treatment is preferentially given to those units with higher $AnyListings$ counterfactuals $\{Y_{ij}(\overline{t}_j)\}$ (i.e., corruption indictments are given to more corrupt-appearing prefectures\textemdash that is, those where a disproportionate number of listings would occur in the absence of indictments), even after controlling for past treatment and measured covariate history. As I expect that treatment is preferentially given to those units with higher (unmeasured) suspected corruption, I focus primarily on positive values of $\varphi$ in my sensitivity analysis. 

In essence, when $\varphi > 0$, the log-odds of the estimated treatment effect drop consistently, suggesting that the estimates derived under sequential ignorability as reported in column (3) are conservative: they underestimate the true effect if unmeasured confounding exists. The same is true for those estimates in column (4). A further discussion (with plots of $\varphi$) is included in Appendix \ref{sensitivity check sequential ignorability}.

For positivity, I investigate the overlap of the propensity score distribution between the treated and untreated groups for the untruncated weights, as well as \textcite{petersen_diagnosing_2012}'s version of the parametric bootstrap which is designed to detect positivity violations. I also check the effective sample size to ensure that the weights are not dominated by a few extreme values; in all three cases, I do not find any evidence of positivity violations. 

Beyond evaluating propensity score overlap, ensuring covariate balance, and checking the ESS, the literature offers no further consistency-focused sensitivity analyses that can be feasibly implemented here. Thus, while the previous sensitivity checks and statistics did not elucidate any blatant consistency violations, I suspect that consistency may nevertheless be violated and discuss the implications of this on the results. If there is in fact spillover and other prefectures witnessing the anticorruption campaign preemptively reduce their corruption\textemdash and the probability of having $AnyListings$ in turn declines\textemdash then this behavior biases the magnitude of the coefficient of interest towards zero. Then, in the presence of a true control group (i.e., where there is no anticipation), the magnitude of the coefficient would be larger and maintain its negative sign, indicating a stronger negative causal effect. Thus, even if consistency is violated by spillover, the presence of the statistically significant negative coefficient on $AnyListings_{ij}$ in column (3) suggests that the effect would remain if consistency were fully met\textemdash and the true coefficient would be more negative. 

\subsection{Multiple treatments in the outcome model}

As the MSM also offers insight into behavior resulting from a certain sequence of treatments, I estimate a series of regressions to discern the impact of having sequential periods of corruption indictments on the likelihood of having any listings. 

I begin with equation (\ref{eq: binary outcome no time, lag 1}), which regresses $AnyListings_{ij}$ against not only $ACI_{ij}$, but also against a one period lag, $ACI_{i,j-1}$, as specified:
\begin{align}
   \text{logit Pr(}AnyListings_{ij} = 1) =  \psi_0 + \psi_1ACI_{ij} + \psi_2ACI_{i,j-1}.
   \label{eq: binary outcome no time, lag 1}
\end{align}
This equation allows me to estimate the effect of treating two periods in sequence on $AnyListings_{ij}$, and these results are shown in column (2) of Table \ref{tab: binary MSM outcome model no time outcome, sequential treatments}. I likewise estimate regressions including $ACI_{ij}$ and two lags ($ACI_{i,j-1}$ and $ACI_{i,j-2}$) in column (3), and three lags ($ACI_{i,j-1}$, $ ACI_{i,j-2},$ and $ACI_{i,j-3}$) in column (4). Regressing all three of these unique specifications allows me to compare coefficients on each lag across the results to understand the stability of the estimates. Table \ref{tab: frequency of seq treatment} lists the number of ``runs" and observations that experience consecutive treatment. A ``run" is an incidence of consecutive treatment, so for example, per the table, there are 352 incidences of a prefecture being treated in exactly two consecutive periods\textemdash and this is comprised of 712 individual monthly observations (as each two-period run yields two observations). 

I also estimate (\ref{eq: cumulative binary outcome no time}), wherein the variable of interest in each time $j$ is in effect the number of periods (up to $j$) with nonzero corruption indictments for a given prefecture $i$, as written below:
\begin{align}
   \text{logit Pr(}AnyListings_{ij} = 1) =  \psi_0 + \psi_1 \sum_{s =1}^jACI_{is}. 
   \label{eq: cumulative binary outcome no time}
\end{align}
This means, for instance, that if 5 months had nonzero corruption indictments in periods 1 through $j$, then $\sum_{s =1}^jACI_{is} = 5$.  Summary statistics on the frequency of cumulative treatment are included in Table \ref{tab: frequency of seq treatment} as well. While prior specifications such as (\ref{eq: binary outcome no time, lag 1}) offered insight into the impact of certain treatment sequences on $AnyListings_{ij}$, this one investigates the impact of the total number of treated periods regardless of sequence; the results are shown in column (5) of Table \ref{tab: binary MSM outcome model no time outcome, sequential treatments}.

\vspace{3mm}
\begin{table}[H] 
\centering
\caption{Frequency of sequential and cumulative treatment}
\label{tab: frequency of seq treatment}
\small
\begin{threeparttable}
\setlength{\tabcolsep}{10pt} 
\renewcommand{\arraystretch}{1.3} 
\begin{tabular}{lcc}
\hline \hline
Periods of treatment  \\
Consecutive & \multicolumn{1}{c}{Number of Runs} & \multicolumn{1}{c}{Number of Observations} \\
\cmidrule(lr){2-3}

\quad 2 & 356 & 712 \\
\quad 3 & 28 & 84 \\
\quad 4+ & 9 & 38 \\[1.5ex]
Cumulative & \multicolumn{1}{c}{Number of Prefectures} & \multicolumn{1}{c}{Percentage of Prefectures} \\
\cmidrule(lr){2-3}
\quad 1-2 & 27 & 7.8\% \\
\quad 3-4 & 99 & 28.6\% \\
\quad 5-6 & 129 & 37.3\% \\
\quad 7-9 & 68 & 19.7\% \\
\quad 10+ & 23 & 6.6\% \\
\hline \hline
\end{tabular}
\end{threeparttable}
\caption*{\footnotesize Note: The maximum number of consecutive periods treated is 5. The maximum number of cumulative treatments is 16, and the minimum is 1.}
\end{table}

\vspace{3mm}
\begin{table}[H] 
\centering
\caption{Presence of corruption impacting $AnyListings$ under sequential treatments}
\label{tab: binary MSM outcome model no time outcome, sequential treatments}
\small
\begin{threeparttable}
\setlength{\tabcolsep}{10pt} 
\renewcommand{\arraystretch}{1.3} 
\begin{tabular}{l*{5}{c}}
\hline \hline
& \multicolumn{5}{c}{AnyListings} \\
\cmidrule(lr){2-6}
                    & (1) & (2) & (3) & (4) & (5)\\
\midrule
ACI$_{ij}$ & -0.1135* & -0.0872 & -0.1009* & -0.1088*  \\
& (0.0490) &  (0.0478)& (0.0482) &  (0.0485) \\
ACI$_{i,j-1}$ & & -0.1921*** & -0.1665*** &  -0.1793*** \\
& & (0.0487) & (0.0470) & (0.0476) \\
ACI$_{i,j-2}$ & & & -0.1888*** & -0.1619*** \\
 & & & (0.0519) & (0.0496) \\
ACI$_{i,j-3}$ &  & & & -0.1976*** \\
 & & & & (0.0564) \\
Cumulative ACI$_{ij}$ & & & & & -0.0035 \\
 & & & & & (0.0181) \\
\hline \hline
\end{tabular}
\end{threeparttable}
\caption*{\footnotesize Note: Standard errors are in parentheses (\sym{*} \(p<0.05\), \sym{**} \(p<0.01\), \sym{***} \(p<0.001\)); they are estimated via a pairs clustered bootstrap, clustered at the prefecture-level, over 500 replications. All columns have weights, prefecture-level fixed effects, and an effective sample size (ESS) of 95.03\%, as the weights do not change across columns, only the versions of the treatments. Each column also has 16,463 observations total, from 343 prefectures and 31 provinces. The intercept is also not reported for all columns. The above results are for weights truncated at the $1^{\text{st}}$ and $99^{\text{th}}$ percentiles. Recall that these results are in log-odds as the logistic regression is used. Note that column (1) is the basic specification of equation (\ref{eq: binary outcome no time}), as is shown in Table \ref{tab: binary MSM outcome model no time outcome}'s column (3).}

\end{table}

As Table \ref{tab: binary MSM outcome model no time outcome, sequential treatments} illustrates, the simple specification of column (1) camouflages a more nuanced effect that is apparent when sequential treatments are examined. Specifically, treatments occurring in sequential periods magnify the drop in the likelihood of having any listings: when 4 periods ($j-3$, $j-2$, $j-1$, and $j$) are treated in sequence, the first period $j-3$ again has the most negative coefficient; there are then two slightly smaller but statistically significant negative coefficients for the second and third periods of treatment ($j-2$, $j-1$), and finally, for the last period, a negative but statistically significant coefficient. These specifications thus illustrate that corruption indictments in a prior period (like $j-3$, $j-2$, or $j-1$) have a stronger negative impact on current listings than corruption indictments in the same period. This result, like that in Table \ref{tab: binary MSM outcome model no time, varied ACI outcome}, is likely related to the time difference between officials deciding a property's sale method and the property actually being sold, as discussed in Section \ref{MSM contextualizing results}. In effect, it echoes the findings of Table \ref{tab: binary MSM outcome model no time, varied ACI outcome}\textemdash that there is often a delay between the period(s) of corruption indictments and the largest measured deterrent effect of the investigations/indictments. 

In terms of magnitude, again using incremental effects, a prefecture having corruption indictments in four sequential periods (the three prior periods and the current period), per column (4), leads to a 6.75 percentage point decrease in the probability of having any listings in that current period. In column (4), each of these coefficients has a 95\% confidence interval exclusive of zero; further, the coefficient on $ACI_{ij}$ is roughly similar in magnitude across specifications, dropping slightly in column (2) before rising again in columns (3) and (4).\footnote{Note that the p-value for $ACI_{ij}$  in column (2) is slightly above 0.05 and is thus not statistically significant.}

Perhaps most interesting is these results in comparison to the coefficient in column (5), which regresses $AnyListings$ on the cumulative count of the number of periods that a unit experienced corruption indictments up to time $j$. Interestingly, this coefficient is very close to 0 with a comparatively large standard error, suggesting it is not the number of corruption indictments but the chronology of them that most impacts the presence of listings. This result suggests that either 1) corruption investigations and indictments provide such a strong deterrent to listings\textemdash as listings give the appearance of corruption, even if there is actually no corruption\textemdash that officials avoid them around the timing of investigations but return to them, when reasonable, afterwards. Alternatively, 2) if most incidences of listings are in fact due to corruption, officials return to listings in the absence of the investigations, which would indicate a weak long-term deterrent effect of corruption investigations in the land sector. Regardless of the cause, these results highlight that the impact of corruption investigations on the likelihood of having any listings is more of a short-term result than a long-term one, as echoed by specifications from Table \ref{tab: binary MSM outcome model no time, varied ACI outcome}.

A set of sensitivity checks are again conducted to ensure that the results can be interpreted causally. The results are conceptually similar to the previous outcome specifications with prefecture-level fixed effects in the IPTW model (equation (\ref{eq: binary outcome no time}), for instance). As before, any violations of positivity, sequential ignorability, or the spillover dimension of consistency would bias the estimates towards zero, suggesting the estimated coefficients in column (4) are lower bounds for the true effect.\footnote{Appendix \ref{appendix: Sensitivity Checks multiple treatments}  includes the full sensitivity checks and their results. Results for the multiple treatments in the outcome model with province-level fixed effects can be found in Appendix \ref{appendix: Multiple treatments in the outcome model with province-level fixed effects}.}

\bigskip

\section{The effects of corruption indictments on price} \label{sec: Regressing corruption indictments on price}

In order to better understand how corruption impacts price in land sale, I also run two additional versions of the outcome model, as detailed below:  
\begin{align}
   \mathbb{E}\text{[log(}AvgPrice)_{ij}] =  \psi_0 + \psi_1ACI_{ij}.
   \label{eq: price MSM outcome}
\end{align}
In this specification, the outcome is the average price per square meter for (residential real estate) land sold in prefecture $i$ at time $j$. 
I also estimate a version of the above specification that uses $\text{log(}MedianPrice)_{ij}$ as the outcome, instead of the average price, but is otherwise identical. Both specifications use the balanced and stabilized weights utilized in Tables \ref{tab: binary MSM outcome model no time outcome} and \ref{tab: binary MSM outcome model no time outcome, sequential treatments}, as the treatment remains the same.\footnote{Note also that I use the log transformation of the outcome because there is a great disparity in prices per square meter: prices per square meter in highly sought-after areas like Beijing and Shanghai have average prices per square meter several times that of more rural prefectures. } The results, using both prefecture-level and province-level fixed effects in the IPTW model, are shown below in Table \ref{tab: Corruption indictments impacting sale price}. 

\vspace{3mm}
\begin{table}[H] 
\centering
\caption{Corruption indictments impacting sale price}
\label{tab: Corruption indictments impacting sale price}
\small
\begin{threeparttable}
\setlength{\tabcolsep}{6pt} 
\renewcommand{\arraystretch}{1.3} 
\begin{tabular}{l*{4}{c}}
\hline \hline
                    & \multicolumn{2}{c}{Log Mean Price/m$^2_{ij}$} & \multicolumn{2}{c}{Log Median Price/m$^2_{ij}$}\\
\cmidrule(lr){2-3} \cmidrule(lr){4-5} 

                    & (1) & (2) & (3) & (4)  \\
\midrule
ACI       &  0.0656***     & 0.0646*** &0.0733*** &  0.0710**   \\
                    & (0.0154)        & (0.0167) & (0.0166)  &(0.0182)      \\
\midrule
Weights             & Yes             & Yes  & Yes & Yes           \\
Fixed Effects       & Prefecture-level & Province-level &Prefecture-level  & Province-level \\
Number of Prefectures & 346     &   346   & 346  & 346           \\
Number of Provinces & 31       &  31     & 31  & 31            \\
Observations        & 16,559          & 16,551 & 16,559   & 16,551        \\
Effective Sample Size & 95.03\%  &  96.78\%   & 95.03\% & 96.78\%      \\
\hline \hline
\end{tabular}
\end{threeparttable}
\caption*{\footnotesize Note: Standard errors are in parentheses (\sym{*} \(p<0.05\), \sym{**} \(p<0.01\), \sym{***} \(p<0.001\)); they are estimated via a pairs clustered bootstrap, clustered at the prefecture-level, over 500 replications. The intercept is not reported for all columns. The above results are for weights truncated at the $1^{\text{st}}$ and $99^{\text{th}}$ percentiles.}
\end{table}

As is apparent above, a statistically significant positive coefficient persists across both prefecture-level and province-level fixed effects in the IPTW model, for both the logarithm of the mean and median price per square meter. Column (1) yields that having corruption indictments in a prefecture in a given month/year causes, on average, a 6.78\% increase (exp$(0.0656)-1$) in the mean price per square meter. Similarly, in terms of the median, column (3) suggests that having corruption indictments in a given month causes a 7.61\% increase (exp$(0.0733)-1$) in the median sale price per square meter for land sold in that prefecture. Columns (2) and (4) yield very similar effects, and given the covariate balance is only borderline met for province-level fixed effects in the IPTW, I focus primarily on the results from the prefecture-level fixed effects. 

The median price per square meter is around 1,400 yuan (US\$212). The mean, however, is much higher given the significant amount of stratification in underlying land demand/value: 2010-2016 was driven by the rapid urbanization of previously underdeveloped areas (such as Tier 3 and 4 cities), whose land prices per square meter were often extremely low. At the same time, land prices per square meter in places like Shanghai, Beijing, and Tianjin could be hundreds of times higher than those of more rural areas. Thus, while the mean is 2,200 yuan per square meter (US\$350), this camouflages a significant amount of underlying heterogeneity. The highest mean price per square meter sold, for instance, was in Shenzhen in June 2016 at 162,309.8 yuan per square meter (US\$24,655).\footnote{Note that while this ratio may seem extremely high, I checked the original data and verified its correctness. Given the extremely high demand in Shenzhen at the time, these were highly coveted properties in prime locations, and the sale prices were extremely high.}

Property sizes tend to be tens of thousands of square meters, if not more, so for the upper echelon of properties, a 6-7\% increase in price per square meter is extremely significant. Even in more rural areas, an increase of 6-7\% in the cost per square meter can cost developers hundreds of thousands to millions of additional yuan. These specifications thus suggest a non-negligible increase in the median and mean price per square meter when corruption indictments occur.\footnote{Given the ambiguity in the timing of indictments described in Section \ref{MSM contextualizing results}, I also sought to investigate whether there is an increase in the logarithm of the average price in neighboring time periods. I sought to test this hypothesis with several versions of the outcome model (equation (\ref{eq: price MSM outcome})) wherein $AvgPrice_{ij}$ is converted to leads to capture future months ($AvgPrice_{i,j+1}$, $AvgPrice_{i,j+2}$, $\dots$) and lags to reflect past months ($AvgPrice_{i,j-1}$, $AvgPrice_{i,j-2}$, $\dots$), all while $ACI_{ij}$ remains as is. To do this analysis, however, I need to ensure that the results persist when the model is re-estimated using weights from period $j+1$ while maintaining treatment timing at $j$. This examines whether the adjustment for confounders up to $j+1$ (rather than $j$) meaningfully alters the effect of $ACI_{ij}$ on the outcomes. I find coefficient instability across specifications, likely catalyzed by potential mediators (like sale method) that I do not control for. Instead, it suggests that time-varying confounders between $j+1$ and $j$\textemdash which are not fully captured by my controls of GRP and population\textemdash meaningfully impact both corruption investigations and price movements. Since future-aligned weights are often problematic because they adjust for post-treatment variables, I focus on the original specification while acknowledging that monthly spillover is possible but uncertain.}
Note also that part of this positive effect is likely due to a slightly higher proportion of auctions that accompany corruption indictments, as auctions are associated with higher prices in the literature. This coefficient is therefore more of a total effect (including this mediator of sale method) than a direct effect of only corruption indictments on the average sale price per square meter.\footnote{Another pathway through which corruption indictments could impact price, for example, is the floor area ratio (FAR). In the presence of corruption, the FAR of a property could be set low at the listing stage, deterring other possible bidders. Then, sometime after sale, the developer may bribe an official to raise the FAR, allowing the developer to build more square footage on the land. In the presence of the anticorruption campaign, officials may reject bribes, setting the FAR higher at the time of sale rather than adjusting it later. In doing so, the sale price would be higher.} 

Sensitivity checks are conducted (Appendix \ref{Sensitivity checks for MSM regressing corruption indictments on price}). While the checks reveal the results are sensitive to violations of sequential ignorability, the lack of obvious (uncontrolled for) time-varying confounders ameliorates this concern. The tests broadly suggest that the specifications are well-fit and do not have glaring violations of positivity. Again, the presence of anticipation threatens consistency, but further analysis (via other identification methods) would be needed to confirm whether this meaningfully changes the direction of the coefficients. 

\section{Conclusion} \label{sec: conclusion}

These results have important implications on the supply and demand dynamics underpinning residential real estate in a market highly susceptible to corruption\textemdash not only does this analysis show that the anticorruption campaign deters listings, but that prefectures investigated for corruption see a short-term increase in prices.

Analyzed in tandem, these results ultimately suggest two potential diverging conclusions about corruption in the Chinese land market. The first possibility is that the incentive to engage in corruption in the land market is so high that in the absence of an immediate credible threat of indictment, corrupt behavior persists, with land being diverted towards listings. In this lens, prices rise towards the level that the market demands when corruption is deterred. Ultimately, though, the anticorruption campaign fails to deter corruption beyond the period immediately surrounding investigation. 

The other pathway is that the anticorruption campaign creates so much fear of indictment that officials substitute away from actions that even have the \textit{appearance} of corruption\textemdash such as selling properties via listing. The departure of the Central Inspection team, then, allows normal behavior to return. In this view, it is not certain whether listings actually foster corruption, but they are definitely associated with the \textit{appearance} of corruption. Prices rise when corruption indictments occur\textemdash but this is counter to government desires to lower prices/price growth and stem the formation of a property bubble (as discussed in Section \ref{Pathway of Corruption in Land Sale}). In this world, the anticorruption campaign inadvertently impedes the government's goal of lowering land sale price growth, and the departure of the campaign means that the sale method can return to its equilibrium. 

At the nexus of these options is the question\textemdash does the campaign feed off of corruption as it is perceived or as it actually exists? I expect that the reality is a combination of both cases: in an environment with investigators dispatched to a province for a short period of time, on the lookout for any whiff of corruption, officials divert towards auctions to avoid trouble\textemdash but at the same time, it is more likely that those who do divert have something to hide. After all, a marked body of literature suggests that listings are hotbeds of corruption, so there are almost certainly some properties being (corruptly) diverted to listings pre-campaign. 

Simultaneously, anticorruption indictments cause higher prices, which goes directly in the face of the government's goal of moderating land sale prices.\footnote{I expect that part of this effect is driven by the shift towards auctions, which have a higher sale price on average. Yet, it cannot be determined whether prices increase beyond the level that the increase in competition causes.} If anything, one would expect the proportion of listings to rise slightly throughout this time period as officials worked to moderate land sale prices, which could sometimes skyrocket in auctions. Thus, in the absence of the anticorruption campaign, corruptly diverting ``hot" properties toward listings would be almost entirely camouflaged and would likely spike further. In this setting, the shift away from listings\textemdash especially apparent in the statistically significant binary case of $AnyListings$ and its leads/lags\textemdash suggests an extremely strong (temporary) shift in behavior, almost an over-reaction. 

I thus expect that the anticorruption campaign does deter actual corrupt behavior of having listings on favorable properties, but also impacts behavior for properties on the margin (i.e., those that could be sold as either auctions or listings). After all, those who have the best idea about whether the sale method selected for a certain property is defensible are those in the local land bureau who are intimately familiar with the government's motivations, aims, and the property sale landscape. The results indicate that these officials do change their sale method selection around the time of the campaign, and it is likely because a certain subset of properties \textit{could} be auctions\textemdash or \textit{should} reasonably be auctions. At the end of the day, each official conducts a utility calculation, weighting the desire to protect oneself with external factors and possible kickbacks, and this calculation may not include whatever broader aims the local government may have about keeping price growth in check. 

The statistically significant negative effect on $AnyListings$, however, does not persist past three leads, nor is the cumulative amount of periods with at least one corruption indictment significant. The shift towards auctions is thus short term. Yet, in the periods after corruption indictments, there is no evidence of an \textit{increase} in the probability of having any listings, as would likely be the case in the absence of the campaign (due to the government's desire to stem price growth). Thus, after the investigation team departs from a province, there is likely a simultaneous decrease in the corrupt diversion of listings to auctions and a general shift towards listings to stem price growth\textemdash these two factors offset each other and create the null effect observed in future periods. Yet, while this is my ultimate reasoning based on the results, the fact that so many factors intersect here means that this conclusion is far from certain. 

On its own, this paper's main implication is thus the clear causal effect of the anticorruption campaign on the probability of having any listings, providing strong evidence of the campaign causing a (short-term) substitution towards auctions. Future research should focus more on the relationship between corruption indictments and price, working to partial out the effect of the sale method itself, to determine whether the price post-corruption indictment is higher when controlling for sale method. If so, this would be a strong indicator that the anticorruption campaign is successful in combating corruption: even if the share of listings still rises after corruption indictments occur, a higher price would indicate that these listings are less susceptible to the kind of pervasive corruption that historically occurred.

\pagebreak

\begin{CJK*}{UTF8}{gbsn} 
\printbibliography
\end{CJK*} 

\pagebreak

\appendix
\section{Appendix}
\subsection{Pathway of corruption in land sale} \label{appendix: pathway of corruption in land sale}

While the State owns all urban land, much of the corruption occurring in land sale occurs at the local government and local land bureau levels, as the former holds the legal rights to dispose of land within its bounds while the latter executes these disposal rights (\cite{zhu_shadow_2012}; \cite{zhu_lessons_2015}). Indeed, there are multiple pathways and incentive structures that facilitate and encourage corruption in land finance\textemdash and many of them operate in opposite directions: officials often want land to be as valuable as possible, but at the same time, one of the most common forms of corruption involves auction manipulation, wherein land is sold to developers for less money. I discuss several possible forms of corruption in the land sale market\textemdash and how they fit together\textemdash below.

First, LUR revenue is a critical part of financing for local governments, sometimes comprising more than half its local budget revenue (\cite{zhou__2004}; \cite{zhou__2007}; \cite{liu_instrumental_2008}). Local governments raise fiscal revenue by selling LURs to private firms, and these costs are then absorbed by firms and passed on to buyers as higher real estate prices (\cite{wu_evaluating_2016}; \cite{liu_formation_2022}). Additionally, most of local governments' ability to finance debt was tied to its current and future land sale profits, heavily influencing borrowing capacity (\cite{pan_developing_2017}).\footnote{Note that these funds are borrowed via local government financing vehicles (LGFVs), also known as local financing platforms (LFPs), to circumvent rules against local government borrowing in China (\cite{pan_developing_2017}; \cite{azuma_examining_2011}).} Over the last two decades, these borrowed funds and land finance profits have been used to support local governments' massive infrastructure investments\textemdash which in turn drive up land sale prices and the local government's revenue potential (\cite{liu_formation_2022}; \cite{tao_land_2010}). This behavior helps create a self-propagating cycle wherein present and future value is created through higher prices, encouraging officials to sell LURs for more money and reap the benefits (\cite{lichtenberg_local_2009}). To that end, local governments ``hoard" land, only releasing limited quantities for sale so that they can sell larger amounts when demand (and therefore prices) rise in the future (\cite{du_land_2014}). 

Further, land transactions are intimately connected with officials' promotions: land is a large influencer on metrics used to assess political performance on career evaluations in two ways. First, local officials can have an incentive to raise land prices as much as possible to increase their opportunities for promotion (\cite{li_political_2005}). In the Chinese Communist Party hierarchy, promotion is typically determined by an official at the level immediately above an individual (the ``one-level-up" policy) (\cite{chen_busting_2019}). Public officials at the same level (for instance, across the province) are ``made to compete with each other under broadly similar economic conditions for promotion to the next level up" (\cite{chen_land_2016}). One of the key metrics of comparison is thus land revenue and its resulting economic growth, and in a system where those who fail to get promoted in their first term overwhelmingly stay in the same level position for the rest of their lives, competition is fierce (\cite{luo_chinas_2021}).\footnote{As reported by \textcite{chen_land_2016}, only 6.94\% of county-level officials ever get promoted, and of those who fail to be promoted in their first term, 90.53\% ``either stayed in the same position or transferred to a different locale of the same level and served in the same capacity" for the rest of their time in public service (p. 89).} Thus, officials have a strong incentive to utilize unsupervised windfall revenue in ways that would increase their chances of promotion: for instance, shifting this windfall revenue to infrastructural projects creates powerful signals of officials' primacy and ``achievements."\footnote{Note that local governments do not have to share all of this ``extra-budgetary income" with provincial-level governments (\cite{chen_land_2016}).} Particularly ostentatious projects are known as ``political achievements" in Chinese and are often strategically timed so as to catalyze the promotion of the official in charge (\cite{chen_land_2016}). Some local officials may even use land revenue ``directly to bribe their way to promotion," paying off their superior or directing the windfall towards the superior's projects (\cite{chen_land_2016}, p. 87). 

Alternatively, depending on the circumstances, officials may instead seek to sell as much land as possible, particularly for industrial uses, since the speedy growth of industry contributes not only to local revenue growth, but directly influences performance evaluations. Even if the official conducts an under-the-table deal and sells the land for a lower price than the market rate, he may be advantaged from doing so if the sale contributes meaningfully to the economic development of the region (\cite{liu_instrumental_2008}).\footnote{This practice was so prevalent in industrial settings that China implemented minimum prices for industrial land in 2007 (\cite{zhao_land_2025}; \cite{wan_research_2016}). Following this policy, the share of industrial land parcels ``priced below the standard decreased from 57.68\% to 10.06\%" (\cite{zhao_land_2025}, p. 1).}

A third and final incentive is for officials to curry favor with elites, who in turn influence their promotion. In China, members of the Politburo (the highest political body of the Chinese Communist Party) wield immense power in everything from policy to personnel appointments and are frequently approached by those seeking favor. In turn, their families tend to become ``extraordinarily wealthy," particularly their offspring, who have been dubbed ``princelings" in Chinese media (\cite{chen_busting_2019}, p. 186). Examining price data and Politburo connections between 1997-2016, \textcite{chen_busting_2019} find that firms connected to princelings (``princeling firms") receive a significant (55.4\%) discount on their land purchases, compared to the land parcels purchased by non-princeling firms in the same 500-meter area with the same usage, controlling for other transaction-level variables. They further find that the more powerful the princeling, the steeper the discount, and that those who have provided discounts are more likely to be promoted\textemdash ``with the likelihood of promotion increasing with the size of the price discount and the quantity (area) of land sold to the princeling firm" (p. 188).\footnote{\textcite{manso_are_2026} replicates \textcite{chen_busting_2019} and finds several issues in the data handling and reliability; these data issues make the magnitude of Chen and Kung's findings suspect. } 

Thus, while officials are generally incentivized to raise land prices, there are some circumstances where they are drawn to under-the-table deals, letting certain individual parcels of land sell for lower prices.

An additional aspect of corruption comes from manipulation of the floor area ratio (FAR, also known as the plot ratio), which is the ratio of the ``total floor area of the buildings on a certain plot of land to the total area of the plot of land" (\cite{zhu_shadow_2012}, p. 256). This means, for instance, that if the total area of the plot is 20,000 feet, and the floor area ratio is 3, the developer could build a building totaling 60,000 square meters (\cite{zhu_shadow_2012}; \cite{ma______2006}). The FAR is usually fixed by the local government before selling a property, but real estate developers will often buy plots of land with lower FARs and use their connections in the local government to increase the FAR before construction. As one CEO interviewed by \textcite{zhu_shadow_2012} detailed, even increasing the ratio slightly can lead to massive increases in profit given the large sizes of the plots and the high price per square meter that buyers pay. Many officials in urban planning departments have indeed been ``found taking bribes and adjusting the [FAR] for [real estate] companies" in the years leading up to Xi's anticorruption campaign (launched in 2012) (\cite{zhu_shadow_2012}, p. 256). 

Those who do not obtain official increases in FAR may take a more covert approach, knowing that detection of ``a one-tenth to two-tenths of a percentage point increase" in FAR is very difficult\textemdash and in the words of a developer, ```government supervision highly depends on the person in charge, or actually depends on how much money we send to them'" (\cite{zhu__2006}; qtd. in \cite{zhu_shadow_2012}, p. 256). 

Other corruption is tacit: developers want the development to be finished as quickly as possible, ideally with no delays, and they have historically bribed many officials who approve various stages of the construction, greasing the various approvals and permit processes to speed up the construction timeline. These bribes take several forms, from cash payments and gift cards to property discounts. For instance, studies like \textcite{fang_gradients_2019} have found that house prices paid by bureaucrat buyers were significantly lower than their non-bureaucrat peers, even when controlling for house, mortgage, and buyer characteristics; they also find evidence of a gradient of discount that depends on the official's rank, how essential the official's government agency is to real estate developers, and geography. 

Thus, corruption is endemic to the land finance and real estate markets of China, often having a tangible impact on the LUR price faced by developers and the final prices faced by buyers. 

\subsection{The anticorruption landscape and President Xi's anticorruption campaign} \label{sec: President Xi's Anticorruption campaign background}

I briefly highlight the anticorruption landscape prior to President Xi's campaign, then discuss the campaign's saliency.

\subsubsection{The anticorruption landscape pre-campaign}

Prior to President Xi's 2012 anticorruption campaign, corruption\textemdash in real estate and beyond\textemdash had been a persistent problem in China largely because of insufficiently equipped anticorruption agencies, which are designed as ``dual-leadership" and ``dual-track" (\cite{deng_national_2018}). In effect, the anticorruption agency is nominally led by both the local Party leaders and the superior anticorruption agency (``dual-leadership"), but the local Party leader ``substantively [controls] the nomination and promotion of the local anticorruption agency leadership," in addition to providing the funds, equipment, and staffing dictates for the local anticorruption agency (\cite{deng_national_2018}, p. 58). Then, for a suspected corruption case, investigation was first led by the superior anticorruption agency\textemdash one of the Party's Discipline Inspection Committees (DICs)\textemdash who would then decide whether to transfer the case to prosecutors. The prosecutors would then investigate and prosecute corruption cases. This interplay between the DICs and the judicial system results in a ``dual-track" system that is highly susceptible to influence from local Party leaders, who could sway whether the case was passed on to the prosecution and then whether it was ultimately prosecuted (\cite{deng_national_2018}). The local anticorruption agencies were thus heavily influenced and often ``captured" by the agencies they were supposed to supervise, deriving not only their enforcement resources and tools, but also their enforcement prerogative, from the local Party officials they were tasked with regulating (\cite{manion_corruption_2004}). 

Anticorruption campaigns instigated from the top down were largely performative: lasting roughly a year, these prior campaigns tended to rely on political pressure to activate the local anticorruption agencies and draw public attention, offering lighter sentences for self-surrendering (\cite{manion_corruption_2004}). Ultimately, each campaign resulted in very few cases actually being prosecuted and did not act as strong deterrents to corruption itself; in short, the underlying incentive structure remained. 

\subsubsection{President Xi's anticorruption campaign}

The anticorruption campaign is so salient in part because it not only seeks to detect current officials actively engaging in corrupt practices, but also those who have committed wrongs in the past. For example, a former researcher and deputy director at the Gongshu District Housing and Construction Bureau, Tan Zhaotu had been retired for six years when he was investigated and indicted in 2015; he was found guilty of embezzling more than 5 million yuan (roughly US\$700,000) of public property, in addition to illegally accepting more than 2 million yuan (nearly US\$300,000) in property on behalf of others, and was sentenced to 19 years in prison \begin{CJK*}{UTF8}{gbsn} 
(\cite{xu_30__2017}). In his trial, Tan urges other officials to learn from his example, urging them to ``do things cleanly and be a down-to-earth person" (trans. \cite{noauthor__2015}). \end{CJK*} 

Each of these investigations, then, is not only important in detecting active corruption, but in providing powerful signals to all officials: corruption, past and present, is no longer tolerated, and all those involved will be prosecuted. Indeed, in similar cases, other local party members were encouraged to attend the trials of their peers, further increasing the salience of the anticorruption campaign \begin{CJK*}{UTF8}{gbsn}(\cite{noauthor__2015-1}).\end{CJK*} Additionally, many of the indictments relied upon tips from the public; the risk of detection was thus high not only because transaction data was being scrutinized by newly-empowered enforcement officers, but because every individual was urged to come forward and report possible corrupt behavior\textemdash especially because knowledge of corruption, coupled with inaction, was incriminating.

\subsection{Data} \label{data cleaning appendix}
\subsubsection{Data cleaning}

As highlighted in Section \ref{sec: data}, I obtain 209,706 individual transaction records from 2010-2017; when limiting the date range to 2010-2016, inclusive\textemdash the date range of the anticorruption campaign\textemdash this yields 193,171 transaction records. I examine several indicators of incorrect duplicates, using the Ministry of Land and Resources' website itself to verify the corresponding plots of the transaction records. Specifically, I look for duplicates in the electronic lookup number, which is the database's main identifier/search key for unique land transactions. Having two ``separate" transactions with the same electronic lookup number is thus impossible; it can only be one transaction listed twice in error.

After filtering these incorrect duplicates out, I investigate the remaining plots with apparent duplicates, searching for patterns in transaction records and their analogs on the Land Transaction Monitoring System website. After identifying the transaction on the website, one can look at a transaction's original sale announcement (the ``Transfer Announcement") and confirm how many parcels of each size are up for sale in the area at the time. With this information, duplicate land transaction listings that refer to the same underlying parcel can be confirmed as erroneous duplicates. With this method, I identify that multiple transactions having the same contract number, which is often (but not always) unique for each listing, is also a signal of an incorrect duplicate. Duplicates in contract number particularly occurred when there were minor discrepancies between listings\textemdash not contradictions, but rather missing information. These omissions are non-essential details (e.g., greening ratio, floor area ratio) rather than key identifiers (e.g., buyer, area, quality). A second listing of the same parcel may thus exist simply to complete the initial record. I examine these transaction records on a case-by-case basis and remove those that are incorrect duplicates. 

On the whole, 3,157 correct duplicates for nearly 200,000 rows passes a commonsense check, reflecting roughly 1.5\% of transactions; the database confirms there are indeed many instances of developers buying neighboring plots at the same time with identical features. Contextually, developers likely have an incentive to purchase two identical neighboring plots of land, able to effectively build an economy of scale in construction and administration with two neighboring plots rather than two disparate ones (for instance, by saving on costs of land surveying, materials transport, inspection, etc.). 

\subsubsection{Defining the treatment and outcome} \label{appendix: defining treatment, outcome}

The outcome used in the fixed effects specifications is $ShareOfListings_{ij}$, which is the share of properties sold in prefecture $i$ at time $j$ that were listings. As such, if $ShareOfListings_{ij} = 1$, all properties sold in the specified prefecture in a given month/year were listings. Conversely, if $ShareOfListings_{ij} = 0$, none of the properties sold in the specified prefecture at the specified time were listings. 

$CorruptionIndictments_{ij}$ is the count of corruption cases for prefecture \( i \) at time \( j \). Note that this counter also works to reflect provincial-level indictments; specifically, if indicted officials of ranks 1-7 in Table \ref{tab:breakdown of officials' rank} have a province but not a prefecture associated with them, I distribute their indictment to prefectures within the province for the specified time period via the formula $weight = 11-rank$. Each of these indictments is weighted by importance, which is proportional to their rank (scaled from 1-10 by \cite{wang_how_2020}). For instance, an official of rank 3 (a provincial governor), would effectively have the weight of 7, meaning it is equivalent to 7 provincial-level officials being indicted. 

I define $CorruptionIndictments_{ij}$ in this way because the indictments of provincial-level officials provide powerful signals for those at the prefecture level, in proportion to their rank. I believe omitting the provincial-level officials from the individual prefecture corruption counts would severely bias the estimates. 

Note, however, that these fully numeric counts are only used for the fixed effects estimation; the MSM utilizes the binary version of $CorruptionIndictments_{ij}$ ($ACI_{ij}$) while the DiD's treatment is the beginning of the investigation wave. For the outcome, the MSM uses a binary version of $ShareOfListings_{ij}$, called $AnyListings_{ij}$, which is 0 if there are no listings in a prefecture in a given month and 1 if there are listings in a prefecture $i$ at time $j$. 

\pagebreak

\subsection{Fixed effects estimation} \label{sec: two-way FE}

To examine the relationship between corruption indictments and the share of listings in a prefecture, I first estimated a set of fixed effects regressions but ultimately found that the assumptions are not met, particularly for two-way fixed effects. I briefly describe the setup and results as follows. 

With a panel data set of $N$ units and $J$ time periods, the outcome variable $Y_{ij}$ is observed for each unit $i$ at time $j$. A basic specification with unit fixed effects is
\begin{align}
    Y_{ij} = \alpha_i + \beta X_{ij} + \epsilon_{ij},
    \label{eq: unit FE eq}
\end{align}
where $\alpha_i$ is a fixed (but unknown) intercept for unit $i$, and $\epsilon_{ij}$ is a disturbance term for unit $i$ at time $j$, with $\mathbb{E}(\epsilon_{ij}) = 0$ (\cite{imai_when_2019}). $X_{ij}$ is the treatment, reported at each time period. As \textcite{imai_when_2019} highlight, each fixed effect can be defined as $\alpha_{i} = h(\text{U}_i)$, where $\text{U}_i$ ``represents a vector of unobserved time-invariant confounders and $h(\cdot)$ is an arbitrary and unknown function" (p. 469). \textcite{imai_when_2019} also note that the strict exogeneity of $\epsilon_{ij}$ is assumed to identify $\beta$, and when this is the case, the least squares estimate of $\beta$ is obtained by regressing 
\begin{align*}
    \hat{\beta} = \operatorname*{argmin}\limits_{\beta} \sum_{i = 1}^{N} \sum_{j = 1}^{J} \{(Y_{ij} - \overline{Y}_{i}) -\beta (X_{ij} - \overline{X}_{i})\}^2,
\end{align*}
where $\overline{Y}_{i} = \sum_{j = 1}^{J} Y_{ij}/J$ and $\overline{X}_{i} = \sum_{j = 1}^{J} X_{ij}/J$ are unit-specific means. 

Two-way fixed effects can then be written as 
\begin{align}
    Y_{ij} = \alpha_i + \gamma_j + \beta X_{ij} + \epsilon_{ij},
    \label{eq: two-way FE eq}
\end{align}
where $\gamma_j$ represents time fixed effects. $\gamma_j = f(\text{V}_j)$, where $\text{V}_j$ reflects a vector of unobserved time-specific (but unit-invariant) unobserved confounders and $f(\cdot)$ is again an arbitrary, unknown function. Following \textcite{imai_use_2021}, the least squares estimate of $\beta$ can now be computed as 
\begin{align*}
    \hat{\beta} = \operatorname*{argmin}\limits_{\beta} \sum_{i = 1}^{N} \sum_{j = 1}^{J} [\{(Y_{ij} - \overline{Y}) - (\overline{Y_i} - \overline{Y}) - (\overline{Y_j}\ - \overline{Y})\} -\beta \{ (X_{ij} - \overline{X})-(\overline{X_i} - \overline{X}) - (\overline{X_j}\ - \overline{X})\}]^2,
\end{align*}
where $\overline{Y}_{j} = \sum_{i = 1}^{n} Y_{ij}/N$ and $\overline{X}_{j} = \sum_{i = 1}^{n} X_{ij}/N$ are time-specific means, and $\overline{Y} = \sum_{i = 1}^{N} \sum_{j = 1}^{J} Y_{ij}/NJ$ and $\overline{X} = \sum_{i = 1}^{N} \sum_{j = 1}^{J} X_{ij}/NJ$ are overall means. 

I estimate specifications with unit-level fixed effects per (\ref{eq: unit FE eq}) and two-way fixed effects per (\ref{eq: two-way FE eq}). In this case, my $X_{ij}$ is the number of corruption indictments in prefecture $i$ at time $j$ ($CorruptionIndictments_{ij}$) while $Y_{ij}$ is the share of listings sold in prefecture $i$ at time $j$ ($ShareOfListings_{ij}$).\footnote{Note that Appendix \ref{appendix: defining treatment, outcome} describes in detail how each of these variables are created/defined.} The results are shown in columns (1) and (2) of Table \ref{tab: FE results}. I also estimate a cumulative anticorruption measure to inspect how the sale method coefficient changed not just for anticorruption indictments in time $t$, but also in times $t-1$, $t-2$, etc. I include the results for a 6-month cumulative value below.\footnote{Note that the cumulative anticorruption indictment measures all have statistically significant negative coefficients for measures ranging from 1-6 months (when excluding time effects). I include only the 6-month cumulative value in the table.} Column (3) thus estimates 
\begin{align*}
    ShareOfListings_{ij} = \alpha_i + \beta_1 CorruptionIndictments_{ij} + CorruptionIndictments_{i,[j-1:j-6]} + \epsilon_{ij}.
\end{align*}
As described, $CorruptionIndictments_{p(t-1:t-6)}$ is the cumulative anticorruption indictment amount for prefecture  \( p \) from time $t-1$ to time $t-6$. Column (4) regresses this equation, but also includes time fixed effects ($\gamma_j$). Standard errors are clustered at the prefecture level for all results. 

\vspace{3mm}
\begin{table}[H]
    \centering
    \caption{\centering Fixed effects specifications relating corruption indictments and the share of listings}
    \renewcommand{\arraystretch}{1.25} 
    \label{tab: FE results}
    \resizebox{\textwidth}{!}{
\begin{tabular}{l*{6}{c}}
\hline\hline
                    & \multicolumn{4}{c}{Share of Listings} \\
                    \cmidrule(lr){2-5}
                    & (1) & (2) & (3) & (4) \\
\hline
Indictments    &       -0.012\sym{***}&      -0.001         &      -0.018\sym{***}&      -0.004         \\
                            &     (0.003)         &     (0.004)         &     (0.004)         &     (0.005)         \\

Indictments in the past 6 months     &                                        &                     &      -0.017\sym{***}&      -0.006         \\
                                        &                     &                     &     (0.004)         &     (0.004)         \\
\hline
Fixed Effects & Prefecture & Prefecture, time & Prefecture & Prefecture, time\\
Number of Prefectures&           298         &         298         &         298         &         298         \\
Observations        &          1541         &        1536         &        1541         &        1536         \\
Adjusted \(R^{2}\)      &       0.280         &       0.311         &       0.291         &       0.311         \\
\hline\hline
\end{tabular}
}
\caption*{\raggedright \footnotesize Standard errors are clustered at the prefecture level and are in parentheses: \sym{*} \(p<0.05\), \sym{**} \(p<0.01\), \sym{***} \(p<0.001\).}

\end{table}

As is apparent above, the coefficient is statistically significant and negative for both specifications with prefecture-level fixed effects (columns (1) and (3)), but the coefficient(s) of interest are no longer statistically significant when time fixed effects are included (columns (2) and (4))\textemdash likely because the time fixed effects are blocking part of the causal pathway, absorbing part of the causal effect and biasing the estimates.

Indeed, linear fixed effects models require two causal identification assumptions: first, that past outcomes do not impact the current treatment and, second, that past treatments do not directly influence the current outcome (\cite{imai_when_2019}). In the case of the Chinese anticorruption campaign, both of these assumptions are violated as there is likely strong feedback between the treatment ($CorruptionIndictments$) and the outcome ($ShareOfListings$): first, corruption investigations, which catalyze corruption indictments, are likely targeted towards provinces with higher perceived initial levels of corruption\textemdash and one of the possible ways that this can manifest is a higher $ShareOfListings$, which signals land sale methods less open to competition. Past measures of $ShareOfListings$ can therefore impact the likelihood of current treatment, violating the first fixed effects causal identification assumption. 

Further, when a province is investigated by central inspectors and receives subsequent feedback\textemdash which often includes strengthening provincial anticorruption infrastructure\textemdash the provincial officials have a strong incentive to implement the feedback, saving face, protecting their positions, and avoiding future visits from the Central Inspection team (\cite{wang_76_2013}; \cite{gong_fighting_2022}). Even beyond the strengthening of the anticorruption infrastructure, on the individual level, an increased fear of detection acts as a strong deterrent to corruption for officials in land sale transactions and beyond, especially given the performative nature of the anticorruption campaign \begin{CJK*}{UTF8}{gbsn} (see, for instance, \cite{noauthor__2015-1}).\end{CJK*} Therefore, past inspections, which are manifested in the data as corruption indictments, very likely impact the $ShareOfListings$ in future periods\textemdash that is, past treatments influence the current outcome, violating the second fixed effects causal identification assumption.

On top of this, I also expect that there are some time-varying confounders like gross regional product (GRP), which may make a unit more likely to be treated\textemdash for instance, if richer areas are treated sooner. At the same time, being treated likely affects the area's GRP starting from the time of treatment. Such variables are both simultaneous confounders and intermediate variables, and including time fixed effects therefore blocks part of the causal pathway, biasing estimates. 

Thus, given the assumptions are not met and time-varying confounders are present, I do not think these fixed effects estimations are particularly informative nor that they should be interpreted causally. 

\pagebreak

\subsection{A more mathematical description of the MSM} \label{appendix: msm more math}

\subsubsection{Basic MSM} 

Following \textcite{robins_marginal_2000}, \textcite{imai_robust_2015}, and \textcite{blackwell_framework_2013}, the setup of the MSM is as follows: suppose we observe $N$ units indexed by $i = 1, 2, \dots , N$ at each of $J$ time periods. At each time period $j = 1, 2,\dots, J$, we observe the time-dependent treatment variable $T_{ij}$ and the time-varying covariates $X_{ij}$ that could be impacted by past treatments. Define treatment $T_{ij}$ to be a binary treatment variable where $T_{ij} =1 $ implies that unit $i$ is treated in period $j$. Conversely, $T_{ij} =0$ suggests that unit $i$ is not treated in period $j$. Assume that $X_{ij}$ is already realized before the treatment at time $j$ and is therefore not impacted by the treatment in period $j$, $T_{ij}$.

The observed treatment history for each unit $i$ up to time $j$ is represented by $\overline{T} _{ij} = \{T_{i1}, T_{i2}, \dots, T_{ij}\}$ while the observed time-invarying covariate history for unit $i$ up to time $j$ is captured as $\overline{X} _{ij} = \{X_{i1}, X_{i2}, \dots, X_{ij}\}$. The set of possible treatment and covariate values at time $j$ is $\overline{\mathcal{T}} _j$ and $\overline{\mathcal{X}}_j$, respectively. The outcome of interest $Y_{ij}$ is observed at the time period $j$. Given the longitudinal nature of the analysis, this outcome is impacted by the entire treatment history up until $j$; thus, $Y_{ij}(\overline{t}_j)$ is used to denote the potential value of the outcome variable for unit $i$ at the time period $j$ under the treatment history $\overline{T}_{ij} = \overline{t}_j$, where $\overline{t}_j \in \mathcal{\overline{T}}_j$. For any unit $i$ and time $j$, only one of these potential outcomes can be observed, as a unit cannot follow multiple treatment paths over the same time window. As described further below, the standard consistency assumption is therefore used to connect the potential outcome to the observed outcome; it states, in effect, that the observed outcome and the potential outcome are the same for the observed history. Following \textcite{cole_marginal_2005} and  \textcite{cole_determining_2007}, the observed outcome for unit $i$ at time $j$ is therefore given as $Y_{ij} = Y_{ij}({\overline{T}_{ij}})$. In this framework, $X_{ij}(\overline{t}_{j-1})$ reflects the potential values of covariates for unit $i$ at each time period $j$ given the treatment history up to time $j-1$\textemdash that is, $\overline{T}_{i,j-1} = \overline{t}_{j-1}$\textemdash and the observed values of covariates for unit $i$ at time $j$ can thus be written as $X_{ij} = X_{ij}(\overline{T}_{i, j-1}$). 

\begin{assumption}{1}{Consistency}
\assumlabel{MSM Assump 1 basic}    
As written above, consistency states $Y_{ij} = Y_{ij}({\overline{T}_{ij}})$. Implicit in this definition of consistency is that the treatment history can impact the outcome via the history of the time-varying covariates. The consistency equation can thus be equivalently rewritten as $Y_{ij} =Y_{ij}(\overline{T}_{ij}, \overline{X}_{ij}(\overline{T}_{i,j-1}))$, where $\overline{X}_{ij}(\overline{T}_{i,j-1})$ reflects the values that the covariate history would take under this treatment history (\cite{blackwell_how_2018}). 
\end{assumption}

\begin{assumption}{2}{Positivity}
\assumlabel{MSM Assump 2 basic}
This assumption states that the conditional probability of treatment assignment is between zero and one, exclusive, for each time period. Mathematically, this is reflected as
\begin{align}
    0 \ < \ \text{Pr}(T_{ij} = 1 \ | \ \overline{T}_{i,j-1} = \overline{t} _{j-1}, \overline{X}_{ij} = \overline{x} _j) \ < \ 1
    \label{positivity}
\end{align}
for any time period $j$ for a given treatment history $\overline{t}_{j-1} \in \overline{\mathcal{T}} _{j-1}$ and covariate history $\overline{x}_j \in \overline{\mathcal{X}} _j$.
\end{assumption}

\begin{assumption}{3}{Sequential Ignorability}
\assumlabel{MSM Assump 3 basic}
Also referred to as ``no unobserved confounders," sequential ignorability effectively states that the treatment assignment of unit $i$ at time $j$ is exogenous given the unit's treatment and covariate history up until that time. This is expressed formally as
\begin{align}
    Y_{ij}(\overline{t} _j) \perp \!\!\! \perp T_{ij} \ | \ \overline{T}_{i,j-1} = \overline{t} _{j-1}, \overline{X}_{ij} = \overline{x} _j
    \label{sequential ignorability}
\end{align}
at any time $j$ for unit $i$ given the treatment history $\overline{t}_j \in \overline{\mathcal{T}} _j$ and a covariate history of $\overline{x}_j \in \overline{\mathcal{X}} _j$. 
\end{assumption} 

\textcite{robins_marginal_1999} proved that under these assumptions, IPTW can be used to ``consistently estimate the marginal mean of any potential outcome" $\mathbb{E}\{Y_{ij}(\overline{t}_j)\}$ for any treatment history $\overline{t}_j \in \overline{\mathcal{T}} _j$ (\cite{imai_robust_2015}, p. 1014). 

\subsubsection{MSM with fixed effects: further technical details} \label{msm with FE}

When using the logistic sigmoid function to model the conditional probability of treatment, as the individual propensity score can be written as $\pi_{ij}(\beta, \alpha_i) = \text{Pr}(T_{ij} = 1|\overline{T}_{i,j-1}, \overline{X}_{ij}, \alpha_i) = \text{expit}(\beta^\top\overline{V}_{ij}^* + \alpha_i)$, joint propensity scores are calculated as
\begin{align}
    \pi_\beta(\overline{T}_{i,[j-k,j]},\overline{X}_{i, [j-k, j]}, \alpha_i)       &= \prod_{s =j-k}^j \text{expit}(\beta^\top\overline{V}_{is}^* + \alpha_i)^{T_{is}}(1-\text{expit}(\beta^\top\overline{V}_{is}^* + \alpha_i))^{1-T_{is}} 
\end{align}
where $\overline{V}_{is}^*$ is $ [\overline{T}_{i, s-1}^\top \ \overline{X}_{is}^\top]^\top$, and $\beta$ is a parameter vector. 
In practice, the parameters $\beta$ and $\alpha$ are estimated with a maximum likelihood approach via the $\texttt{feglm}$ function from the $\texttt{fixest}$ R package.\footnote{Using the language of \textcite{berge_efficient_2018}, I focus here on a one ``cluster" fixed-effect model, as there is only one dimension of fixed effects (i.e. by unit) in my implementation. Having further dimensions of clusters further complicates the estimation procedure and described equations.} Developed by \textcite{berge_efficient_2018} $\texttt{feglm}$ uses concentrated log-likelihood, as calculating the log-likelihood  directly is computationally difficult. Implicitly, the full log-likelihood for unit $i$ at time $j$ for the fixed-effects logistic sigmoid model is
\begin{align}
    \ell_{ij}(\beta, \alpha) = \sum_{s=1}^j [T_{is}\text{log(expit(}\beta^\top\overline{V}_{is}^* + \alpha_i)) + (1-T_{is})\text{log(}1-\text{expit(}\beta^\top\overline{V}_{is}^* + \alpha_i))], \nonumber 
\end{align}
which can be computationally intractable for large $N$. According to \texttt{feglm}'s source code, the concentrated likelihood approach implemented optimizes $\beta$ via iteratively weighted least squares such that in each iteration, the generalized linear model is approximated as a weighted least squares problem; in the process of solving this, the fixed effects $\alpha_i$ are ``concentrated out" during weighted OLS steps via demeaning (\cite{berge_fixest_2019}). This can equivalently be written as solving the following first order condition for the $\alpha_i(\beta)$ which maximizes the log-likelihood for each unit $i$ at time $j$ (given $\beta$), such that
\begin{align}
    \frac{\partial\ell_{ij}}{\partial\alpha_i} = \sum^j_{s = 1} [T_{is} - \text{expit}(\beta^\top\overline{V}_{is}^* + \alpha_i)] = 0. 
    \label{derivative of log likelihood}
\end{align}
After ``concentrating out" $\alpha_i$, $\alpha_i(\beta)$ is effectively substituted back into the log-likelihood such that the concentrated likelihood $g(\beta)$ is estimated as 
\begin{align}
    g(\beta) = \sum_{i = 1}^N \sum_{j = 1}^J \ell_{ij} (\beta, \alpha_i(\beta)).
    \label{g(beta)}
\end{align}
which depends on $\beta$. \texttt{feglm} does not explicitly evaluate this equation as it iteratively refines $\beta$ and $\alpha_i$ through weighted least squares (wherein demeaning automatically enforces the first-order condition for $\alpha_i$), but equations (\ref{derivative of log likelihood}) and (\ref{g(beta)}) effectively illustrate the quantities being successively estimated in the model. 

Then, after these weights are estimated via these approximations of $\alpha_i$ and $\beta$, they are passed into the following moment condition to obtain estimates of $\gamma$, the parameter of interest: 
\begin{align}
    0 &= \mathbb{E}\{ \sum_{j = 1}^Jw_{ij}(\overline{T}_{ij},\overline{X}_{ij}, \alpha_i) h(\overline{T}_{i,[j-k, j]})(Y_i - g(\overline{T}_{i,[j-k, j]};\gamma)))\} 
    \label{IPTW Estimates FE}
\end{align}
$h(\cdot)$ is a specified vector of dim($\gamma) \times 1$ of fixed functions of $\overline{t}_j$.\footnote{Note that $\overline{T}_{ij}$ is used in the weights formula, but as the weights formula itself is only estimated for the time period $[j-k,j]$, this $\overline{T}_{ij}$ is effectively $\overline{T}_{i,[j-k,j]}$ It is written as-is in (\ref{IPTW Estimates FE}) for comparability across equations.} With these population moment conditions, the estimator can be found with generalized method of moments (GMM).

With this framework, Blackwell and Yamauchi prove not only that this MSM with fixed effects is asymptotically normal but that including unit fixed effects in the IPTW model for the MSM in this way can adjust for unmeasured baseline confounding under a sufficiently long timescale. 

To implement this MSM with fixed effects in the IPTW weights, I use IPTW with stabilized weights, per (\ref{stabilized_weights FE}). Unit-based fixed effects are implemented by prefecture and then by province, for comparison. Blackwell and Yamauchi (2021, 2024)'s model only applies for unit-level fixed effects, not time-level fixed effects, as time-level fixed effects would likely require a different set of assumptions. I thus implement only unit-based fixed effects, as described below.  

\begin{figure}[H]
    \centering
    \caption{Directed Acyclic Graph for the MSM with Fixed Effects}
    \includegraphics[width=1\linewidth]{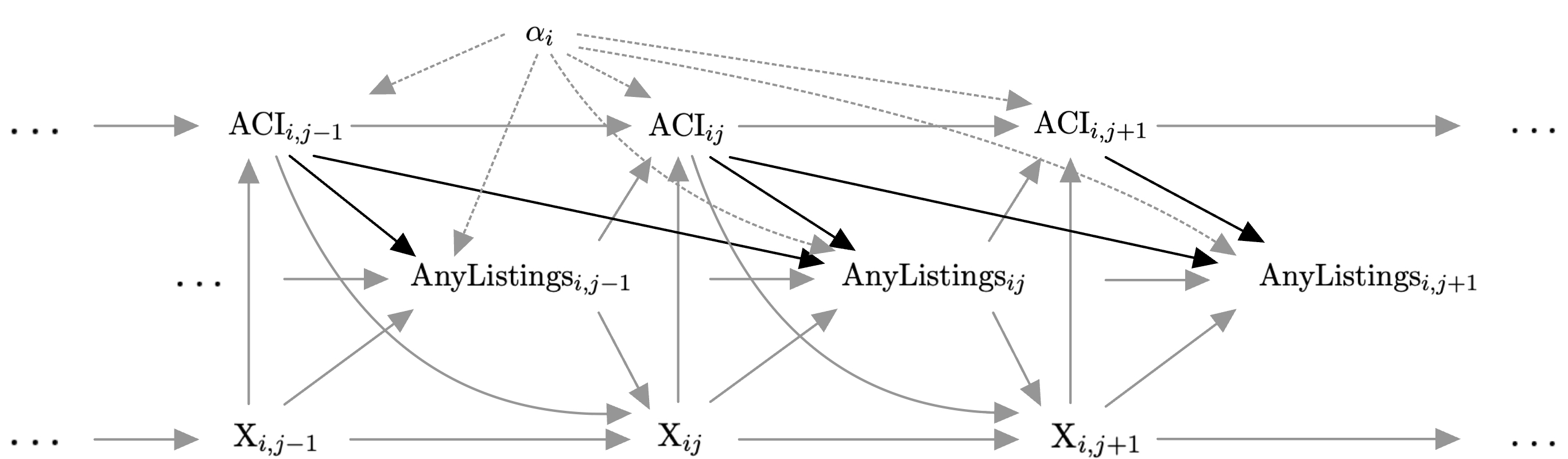}
    \label{fig:MSM DAG FE}
    \caption*{\raggedright \footnotesize Note: Solid gray arrows represent relationships among observed variables while black arrows are the causal relationships of interest. Dotted gray arrows reflect relationships that involve unobserved variables. In this DAG, sequential ignorability holds (although strict exogeneity does not). This DAG follows the style of \textcite{xu_causal_2024} and \textcite{blackwell_how_2018}.}
\end{figure}

It is helpful to recall the relationship being modeled under the assumption of sequential ignorability, as shown in Figure \ref{fig:MSM DAG FE}, with the fixed effects term $\alpha_i$ included. These time-invariant confounders directly influence both the probability of the treatment, $ACI_{ij}$, and the outcome, $AnyListings_{ij}$, for all periods $j$ for a given $i$, hence the arrow pattern shown in Figure \ref{fig:MSM DAG FE} below. 

\vspace{3mm}

To determine the probability of unit $i$ receiving a given treatment in time $j$, as needed to calculate the stabilized weights per (\ref{stabilized_weights FE}), I estimate the following models for the numerator and denominator of the weights, respectively. Probabilities for the numerator are calculated from the equation 
\begin{align}
    \text{ logit Pr}(ACI_{ij} = 1) = \ \beta_0 +  \beta_1 ACI_{i, j-1} + \beta_2 ACI_{i, j-2} \ + \beta_3 ACI_{i, j-3} + \theta^\top C_i + \delta_1 j + \delta_2 j^2.
\end{align}
The treatment is regressed on the lagged terms of the treatment ($ACI_{i, j-1}$, $ACI_{i, j-2}, ACI_{i, j-3}$). I include $C_i$ to reflect time-invariant covariates, but prefecture is the only one I include here as it is essential in modeling the baseline probability of treatment.\footnote{While the denominator model absorbs prefecture-level fixed effects to control for unobserved time invariant confounding, omitting prefecture from the numerator resulted in highly unstable weights and failed to improve the poor covariate balance, suggesting it is essential in predicting the baseline probability of treatment. I investigated whether including further time-invariant covariates in the numerator had an impact, and the coefficient estimates are nearly identical to that when further time-invariant covariates are omitted. I exclude them to avoid over-adjustment (as all of these time-invarying covariates are prefecture-level characteristics that should be captured in the denominator's fixed effects). 

Note also that absorbing unit-level fixed effects $\alpha_i$ are not included in the numerator because the numerator's primary purpose is to stabilize the weights. The denominator adjusts for confounding by modeling the treatment probability given both time-varying and time-invarying covariates.} The sensitivity of the results to the number of lagged treatments included is examined, and the results do not seem particularly sensitive to the number of lags included in the numerator (and denominator) models, and the interpretation did not substantively change when further lags were included.

This numerator model yields the probability of receiving the given treatment for each unit $i$ in each time period $j$ as a function of its past treatment history (for periods $j-3$ to $j-1$).

Then, the denominator model is calculated as
\begin{align}
   \text{logit Pr(}ACI_{ij} = 1 \ | \ \overline{X}_{i,[j-k, j]}(\overline{t}_{i, [j-k-1, j-1]}), \alpha_i) \ = \ & \beta_0+ \beta_1 ACI_{i, j-1} + \beta_2 ACI_{i, j-2}  \ + \\
    & \beta_3 ACI_{i, j-3} + \lambda^\top X_{ij} + \delta_1 j + \delta_2 j^2 +  \alpha_i. \nonumber 
\end{align}
$X_{ij}$ represents time-varying covariates (GRP and population), and $\delta$ is the time trend coefficient. $\alpha_i$ represents the absorbing unit-based fixed effects. GRP has a linear term and a quadratic term, as the suspected quadratic relationship persists across all estimations. Given unit-based fixed effects are included, therein capturing all time-invarying confounding, the time-invariant confounders $C_i$ are no longer included to avoid multicollinearity. 

This denominator model yields the probability of receiving the given treatment for each unit $i$ in each time period $j$ as a function of its past treatment history and time-varying confounders, with prefecture-level fixed effects now included. Then, because weights in time $j$ for prefecture $i$ depend not only on the probability of treatment at time $j$, but that in the periods before it\textemdash starting from $j-k$ as the product of equation (\ref{stabilized_weights FE}) indicates\textemdash I multiply the weights of each of these periods $[j-k:j]$ together to obtain the final stabilized weight at each time period $j$. As described above, I use $k = 4$ here, but experiment with other values of $k$ and do not see significant differences in the results. This behavior is expected given that Blackwell and Yamauchi (2021, 2024) posit that this truncated treatment history treats the treatment history before $k$ lags effectively as a baseline confounder. 
\pagebreak

\subsection{Further details on covariate balance and outliers} \label{appendix: further details on covar balance}

\subsubsection{Covariate balance} \label{covariate balance}

Before discussing the results, it is first important to evaluate whether the IPTW successfully created a pseudopopulation wherein the treated and control groups are comparable with respect to observed confounders; if the covariates are imbalanced, the MSM is likely misspecified or includes covariates with little to no effect on the outcome (\cite{imai_matching_2023}; \cite{austin_moving_2015}). Covariate balance is thus important in justifying sequential ignorability on the observed covariates (and therein whether the model is not misspecified, yielding valid causal conclusions). I examine the covariate balance, which involves assessing the absolute standardized mean differences (SMD) of the covariates/baseline characteristics before and after weighting (\cite{ho_matching_2007}). I use the following formula following \textcite{greifer_cobaltrfunctions_for_processingr_2025} implementation in R's \texttt{cobalt} package 
\begin{align}
   SMD = \frac{\tilde{X}_{nw, treated}-\tilde{X}_{nw, control}}{\sigma_{nw, pooled}},
\end{align}
where $X$, as before, signifies observed covariates. Here, the subscript $nw$ denotes that all of these components are weighted with the normalized version of the stabilized weights $w^*$ (of equation (\ref{stabilized_weights FE})); all weights and weighting conducted in the estimation of standardized mean differences use these normalized versions of the stabilized weights. The $\tilde{X}$ represents the weighted mean of the observed covariate.\footnote{This normalized weight is calculated as $nw_{ij} = \frac{w^*_{ij}}{\sum_{g \in group} \sum_j w^*_g}$, where ``group" is either the treated or control group, and $g$ is a dummy variable used to denote this index. Thus, $\sum_{i \in treated} \sum_j nw_{ij} = 1$ and $\sum_{i \in control} \sum_j nw_{ij} = 1$.} As such, $\Tilde{X}_{nw, treated}$ is the weighted mean of the covariate for the treated group, $\Tilde{X}_{nw, control}$ is the weighted mean of the covariate for the control group, and $\sigma_{nw,pooled}$ is the pooled standard deviation, adjusted for weights such that $\sigma_{nw,pooled} = \sqrt{\frac{Var_{nw}(X_{treated}) + Var_{nw}(X_{control})}{2}}$. In this equation, $Var_{nw}$ is the weighted variance, calculated by \texttt{cobalt} as $Var_{nw}(X_{group}) = \frac{\sum_{i \in group} \sum_j nw_{ij}^*(X_{ij}-\Tilde{X}_{nw, group})^2}{1-\sum_{i \in group} \sum_j (nw_{ij}^*)^2}$ where $nw_{ij}^*$ represents the normalized version of the stabilized weights for unit $i$ at time $j$, and the $group$ is either the treatment or control group (as variances are calculated separately for each group). Likewise, the weighted mean used here ($\Tilde{X}_{nw, group})$ is computed separately for each group.

Figure \ref{fig:MSM Covar Balance FE binary} illustrates the covariate balance plots for the MSM with fixed effects. As highlighted in the literature (for instance, \cite{chesnaye_introduction_2022} and \cite{zhu_boosting_2015}), the standardized differences should be less than 0.10 for the weighted sample for all characteristics/covariates, although 0 is the ultimate target. For each case, I also include the pre-truncation mean IPTW weight and discuss whether there are outlier weights; as highlighted by \textcite{cole_constructing_2008}, the mean of the stabilized weights should be close to 1, as a mean divergent from this can be an indicator of poorly fit IPTW weights and invalidate the results. Similarly, the maximum and minimum weights should not be very extreme. As \textcite{thoemmes_primer_2016} note, there is no fixed threshold of what counts as ``extreme," but weights ``in the hundreds or even higher can be considered quite large" (p. 43). 

As was apparent in Figure \ref{fig:MSM Covar Balance FE binary} of the main text, weighting significantly improves the standardized mean differences, particularly in the presence of prefecture fixed effects (Panel A). Now, all covariates have mean differences below 0.1, confirming that the model is balanced with respect to observed confounders. In Panel B, the standardized mean difference is slightly above the threshold for the lag of GRP, measuring 0.1013. This indicates that the covariates are slightly too imbalanced when province fixed effects are used, but the difference from the threshold (.0013) is rather minor. While this result does not prove sequential ignorability to be true (as it does not evaluate whether there are unmeasured confounders), it does strongly suggest that $T_{ij} \perp \!\!\! \perp \overline{X}_{ij}$ in the prefecture fixed effects case, which is needed for sequential ignorability to hold.

For both cases, the mean pre-truncated IPTW weight is close to 1 (0.991 and 0.992, respectively), and the largest outlier is much closer to the mean (8.22 and 5.42, respectively). When truncated, weights remain close to 1 while the maximum IPTW weight at or below 2 and the minimum is approximately 0.4 for both prefecture-level and province-level fixed effects. With the mean being consistently near 1 and no major outliers in either direction, the IPTW estimation for both fixed effects models appears well-fit.

\begin{figure}[H]
    \centering
    \caption{Box plots for IPTW with prefecture-level fixed effects}
    \includegraphics[width=1\linewidth]{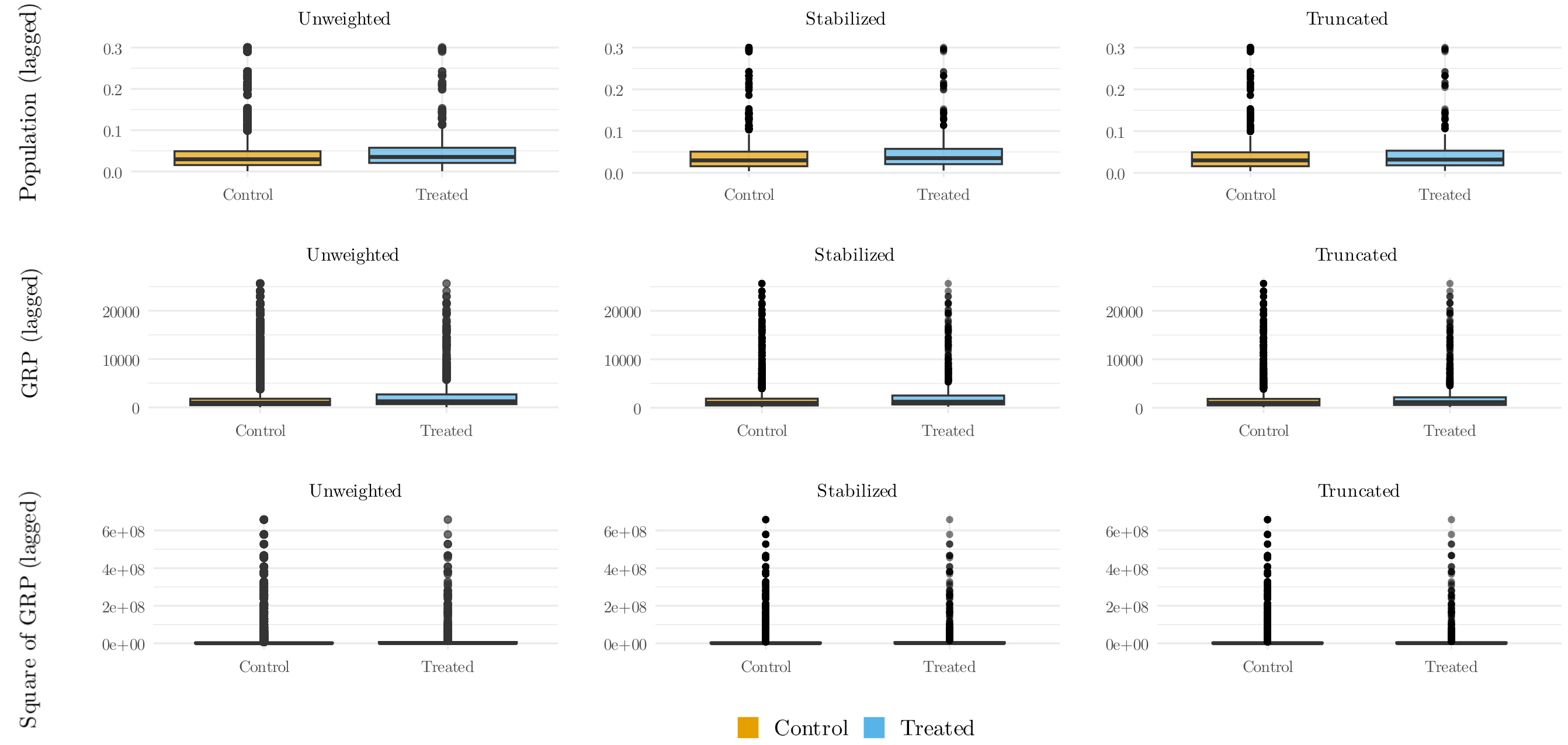}
    \caption*{\footnotesize Note: For scale, population and GRP (lagged) are each divided by 100,000,000; the square of GRP lagged is calculated by squaring this (scaled) lagged value of GRP. The units of GRP are Chinese Yuan. I include the square of GRP, lagged, even though the scale makes the box plot not particularly informative; the empirical CDF is more useful. The truncated weights mentioned here are truncated at the $1^{\text{st}}/99^\text{{th}}$ percentiles.}
    \label{fig: boxplot binary fe pref}
\end{figure}

I also perform the more in-depth checks of evaluating the higher-order moments of \textcite{austin_moving_2015}; as Austin and Stuart highlight, reweighting seeks to balance ``not only means and prevalences but also other characteristics of the distribution... [particularly] higher-order moments" (p. 3666). I thus also use standardized differences to compare the mean of higher-order moments, such as squares. As Austin and Stuart highlight, ``comparing the mean of squares of continuous variables is equivalent to comparing the variance of that variable between treatment groups" (p. 3667). Graphical methods also offer a comparison of continuous variable behavior outside of the mean and higher-order terms, and I thus include side-by-side box plots and empirical cumulative distribution functions (CDFs) comparing the distribution of the continuous variables between treated and control subjects in Appendix \ref{appendix binary aggressive truncation}, Table \ref{tab: binary MSM squared covariate results} and Figures \ref{fig: boxplot binary fe pref}-\ref{fig: empirical CDF FE prov}) (\cite{joffe_model_2004}). These results are consistent with the findings of this section; truncation does not appear to mask any signs of covariate imbalance.

\subsection{Further IPTW results} \label{appendix binary aggressive truncation}

\vspace{3mm}
\begin{table}[H] 
\centering
\caption{Presence of corruption impacting $AnyListings$, aggressive truncation}
\label{tab: MSM outcome model no time binary outcome}
\small
\begin{threeparttable}
\setlength{\tabcolsep}{6pt} 
\renewcommand{\arraystretch}{1.3} 
\begin{tabular}{l*{4}{c}}
\hline \hline
& \multicolumn{4}{c}{AnyListings} \\
\cmidrule(lr){2-5}
                    & (1) & (2) & (3) & (4) \\
\midrule
ACI & 0.0461 & -0.0850 & -0.1258* & -0.0965  \\
& (0.0796) & (0.0840) & (0.0539) & (0.0745) \\
\midrule
Weights & No & Yes & Yes & Yes \\
P-value & 0.88164 & 0.312& 0.0199 & 0.195 \\

Fixed Effects & None & None & Prefecture-level & Province-level \\
Number of Prefectures & 343 & 343 & 346 & 346 \\
Number of Provinces & 31 & 31 & 31 & 31 \\
Number of Observations & 17,986 & 15,907 & 15,142 & 15,138 \\
Effective Sample Size & 100\% & 99.97\% & 97.61\% & 98.46\% \\
\hline \hline
\end{tabular}
\end{threeparttable}
\caption*{\footnotesize Note: Standard errors are in parentheses (\sym{*} \(p<0.05\), \sym{**} \(p<0.01\), \sym{***} \(p<0.001\)); they are estimated via a pairs clustered bootstrap, clustered at the prefecture-level, over 500 replications. As is highlighted in the table, columns differ by the type of weights and fixed effects included. Unweighted estimates (column (1)) do not have a causal interpretation and are shown for comparison purposes only. Column (2) includes several other time-invariant covariates, but these are not reported above. The covariates of column (2) are not balanced, and the column should not be interpreted causally. The intercept is also not reported for all columns.}
\end{table}

Under the aggressive truncation (truncated at $5^{\text{th}}$ and $95^{\text{th}}$ percentiles) shown above, the results are similar to that under regular truncation. The first column is unweighted and has no causal interpretation, and the second column should also not be interpreted causally. Column (3), which has prefecture-level fixed effects in the IPTW model, is again statistically significant and negative, like in Table \ref{tab: MSM outcome model no time binary outcome}. Column (4), with its province-level fixed effects in the IPTW model, is negative but not statistically significant. Ultimately, the results closely mirror those in Table \ref{tab: MSM outcome model no time binary outcome} and do not have a substantively different interpretation. 

\medskip

\subsubsection{Higher-order robustness checks}

\begin{table}[H]
\centering
\caption{ \centering Absolute standardized mean differences for squared covariates for fixed effects models}
\label{tab: binary MSM squared covariate results}
\small
\begin{threeparttable}
\setlength{\tabcolsep}{10pt} 
\renewcommand{\arraystretch}{1.3} 
\begin{tabular}{l*{6}{c}}
\hline \hline
\multicolumn{1}{c}{ } & \multicolumn{5}{c}{Weighting Method} \\
\cmidrule(l{3pt}r{3pt}){2-6}
 & \multicolumn{1}{c}{} & \multicolumn{2}{c}{Prefecture FE} & \multicolumn{2}{c}{Province FE} \\
 \cmidrule(lr){3-4} \cmidrule(lr){5-6}
Covariate & Unweighted & Stabilized & Truncated & Stabilized & Truncated\\
\midrule
(Population (lagged))$^2$ & 0.116 & 0.035 & 0.042 & 0.035 & 0.043 \\
(GRP (lagged))$^2$ & 0.070 & 0.030 & 0.033 & 0.036 & 0.041\\
(Square of GRP (lagged))$^2$ & 0.070 & 0.030 & 0.033 & 0.036 & 0.041 \\
\hline \hline
\end{tabular}
\end{threeparttable}
\caption*{\footnotesize Results show absolute standardized mean differences comparing unweighted, stabilized weights (with no truncation), and stabilized weights calculated at the $1^{st}/99^{th}$ percentiles, for both prefecture-level and province-level fixed effects specifications. }
\end{table}

The above Table \ref{tab: binary MSM squared covariate results} illustrates absolute standardized mean differences for squared versions of the covariates (i.e., $covariate^2$), as part of \textcite{austin_moving_2015}'s suggested robustness checks for covariate balance, comparing stabilized untruncated weights and stabilized truncated weights. Generally, the absolute standardized mean differences of the square of the covariates seem balanced across the specifications, and the truncation does not appear to mask a covariate imbalance. For both prefecture-level and province-level fixed effects, stabilized but untruncated standardized mean differences are all below the threshold of 0.1. Similarly, truncated stabilized weights are far below the threshold, and there is no indication of covariate imbalance.

\vspace{3mm}

As discussed in Section \ref{covariate balance}, the below box plots and empirical CDFs are conducted in line with \textcite{austin_moving_2015}'s suggested robustness checks. I utilize side-by-side box plots to compare the distribution of continuous baseline covariates between the treated and control groups under different levels of weighting. I include stabilized, untruncated weights and stabilized, truncated weights to ensure that the truncation is not disguising any indicators of poor balance. A very slight imbalance between weighted groups appears for the lag of population and of GRP for the stabilized weight (particularly in the location of the $75^{\text{th}}$ percentiles for both), but the magnitude of this difference is not particularly alarming and persists across both the truncated and untruncated stabilized weights. For the lag of GRP squared, the presence of very large outliers makes the box plot difficult to interpret, necessitating the CDF below. 

\medskip
\begin{figure}[H]
    \centering   
    \caption{Empirical CDF for IPTW with prefecture-level fixed effects}
    \includegraphics[width=1\linewidth]{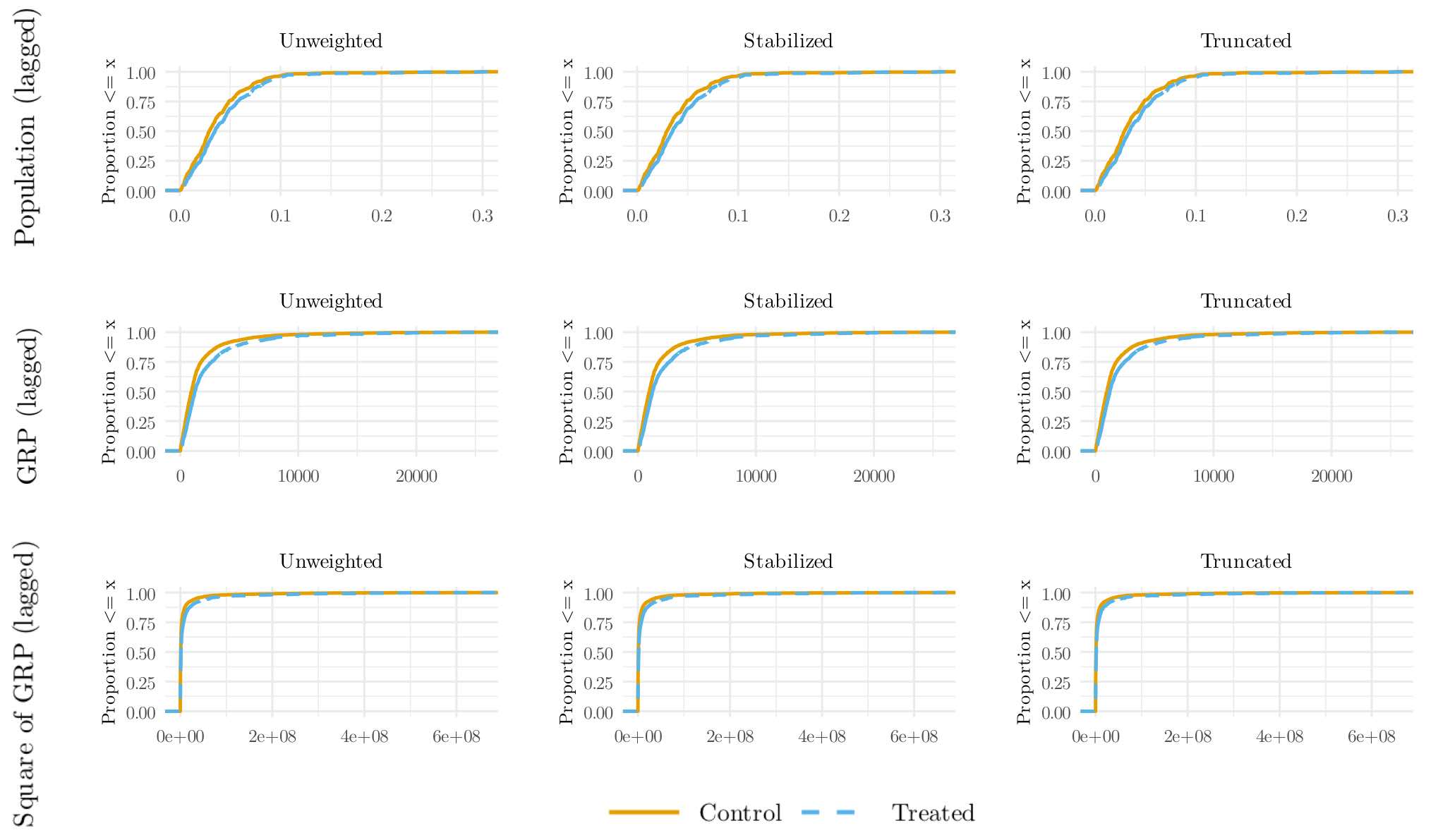}
    \caption*{\raggedright \footnotesize Note: Here, for the truncated weights, truncation occurs at the $1^{\text{st}}/99^\text{{th}}$ percentiles.}
    \label{fig: empirical CDF FE pref}
\end{figure}
\begin{figure}[H]
    \centering
    \caption{Box plots for IPTW with province-level fixed effects}
    \includegraphics[width=1\linewidth]{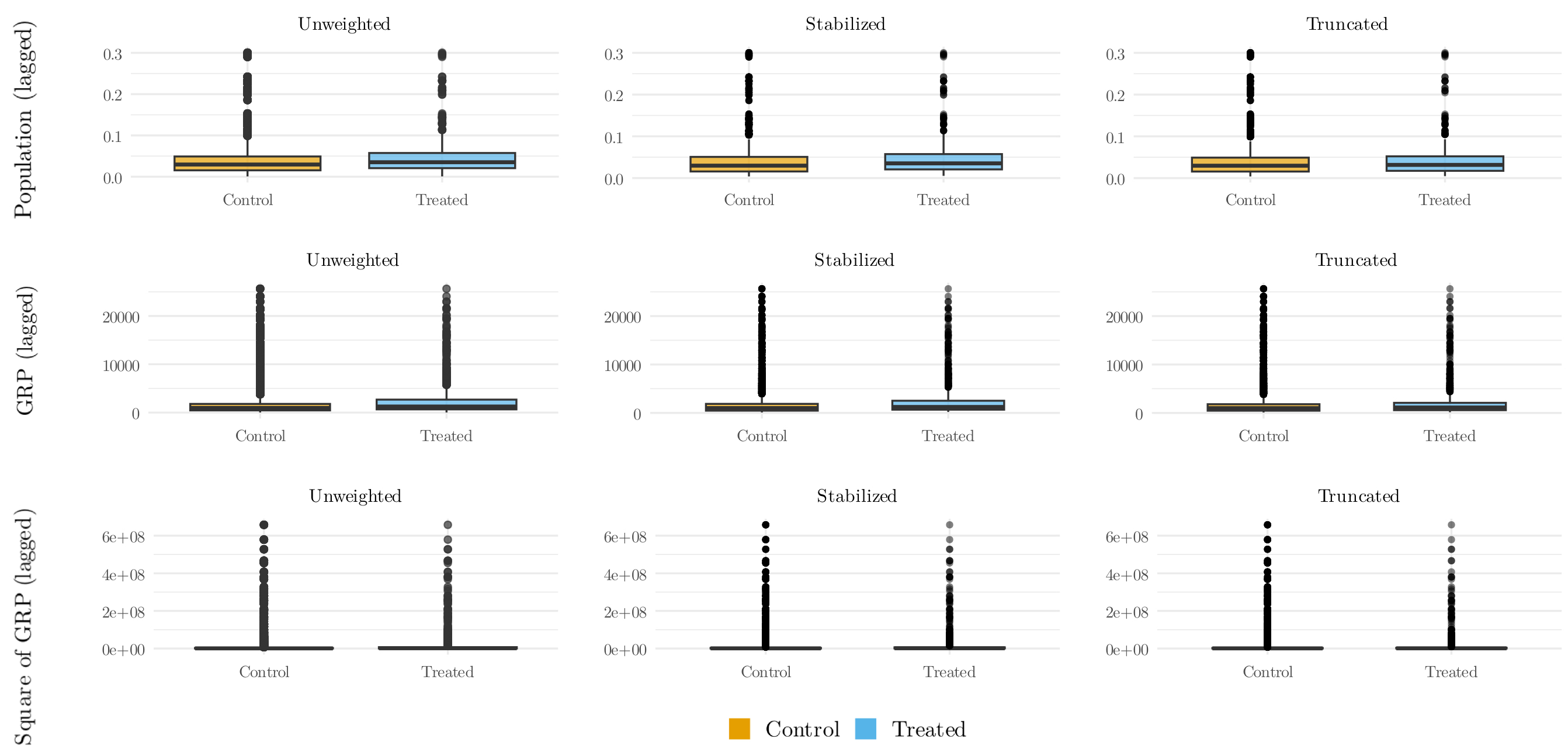}
    \caption*{\raggedright \footnotesize Note: For scale, population and GRP (lagged) are each divided by 100,000,000; the square of GRP lagged is calculated by squaring this (scaled) lagged value of GRP. The units of GRP are Chinese Yuan. I include the square of GRP, lagged, even though the scale makes the box plot not particularly informative; the empirical CDF is more useful. The truncated weights mentioned here are truncated at the $1^{\text{st}}/99^\text{{th}}$ percentiles.}
    \label{fig: boxplot binary fe prov}
\end{figure}
\begin{figure}[H]
    \centering   
    \caption{Empirical CDF for IPTW with province-level fixed effects}
    \includegraphics[width=1\linewidth]{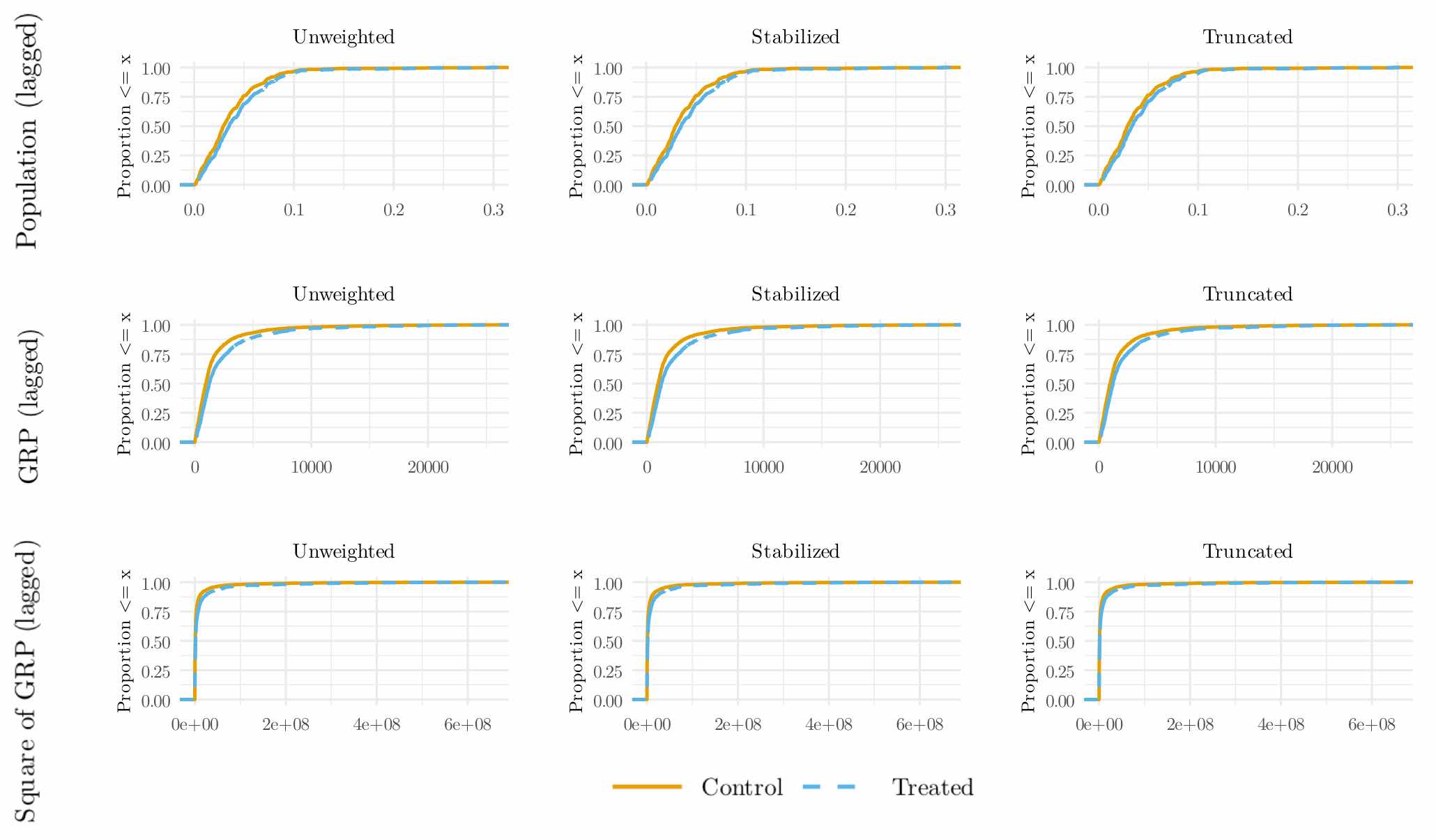}
    \caption*{\raggedright \footnotesize Note: Here, for the truncated weights, truncation occurs at the $1^{\text{st}}/99^\text{{th}}$ percentiles.}
    \label{fig: empirical CDF FE prov}
\end{figure}

As apparent in the CDF results of Figure \ref{fig: empirical CDF FE pref}, the slight divergence between the treatment and controls persists for both the lag of population and the lag of GRP, although the square of GRP lagged seems well fit. Importantly, the slight divergences between the treatment and control are improved with weighting. Thus, while the time-varying covariates do not appear fully balanced between the control and the treated group, the divergence is slight and, on its own, is not enough to suggest the results should not be interpreted causally. I next perform the same analysis for the MSM with province-level fixed effects  in the IPTW model (Figures \ref{fig: boxplot binary fe prov} and \ref{fig: empirical CDF FE prov}).

For the model with province-level fixed effects, the slight imbalance between the treated and untreated groups persists for the lag of population and the lag of GRP, as is apparent in the box plots (Figure \ref{fig: boxplot binary fe prov}). It seems that the biggest difference between the treatment and control for these two variables lies in the location of the $75^{\text{th}}$ percentile, although the median and the $25^{\text{th}}$ percentile, as well as the outlier spread, look roughly even across both. Critically, the differences between the treated and control group are significantly reduced with the weighting; the magnitude of the remaining difference is not particularly alarming and persists across both the truncated and untruncated stabilized weights. For the lag of GRP squared, the presence of very large outliers makes the box plot difficult to interpret, necessitating the CDF below.

\medskip

In these CDF results of Figure \ref{fig: empirical CDF FE prov}, the slight divergence between the treatment and controls persists for both the lag of population and the lag of GRP, although the square of GRP lagged seems well fit. Again, the slight divergences between the treatment and control seem to be improved slightly with weighting. Hence, while the time-varying covariates do not appear fully balanced between the control and the treated group, the divergence is slight and, on its own, is not enough to suggest the results should not be interpreted causally, especially in light of the very low standardized mean differences reported in Table \ref{tab: binary MSM squared covariate results}.

\pagebreak

\subsection{Technical details on sensitivity checks} \label{appendix: technical details on sensitivity checks}

While a high-level overview of these checks was given in the main text, this section provides technical details and relevant figures regarding sensitivity checks.

\subsubsection{Sensitivity check: Sequential ignorability} \label{sensitivity check sequential ignorability}

First, I conduct \textcite{robins_association_1999}'s sensitivity check, as implemented by \textcite{ko_estimating_2003}. If sequential ignorability holds per (\ref{sequential ignorability}), we have that $\mathbb{E}\{Y_{ij}(\overline{t}_j) \ | \ \overline{X}_{ij}, \overline{T}_{i, t-1} = \overline{t}_{i,j-1}, T_{ij} \}$ = $\mathbb{E}\{Y_{ij}(\overline{t}_j) \ | \ \overline{X}_{ij}, \overline{T}_{i, t-1} = \overline{t}_{i,j-1}\}$ since sequential ignorability implies that $Y_{ij}(\overline{t}_j)$ is mean-independent of $T_{ij}$ given the past $\overline{X}_{ij}, \overline{T}_{i, t-1}$. 

However, if unmeasured confounders are present, this equality no longer holds. The function 
\begin{align}
    q_{ij}(t_j, t_j') = \mathbb{E}\{Y_{ij}(\overline{t}_j) \ | \ \overline{X}_{ij}, \overline{T}_{i, t-1} = \overline{t}_{j-1}, T_{ij} = t_j\} - \mathbb{E}\{Y_{ij}(\overline{t}_j) \ | \ \overline{X}_{ij}, \overline{T}_{i, t-1} = \overline{t}_{j-1}, T_{ij} = t_j'\}
\end{align}
will be nonzero when $t_j \neq t_j'$, where $t_j'$ is an alternate treatment (\cite{ko_estimating_2003}); $q_{ij}(t_j, t_j')$ is the natural measure of the ``magnitude of noncomparability with respect to the mean of" [$Y_{ij}(\overline{t}_j)$] of the two groups due to unmeasured confounding (\cite{robins_association_1999}, p. 168). While this magnitude of confounding (selection bias) function cannot be identified without further knowledge of what these confounders are, a version of $Y_{ij}(\overline{t}_j)$ corrected for selection bias (i.e., $Y_{ij}(\overline{t}_j)$'s expectation under $q$) is identifiable under a fixed nonzero $q$ function (\cite{robins_association_1999}; \cite{ko_estimating_2003}).

\textcite{robins_association_1999}, in Lemma 3.1, thus proposes a sensitivity analysis based on interpretable parameterizations of $q$. In a binary case, a simple parameterized form for $q_{ij}(t_j, t_j')$ is chosen; following \textcite{robins_association_1999}, \textcite{ko_estimating_2003} choose $q_{ij}(t_j, t_j') = \varphi\times(t_j - t_j')$ where $t_j' = 1- t_j$ and $\varphi$ is the expected difference between $Y_{ij}(\overline{t}_j)$ given $T_{ij} = t_j$ versus  $T_{ij} = t_j'$ (and conditional on $\overline{X}_{ij}, \overline{T}_{i, t-1}$). Under this definition, $\varphi = 0 $ corresponds to the assumption of no unmeasured confounders. Unmeasured confounders are marked by nonzero $\varphi$ values: if $\varphi > 0$, then on average, treatment is preferentially given to those units with higher $AnyListings$ counterfactuals $\{Y_{ij}(\overline{t}_j)\}$ (i.e., corruption indictments are given to more corrupt-appearing prefectures\textemdash that is, those where a disproportionate number of listings would occur in the absence of indictments), even after controlling for past treatment and measured covariate history. Conversely, $\varphi < 0$ would imply that treatment was preferentially given to prefectures with lower $AnyListings$ counterfactuals, such that indictments were disproportionately given to less corrupt-appearing prefectures, after controlling for past treatments and measured covariate history. As I expect that treatment is preferentially given to those units with higher (unmeasured) suspected corruption, I focus primarily on positive values of $\varphi$ in my sensitivity analysis, although I also include a small range of negative $\varphi$ values for comparison. For each of these fixed $\varphi$ values, I estimate the weighted least squares normal equation with each prefecture's observed outcome replaced by their selection-bias-corrected outcome. That is, 
\begin{align*}
    Y_{ij}(\overline{t}_j;\varphi) = Y_{ij}(\overline{t}_j) - \varphi \sum_{s = 1} ^j \sum_{t_s' = 0}^1 (T_{is} - t_s') \text{Pr(}T_{is} = t_s' \ | \ \overline{T}_{i,s-1}, \overline{X}_{is}). 
\end{align*}
When $\varphi > 0$, the correction replaces $Y_{ij}(\overline{t}_j)$ with a lesser value when treatment is received ($T_{ij})$ and with a greater value when it is not. Following \textcite{ko_estimating_2003}, the probabilities used in this equation are generated from the same model used to calculate treatment probabilities as when the sequential ignorability condition is met\textemdash that is, as used in the denominator of the IPTW weights, per (\ref{stabilized_weights FE}).  The MSM is then refit with this new $Y_{ij}(\overline{t}_j;\varphi)$ as the outcome variable to obtain $\hat{\beta}(\varphi)$. The final result of this analysis is a graph of $\hat{\beta}(\varphi)$ as a function of $\varphi$, with the 95\% confidence interval surrounding each point estimate to capture uncertainty due to sampling variability (\cite{robins_association_1999}). The plot thus ``characterizes the range of selection bias due to unmeasured confounding" (\cite{ko_estimating_2003}, p. 156).

\begin{figure}[H]
    \centering
    \caption{Sensitivity to $\varphi$ for Table \ref{tab: binary MSM outcome model no time outcome}'s column (3)}
    \begin{subfigure}[t]{0.47\linewidth}
        \centering
        \includegraphics[width=\linewidth]{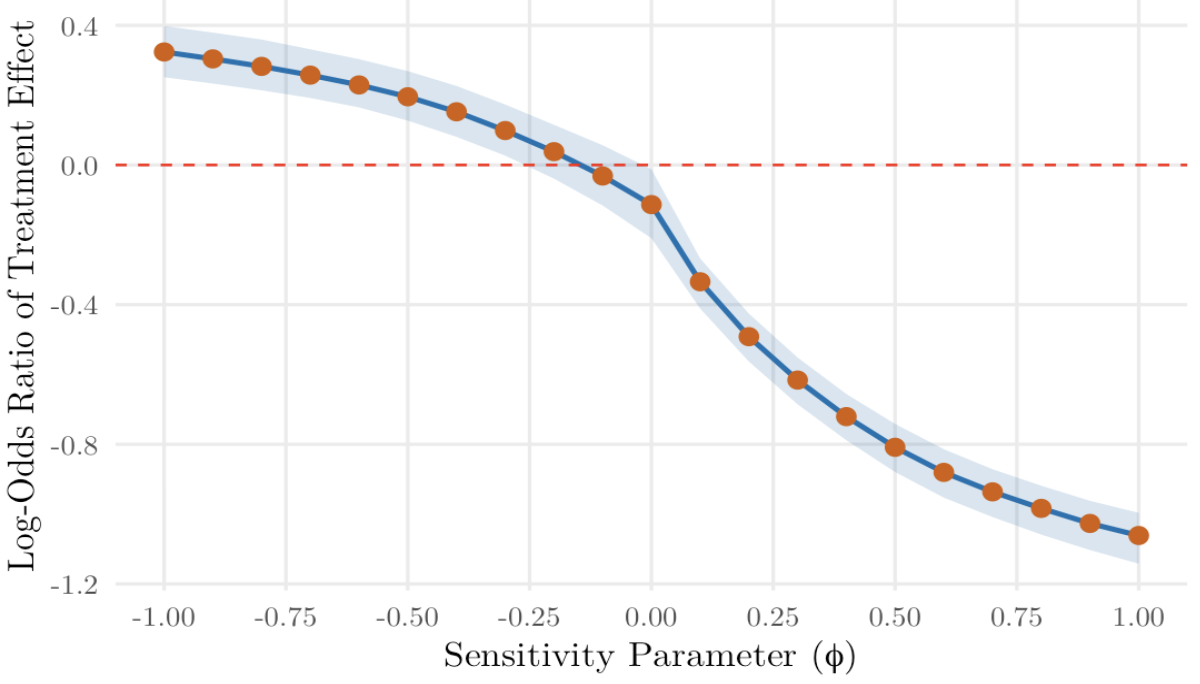}
        \caption{A: Log-Odds Ratio of Treatment Effect vs. $\varphi$}
    \label{fig:sensitivity Log-Odds Ratio of Treatment Effect vs. varphi}
    \end{subfigure}%
    \hspace{0.05\linewidth} 
    \begin{subfigure}[t]{.47\linewidth}
        \centering
        \includegraphics[width=\linewidth]{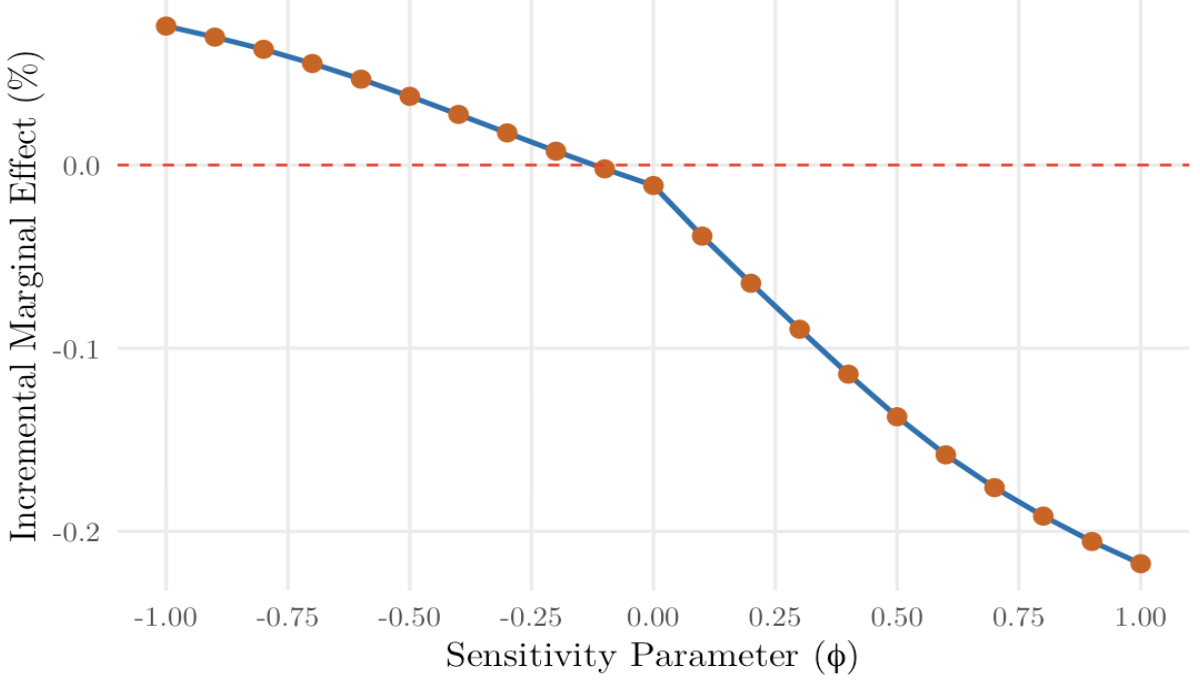}
        \caption{B: Incremental Marginal Effect vs. $\varphi$}
        \label{fig:sensitivity Incremental Marginal Effect vs. varphi}
    \end{subfigure}
    \caption*{\footnotesize Note: The above results are a sensitivity check for the prefecture-level fixed effects estimation (column (3)). The shaded blue band of Panel A reflects the 95\% confidence interval for the log-odds ratio of the treatment effect. Note also that pairs clustered bootstrap standard errors are used throughout these calculations. For Panel A, I confirm that the estimate at $\varphi = 0$ is equal to the estimate in Table \ref{tab: binary MSM outcome model no time outcome}'s column (3); this helps confirm a correct estimation. The estimated incremental marginal effect at $\varphi = 0$ is also that reported for column (3).}
    \label{fig: sensivity to varphi}
\end{figure}

In the below Figure \ref{fig: sensivity to varphi}, Panel A illustrates the estimated treatment effects and 95\% confidence intervals (in terms of the log-odds ratio) as a function of $\varphi$ on the interval $\varphi \in [-1,1]$. Panel B converts these estimates to incremental marginal effects. 

\vspace{3mm}

As is apparent in the above graph,  for $\varphi> 0$, the log-odds of the estimated treatment effect drop consistently, suggesting that the estimates derived under sequential ignorability as reported in column (3) are conservative\textemdash in the sense that they underestimate the true effect if unmeasured confounding exists. A positive $\varphi$ value would correct for the fact that indictments were over-assigned to high-corruption prefectures, and by accounting for this bias, it estimates that the true deterrent effect of indictments on having any listings is larger than originally estimated. At $\varphi = 1$, for instance, the effect (in log-odds) reaches -1.06, which is equivalent to a -21.8 percentage point drop using incremental marginal effects. 

While a negative $\varphi$ value would mean the effect estimated in column (3) is negligible or even reversed, I expect a negative $\varphi$ value is unrealistic as it would necessitate that treatment is preferentially given to less corrupt prefectures. The main aim of the anticorruption campaign is to detect as much corruption as possible, performatively indict corrupt officials, and deter future corruption, and the results are best if there are large numbers of officials to indict, at least in the first stages, which suggests confounding would occur if $\varphi >0$. I include $\varphi < 0$ only for comparison.

The results of the sensitivity analysis for column (4), for the weights with province-level fixed effects, are included below in Figure \ref{fig: sensivity to varphi col 4}. Panel A again illustrates the estimated treatment effects and 95\% confidence intervals (in terms of the log-odds ratio) as a function of $\varphi$ on the interval $\varphi \in [-1,1]$; panel B converts these estimates to incremental marginal effects. As is apparent, the results are largely similar to those in Figure \ref{fig: sensivity to varphi}; as phi increases, the estimated coefficient becomes increasingly negative. Thus, under a positive $\varphi$ value, the check suggests that in the presence of unmeasured confounders, the true deterrent effect of indictments on having any listings would be larger than originally estimated. At $\varphi = 1$, for instance, the effect in log-odds reaches -1.11, which, when using incremental marginal effects, is equivalent to a -23.2 percentage point drop in the probability of having any listings in the presence of corruption indictments. 

\vspace{3mm}
\begin{figure}[H]
    \centering
    \caption{Sensitivity to $\varphi$ for Table \ref{tab: binary MSM outcome model no time outcome}'s column (4)}
    \begin{subfigure}[t]{0.47\linewidth}
        \centering
        \includegraphics[width=\linewidth]{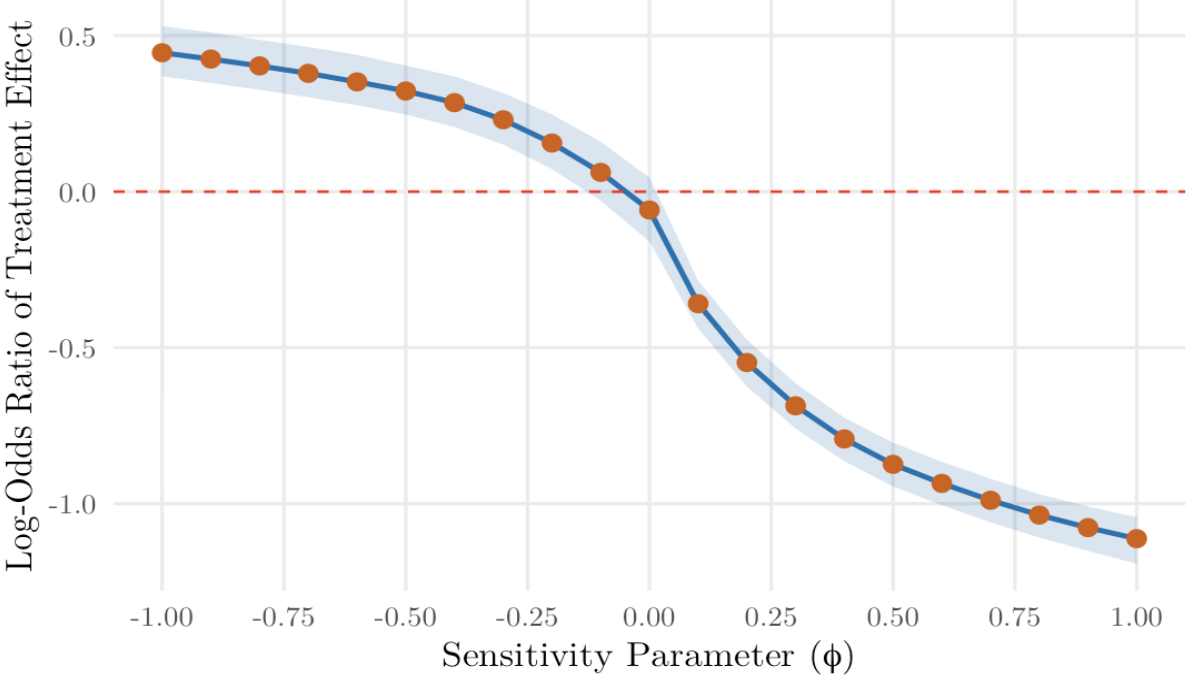}
        \caption{A: Log-Odds Ratio of Treatment Effect vs. $\varphi$}
    \label{fig:sensitivity Log-Odds Ratio of Treatment Effect vs. varphi col 4}
    \end{subfigure}%
    \hspace{0.05\linewidth} 
    \begin{subfigure}[t]{.47\linewidth}
        \centering
        \includegraphics[width=\linewidth]{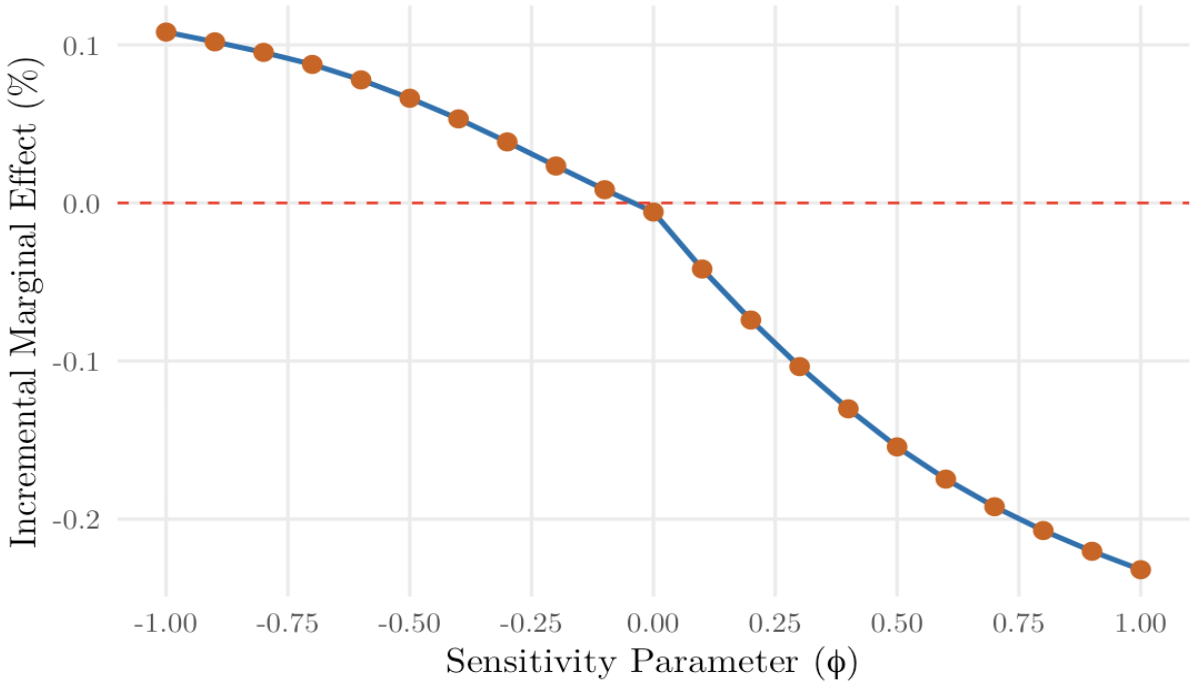}
        \caption{B: Incremental Marginal Effect vs. $\varphi$}
        \label{fig:sensitivity Incremental Marginal Effect vs. varphi col 4}
    \end{subfigure}
    \caption*{\footnotesize Note: The above results are a sensitivity check for the province-level fixed effects estimation (column (4)). The shaded blue band of Panel A reflects the 95\% confidence interval for the log-odds ratio of the treatment effect. Note also that pairs clustered bootstrap standard errors are used throughout these calculations. For Panel A, I confirm that the estimate at $\varphi = 0$ is equal to the estimate in Table \ref{tab: binary MSM outcome model no time outcome}'s column (4); this helps confirm the sensitivity correct is correctly implemented. The estimated incremental marginal effect at $\varphi = 0$ is also that reported for column (4).}
    \label{fig: sensivity to varphi col 4}
\end{figure}

Ultimately, the goal of this sensitivity analysis is to understand which way the coefficient would move if unmeasured confounding was indeed present. As there are fixed effects which absorb all prefecture-level time invariant confounding, the main concerns for unmeasured confounding come from time-varying confounders not directly controlled for, but there are likely not many of these in this prefecture-level case. Regardless, this sensitivity check is still useful in offering insight into what such confounding would imply for the results of Table \ref{tab: binary MSM outcome model no time outcome}.

\subsubsection{Sensitivity checks: Positivity and effective sample size} 

As positivity seems to be well-met in the data on a conceptual level, I briefly report the empirical checks I conducted confirmed this. As recommended by \textcite{zhu_core_2021}, I preliminarily check the overlapping area of the propensity score distribution between treated and untreated groups for the untruncated weights, finding it to be 62.77\% for column (3) and 63.15\% for column (4). While this is a solid measure of overlap, to confirm positivity is met, I also implement \textcite{petersen_diagnosing_2012}'s version of the parametric bootstrap, which is designed to be a diagnostic tool for positivity violations. In essence, they use the parametric bootstrap to sample from ``an estimate of the true data generating distribution, resulting in multiple simulated data sets" (p. 11-12). Inherently, the true data generating distribution and target parameter value are known in the bootstrapped data, and after the ``candidate estimator is then applied to each bootstrapped data set, ... the mean of the resulting estimates across data sets [can be] compared with the known `truth'" (p. 12). In particular, their algorithm uses a data generating distribution that is chosen such that when the estimator is calculated from the bootstrapped samples, it is guaranteed to be consistent unless the propensity scores fail to satisfy the positivity assumption or are truncated. As such, rather than quantifying bias, this sensitivity check warns whether positivity bias poses a threat to inference. 

For column (3), I estimate this coefficient across 500 replicates and find that the estimated bias is 0.0075. This falls below the ``red flag" metrics of 1) the bias being of similar magnitude (or larger than) the estimator's standard error, which was 0.0490, per column (3), or 2) the bias changing the interpretation of the bias-corrected confidence interval. The recalculated confidence interval for column (3) (in log-odds) when adjusted for this bias is now (-0.217, -0.025), again remaining consistently negative. Thus, based on both conceptual groups and these checks, positivity seems to be broadly met for this specification. For column (4), the estimated bias is -0.0315, which again does not reach the ``red flag" level as it is below the estimated standard error. Likewise, the corrected confidence interval of (-0.132, 0.078) again does not have a substantively different interpretation as it still includes 0. Thus, based on these results, there does not appear to be a positivity violation that would threaten causal estimates or interpretation.  

Before discussing consistency, I briefly highlight one more statistic that helps confirm that the model is not misspecified, the effective sample size (ESS). The ESS is a measure of the sample size that a non-weighted sample would have to be in order to achieve the same level of precision as the weighted sample (\cite{ridgeway_toolkit_2024}; \cite{greifer_covariate_2025}). As such, a low ESS could indicate that a few extreme values dominate the weights, implying high variance, unreliable estimates, and likely positivity violations.  Following the standard equation used in the \texttt{cobalt} R package used to create the covariate balance plots, ESS is calculated as
\begin{align} 
ESS = \frac{(\sum_{i=1}^N \sum_{s=1}^j w^*_{is})^2} {\sum_{i=1}^N \sum_{s=1}^j w^{*2}_{is}}
\label{ESS formula}
\end{align}
where $w_{ij}^*$ are the stabilized weights estimated via equation (\ref{stabilized_weights FE}) for the IPTW with fixed effects, respectively. The effective sample sizes illustrated in Table \ref{tab: binary MSM outcome model no time outcome} are close to 100\%, with the lowest being 95.03\% (column (3)). A low ESS would indicate likely model misspecification, positivity violations, and high variance in weights, so such high ESS values provide more evidence that none of these violations are present.

\subsubsection{Sensitivity check: Consistency} \label{consistency discussion}

Finally, I discuss the consistency assumption: as highlighted in Section \ref{sec: MSM with fixed effects in the IPTW}, consistency requires that the outcome for unit $i$ when exposed to the treatment $T_{ij}$ will be the same regardless of what mechanism is used to assign treatment to unit $i$ and regardless of the treatments received by the other units.\footnote{Note that some portions of the literature argue most on clarifying the treatment as much as possible (for instance, by moving the means of exposure to the treatment into the definition of the treatment), but as \textcite{cole_consistency_2009} posit, some residual components of the means of exposure will persist regardless of how ``well" the exposure is defined.} Beyond the checks already conducted (like evaluating propensity score overlap, ensuring covariate balance, and checking the ESS), the literature offers no further consistency-focused sensitivity analyses that can be feasibly implemented here. My concerns about consistency are chiefly that the exact ``version" of each treatment is unique, as different numbers of officials of various ranks and bureaus/divisions are indicted in each prefecture. Thus, the exact ``dose" of the treatment is likely not perfectly equivalent across prefectures and time periods. Second, the possibility of anticipation\textemdash wherein the treatment of one prefecture spills over into another\textemdash is very real as the government sought for the anticorruption campaign to transcend each prefecture and create ripple effects across the country.

Thus, while the previous sensitivity checks and statistics did not elucidate any blatant consistency violations, I suspect that consistency may nevertheless be violated and discuss the implications of this on the results. If there is in fact spillover and other prefectures witnessing the anticorruption campaign preemptively reduce their corruption\textemdash and the probability of having $AnyListings$ in turn declines\textemdash then this behavior biases the magnitude of the coefficient of interest towards zero. Then, in the presence of a true control group (i.e., where there is no anticipation), the magnitude of the coefficient would be larger and maintain its negative sign, indicating a stronger negative causal effect. Thus, even if consistency is violated by spillover, the presence of the statistically significant negative coefficient on $AnyListings_{ij}$ in column (3) suggests that the effect would remain if consistency were fully met\textemdash and the true coefficient would be more negative. 

From the treatment angle, the impact of different treatment versions is less clear-cut. The difference in ``dose" rather than ``route" is the bigger threat to consistency here: even if the indictments in a prefecture occur across different divisions or for different crimes, the impact of this heterogeneity on the results will be minor as investigations/indictments are designed to effectively ``neutralize" corruption across the whole prefecture, relying on unpredictable investigations and performative indictments for deterrence. Thus, I believe that all indictment types have relatively similar effects on listings. In terms of the ``dose," I expect that differing amounts of indictments have different impacts on the decline in $AnyListings$\textemdash that is, the magnitude of the indictments matters since it is often a proxy for the severity of the anticorruption campaign in a prefecture. In this sense, the binary treatment masks dose heterogeneity: if the effect on listings is stronger in prefectures with a stronger dose (high-dose units), the estimated average treatment effect likely underestimates effects for these high-dose prefectures and overestimates for low-dose prefectures. The MSM's coefficient for $AnyListings$ thus represents the average effect of switching from no indictments to any indictments across prefectures, weighted by their probability of treatment. While this result is still an interpretable marginal effect and offers insight into an important policy question, I expect that the number of corruption indictments\textemdash rather than the mere presence of corruption indictments\textemdash has a larger influence on whether there are any listings in a prefecture. 

Thus, after completing the described sensitivity checks, we have more insight into the stability and interpretability of the results: any violations of positivity, sequential ignorability, or the spillover dimension of consistency would bias the estimate towards zero, suggesting the estimated coefficient of interest in column (3) is a lower bound for the true effect. Further, with the spillover across periods\textemdash wherein statistically significant coefficients were apparent for leads and lags of the outcome model for time periods $j-1$, $j, j+1, j+2,$ and $j+3$\textemdash the effect seems to persist across multiple periods before and after indictments, suggesting it is larger than the single 1.16 percentage point drop of column (3). Ultimately, while the magnitude of the coefficient depends on which periods are examined and whether the assumptions hold, the sign seems to be firmly negative when prefecture-level fixed effects are used, a significant result in itself.

\subsection{Multiple treatments in the outcome model} \label{appendix More Advanced Specifications}

\subsubsection{Sensitivity checks for prefecture-level fixed effects with multiple treatments} \label{appendix: Sensitivity Checks multiple treatments}

\begin{figure}[H]
    \centering
    \caption{Sensitivity to $\varphi$ for Table \ref{tab: binary MSM outcome model no time outcome, sequential treatments}'s column (4)}
    \label{fig:sensitivity Log-Odds Ratio of Treatment Effect vs. varphi- 3 lags}
    \begin{subfigure}[t]{0.77\linewidth}
        \centering
        \includegraphics[width=\linewidth]{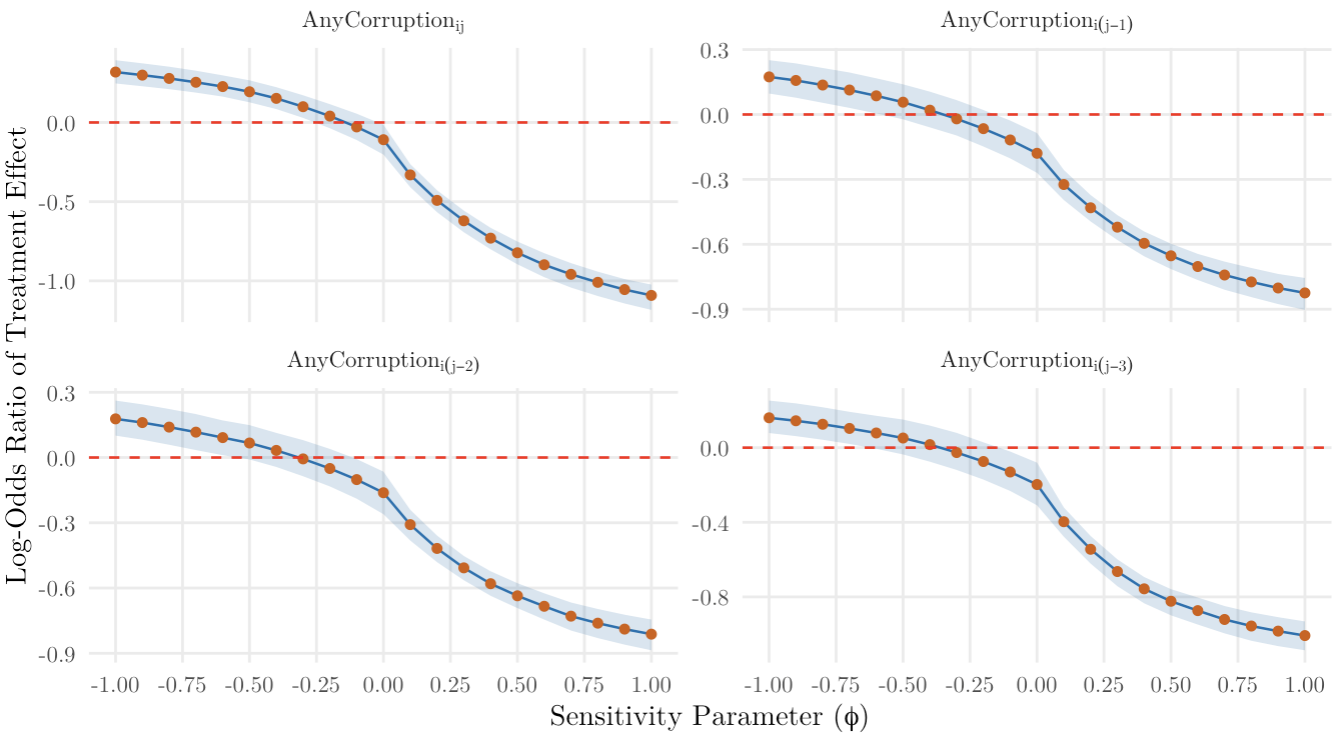}
    \end{subfigure}%
    \caption*{\footnotesize Note: The above results are a sensitivity check for column (4) when the prefecture-level fixed effects are included in the IPTW model. The shaded blue band reflects the 95\% confidence interval for the log-odds ratio of the treatment effect. Points are estimated in increments of 0.25 for the interval [-1, 1]. Note also that pairs clustered bootstrap standard errors are used throughout these calculations. 
    I confirm that the estimate at $\varphi = 0$ is equal to the coefficient estimate in Table \ref{tab: binary MSM outcome model no time outcome, sequential treatments}'s column (4) for each variable; this helps confirm a correct estimation.}
    \label{fig: sensivity to varphi- column (4)}
\end{figure}

\begin{figure}[H]
    \centering
    \caption{\centering Sensitivity to $\varphi$ for Table \ref{tab: binary MSM outcome model no time outcome, sequential treatments}'s column (4) for IPTW with prefecture-level fixed effects: Incremental effects}
    \begin{subfigure}[t]{.75\linewidth}
        \centering
        \includegraphics[width=\linewidth]{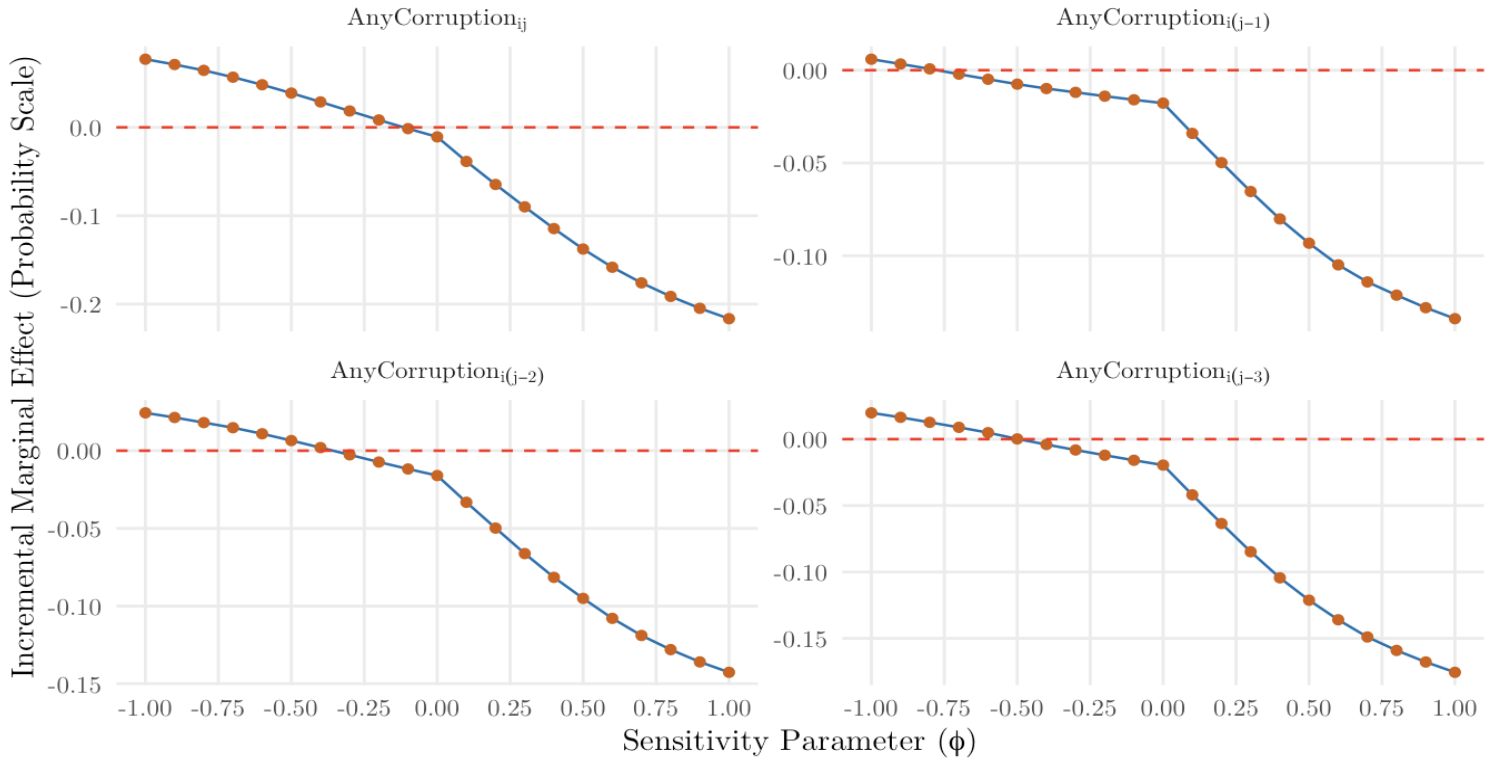}
    \end{subfigure}
    \label{fig:sensitivity Incremental Marginal Effect vs. varphi- 3 lags}
    \caption*{\footnotesize \raggedright Note: The above results are a sensitivity check for Table \ref{tab: binary MSM outcome model no time outcome, sequential treatments}'s column (4) with prefecture-level fixed effects in the IPTW model. Pairs clustered bootstrap standard errors are used throughout these calculations.}
\end{figure}

I then perform sensitivity checks on these specifications, as in Appendix \ref{appendix: technical details on sensitivity checks}. Since the weights are the same as those used in the simple outcome model, I do not need to re-examine covariate balance, ESS, or propensity score overlap, for these all remain unchanged despite the different outcome models. I begin by performing the \textcite{ko_estimating_2003}-style implementation of \textcite{robins_association_1999}'s sensitivity check and then conduct \textcite{petersen_diagnosing_2012}'s parametric bootstrap to check for positivity violations. The sensitivity checks are focused on columns (4) and (5)\textemdash that is, the specifications with 3 lags of treatment (column (4)) and with the cumulative count of the number of treated periods up to period $j$ (column (5)). For the Ko, Hogan, and Mayer check, I examine the behavior of each coefficient of column (4) in the presence of $\varphi$ values [-1, 1], as shown in Figure \ref{fig:sensitivity Log-Odds Ratio of Treatment Effect vs. varphi- 3 lags}. The figure with the log-odds ratio of the treatment effect converted to incremental marginal effects is included afterwards as Figure \ref{fig:sensitivity Incremental Marginal Effect vs. varphi- 3 lags}. 

\vspace{3mm}

As is apparent in the above graph,  for $\varphi> 0$, the log-odds of the estimated treatment effect drop consistently, suggesting that the estimates derived under sequential ignorability, as reported in column (4), underestimate the true effect if unmeasured confounding exists. Generally, the shapes of the curves are quite similar to that in Figure \ref{fig:sensitivity Log-Odds Ratio of Treatment Effect vs. varphi}, with a positive $\varphi$ value correcting for the fact that indictments were over-assigned to high-corruption prefectures. By accounting for this bias, it estimates that the true deterrent effect of indictments on having any listings is larger than originally estimated. Previously, the log-odds of the treatment effect became negative just before $\varphi =0$; this is still the case for the plot of $ACI_{ij}$, but for the three lags, $\varphi = -0.25$ is still negative. Thus, in the (unlikely) event that $\varphi$ is slightly less than 0 (i.e., $-0.25< \varphi <0$), the treatment effect will still be negative for the three lagged terms. 

\vspace{3mm}

Recall that the specification of Table \ref{tab: binary MSM outcome model no time outcome, sequential treatments}'s column (4) is  
\begin{align*}
   \text{logit Pr(}AnyListings_{ij} = 1) =  &\psi_0 + \psi_1ACI_{ij} + \psi_2ACI_{i,j-1} \ + \\ 
   &\psi_3ACI_{i,j-2} + \psi_4ACI_{i,j-3}. \nonumber 
    \label{eq: binary outcome no time lag 3}
\end{align*}
Note that the apparent sharpness of the curve in the plot for $ACI_{i,j-1}$ (the top right panel) is because points are only plotted for $\varphi$ values in increments of 0.25. Plotting at smaller increments smooths out this apparent jump. As the plot highlights, even a small $\varphi$ value, such as 0.25, could have a significant effect on the probability of having any listings since the slope of the curve is quite steep near $\varphi = 0$, before stabilizing after $\varphi=2$.

\begin{figure}[H]
    \centering
    \caption{Sensitivity to $\varphi$ for Table \ref{tab: binary MSM outcome model no time outcome, sequential treatments}'s column (5)}
    \begin{subfigure}[t]{0.47\linewidth}
        \centering
        \includegraphics[width=\linewidth]{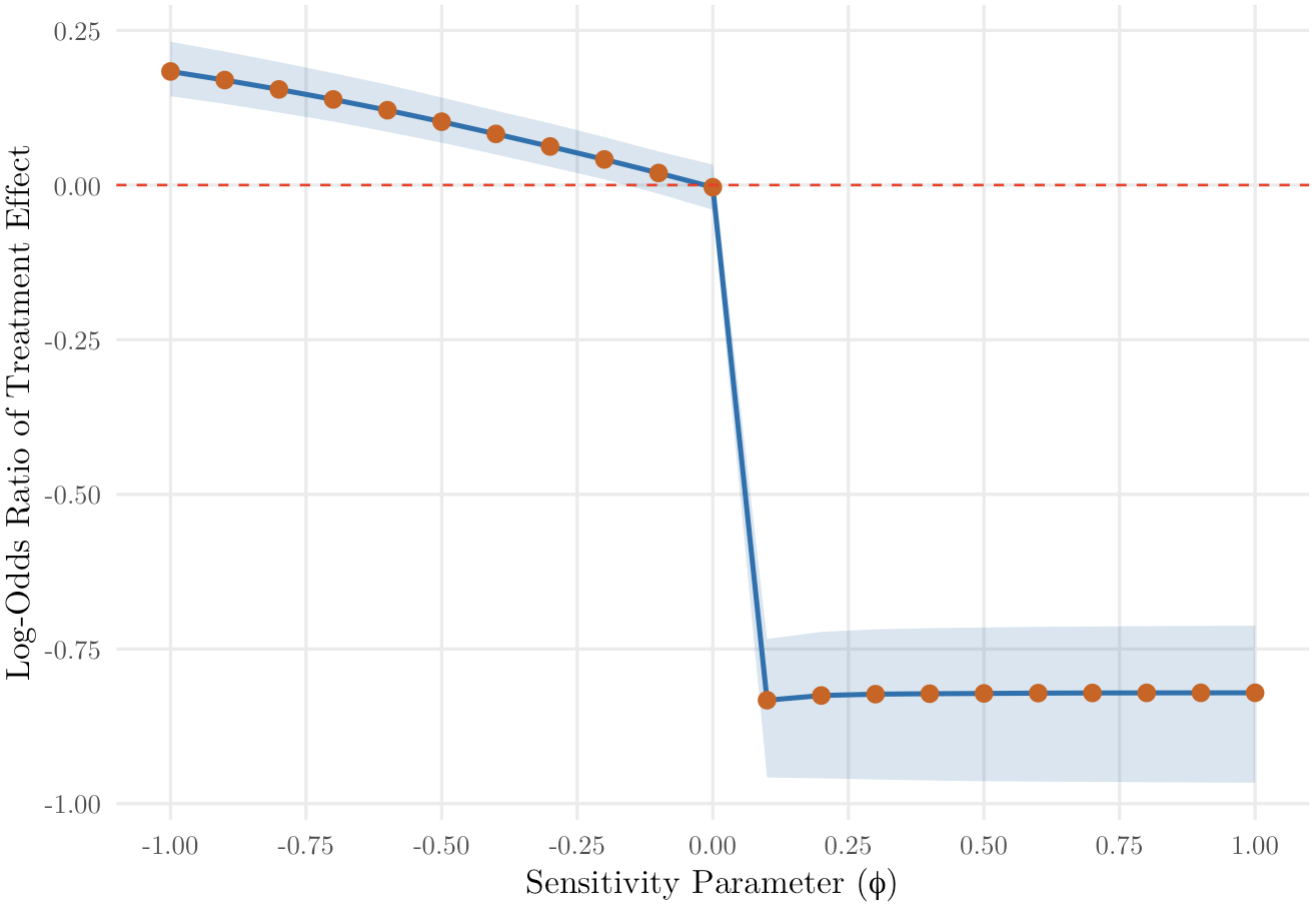}
        \caption{A: Log-Odds Ratio of Treatment Effect vs. $\varphi$}
    \label{fig:sensitivity Log-Odds Ratio of Treatment Effect vs. varphi- cumative outcome}
    \end{subfigure}%
    \hspace{0.05\linewidth} 
    \begin{subfigure}[t]{.47\linewidth}
        \centering
        \includegraphics[width=\linewidth]{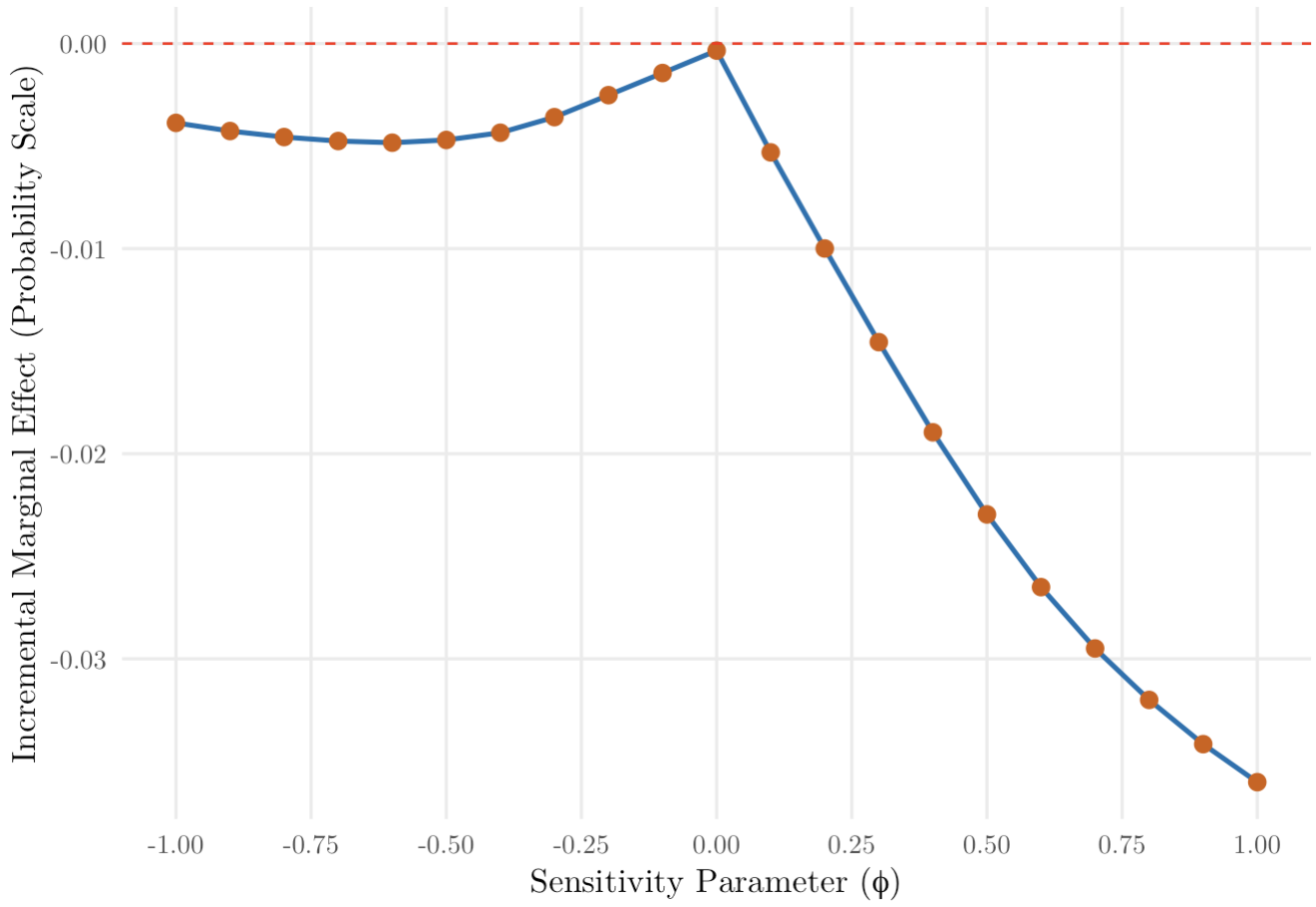}
        \caption{B: Incremental Marginal Effect vs. $\varphi$}
        \label{fig:sensitivity Incremental Marginal Effect vs. varphi- cumative outcome}
    \end{subfigure}
    \caption*{\footnotesize The above results are a sensitivity check for column (5) with prefecture-level fixed effects in the IPTW model. The shaded blue band of Panel A reflects the 95\% confidence interval for the log-odds ratio of the treatment effect. Points are estimated in increments of 0.25 for the interval [-1, 1]. Note also that pairs clustered bootstrap standard errors are used throughout these calculations. For Panel A, I confirm that the estimate at $\varphi = 0$ is equal to the estimate in Table \ref{tab: binary MSM outcome model no time outcome, sequential treatments}'s column (5); this helps confirm a correct estimation. The estimated incremental marginal effect at $\varphi = 0$ is also that reported for column (5).}
    \label{fig: sensivity to varphi- column 5}
\end{figure}

I next examine the behavior of the coefficient for the cumulative measure of $ACI_{ij}$, as captured in column (5), in the presence of different $\varphi$ values; the results are shown in Figure \ref{fig: sensivity to varphi- column 5}.

\vspace{3mm}

As is apparent in the figure above, these results are notably less smooth than that in Figures \ref{fig: sensivity to varphi}, \ref{fig: sensivity to varphi col 4}, and \ref{fig: sensivity to varphi- column (4)}, and notably, the behavior in the incremental marginal effects plot (Panel B) suggests that $\varphi$ is a local maximum. First, the sharp drop in Panel A, which illustrates the log-odds ratio of the treatment effect, may reflect that the model is highly sensitive to the correction term near $\varphi = 0$ or that the outcome model is reacting nonlinearly to the corrected outcomes. Then, in Panel B with incremental marginal effects on the y-axis, as the marginal effects are a function of the coefficients and underlying predicted probabilities, it is apparent that any unmeasured confounders will cause a sharp drop in the probability of having any listings. Notably, though, the scale of the incremental marginal effect is comparatively much smaller than that in Figures \ref{fig: sensivity to varphi}, \ref{fig: sensivity to varphi col 4}, or \ref{fig: sensivity to varphi- column (4)}\textemdash while the incremental marginal effects for the coefficients of these other plots is around 0.15-0.2 when $\varphi = 1$, it is only approximately 0.035 in Figure \ref{fig: sensivity to varphi- column 5}. The ``steep" decline of Panel B is therefore comparatively quite small in size. Thus, with the unusual shapes of these plots likely being due to nonlinear interactions between the corrected outcome and the logistic model, they should not be interpreted as precise reflections of behavior but do suggest an extremely small change in the size of the coefficient (in terms of incremental effects) in the presence of unmeasured confounders.

I next implement \textcite{petersen_diagnosing_2012}'s version of the parametric bootstrap to diagnose possible positivity violations. For column (4), I estimate the coefficients across 500 replicates and find that the estimated bias for $ACI_{ij}$ is 0.0268, which is well below the standard error of 0.0485. Since the direction of the bias is now positive, this signals an overestimation, implying the original estimate was attenuated. The 95\% confidence interval is thus (-0.231, -0.041), which is again exclusive of zero. I then perform similar checks on the other coefficients ($ACI_{i,j-1}$, $ACI_{i,j-2}$, $ACI_{i,j-3}$) and find similar results, with the bias estimate consistently below the standard error and the confidence interval negative and exclusive of zero. For column (5), the estimated bias across 500 replicates is 0.0023. This number is again far below the standard error, and the 95\% confidence interval does not change substantively when bias is adjusted for, still spanning 0. Thus, there does not appear to be any worrying positivity violations that would threaten a causal interpretation in either case, per Petersen et al.'s check. 

Ultimately, interpreting these sensitivity checks in line with the covariate balance statistics and other weight-based checks performed earlier yields conceptually similar results to the previous outcome specifications with prefecture-level fixed effects in the IPTW model (equation (\ref{eq: binary outcome no time}), for instance). As before, any violations of positivity, sequential ignorability, or the spillover dimension of consistency would bias the estimates towards zero, suggesting the estimated coefficients in column (4) are lower bounds for the true effect. Further, these specifications appear to confirm the idea that corruption indictments in period $j$ impact the probability of having any listings in the periods immediately after (as column (4) confirms that having indictments in periods directly before $j$ have a statistically significant negative effect on the probability of having any listings at time $j$).  Column (4)'s coefficients further suggest that the decline in the probability of having any listings is magnified when there are corruption indictments in sequential periods. At the same time, these results confirm that the temporal element/chronology of the indictments matters, with the negative effect on $AnyListings$ vanishing when only the total number of periods up to $j$ with indictments is regressed (Table \ref{tab: binary MSM outcome model no time outcome, sequential treatments} Column (5)). While the sensitivity test for unmeasured confounders for column (5) is somewhat inconclusive, it suggests that even if there were unmeasured confounders, the magnitude of the coefficient (in terms of incremental effects) is quite small. Ultimately, while the magnitude of the coefficient depends on which periods are examined and whether the assumptions hold, the sign seems to be firmly negative in the presence of sequential treatments when prefecture-level fixed effects are used, an informative result in itself. 

\subsubsection{Multiple treatments in the outcome model with province-level fixed effects} \label{appendix: Multiple treatments in the outcome model with province-level fixed effects}

I briefly discuss the results under the same outcome models, but now with province-level fixed effects in the IPTW model (rather than prefecture-level ones), per column (4) of Table \ref{tab: binary MSM outcome model no time outcome}. Table \ref{tab: binary MSM outcome model no time outcome, sequential treatments FE} below illustrates the results when several of these sequential treatment specifications are estimated.

\vspace{3mm}
\begin{table}[H] 
\centering
\caption{\centering Presence of corruption impacting $AnyListings$, for IPTW with province-level fixed effects}
\label{tab: binary MSM outcome model no time outcome, sequential treatments FE}
\small
\begin{threeparttable}
\setlength{\tabcolsep}{10pt} 
\renewcommand{\arraystretch}{1.3} 
\begin{tabular}{l*{5}{c}}
\hline \hline
& \multicolumn{5}{c}{AnyListings} \\
\cmidrule(lr){2-6}
                    & (1) & (2) & (3) & (4) & (5)\\
\midrule
ACI$_{ij}$ & -0.0588 & -0.0454 & -0.0546 & -0.0584\\
& (0.0536) & (0.0518) & (0.0525) & (0.0528) \\
ACI$_{i,j-1}$ & & -0.1032* & -0.0839 & -0.0908 \\
& &  (0.0521) & (0.0502) & (0.0510) \\
ACI$_{i,j-2}$ & & & -0.1373** & -0.1230** \\
 & & &  (0.0525) & (0.0501) \\
ACI$_{i,j-3}$ &  & & & -0.1105 \\
 & & & & (0.0608) \\
Cumulative ACI$_{ij}$ & & & & & 0.0043 \\
 & & & & & (0.0187) \\
\hline \hline
\end{tabular}
\end{threeparttable}
\caption*{\footnotesize Note: Standard errors are in parentheses (\sym{*} \(p<0.05\), \sym{**} \(p<0.01\), \sym{***} \(p<0.001\)); they are estimated via a pairs clustered bootstrap, clustered at the prefecture-level, over 500 replications. All columns have weights, province-level fixed effects, and an effective sample size (ESS) of 96.78\%, as the weights do not change across columns, only the versions of the treatments. Each column also has 16,455 observations total, from 343 prefectures and 31 provinces. The intercept is not reported for all columns. The above results are for weights truncated at the $1^{\text{st}}$ and $99^{\text{th}}$ percentiles. Recall that these results are in log-odds as the logistic regression is used. Note that column (1) is the basic specification of equation (\ref{eq: binary outcome no time}), as is shown in Table \ref{tab: binary MSM outcome model no time outcome}'s column (4).}

\end{table}

As is apparent, when (\ref{eq: binary outcome no time, lag 1}) and its counterparts with further lags are estimated, the coefficient on $ACI_{ij}$ is consistently negative but never statistically significant. Only the coefficient on $ACI_{i,j-2}$ is statistically significant in columns (3) and (4), and $ACI_{i,j-1}$ is statistically significant only in column (2). These results confirm the idea that treatment is more a function of prefecture-level unobservables than province-level unobservables. While the statistically significant coefficient on $ACI_{i,j-2}$ suggests that this result appears to be modified slightly based on how many sequential periods of corruption indictments occur, the effect is certainly not clean or uniform, given $ACI_{ij}$ and all other lags do not reject a null effect. Further, the coefficient for column (5)\textemdash wherein $AnyListings$ is regressed on the cumulative measure of the number of periods with nonzero corruption indictments up to time $j$\textemdash is again effectively 0 with a comparatively large standard error. This behavior again suggests that the number of total corruption indictments (up to period $j$) in a prefecture is not causally relevant to the probability of having any listings. 

I next conduct the sensitivity check suggested by \textcite{robins_association_1999} and  implemented by \textcite{ko_estimating_2003}. 

\vspace{3mm}
\begin{figure}[H]
    \centering
    \caption{\centering Sensitivity to $\varphi$ in Table \ref{tab: binary MSM outcome model no time outcome, sequential treatments FE}'s column (4) for IPTW with province-level fixed effects, in log-odds}
    \begin{subfigure}[t]{0.75\linewidth}
        \centering
        \includegraphics[width=\linewidth]{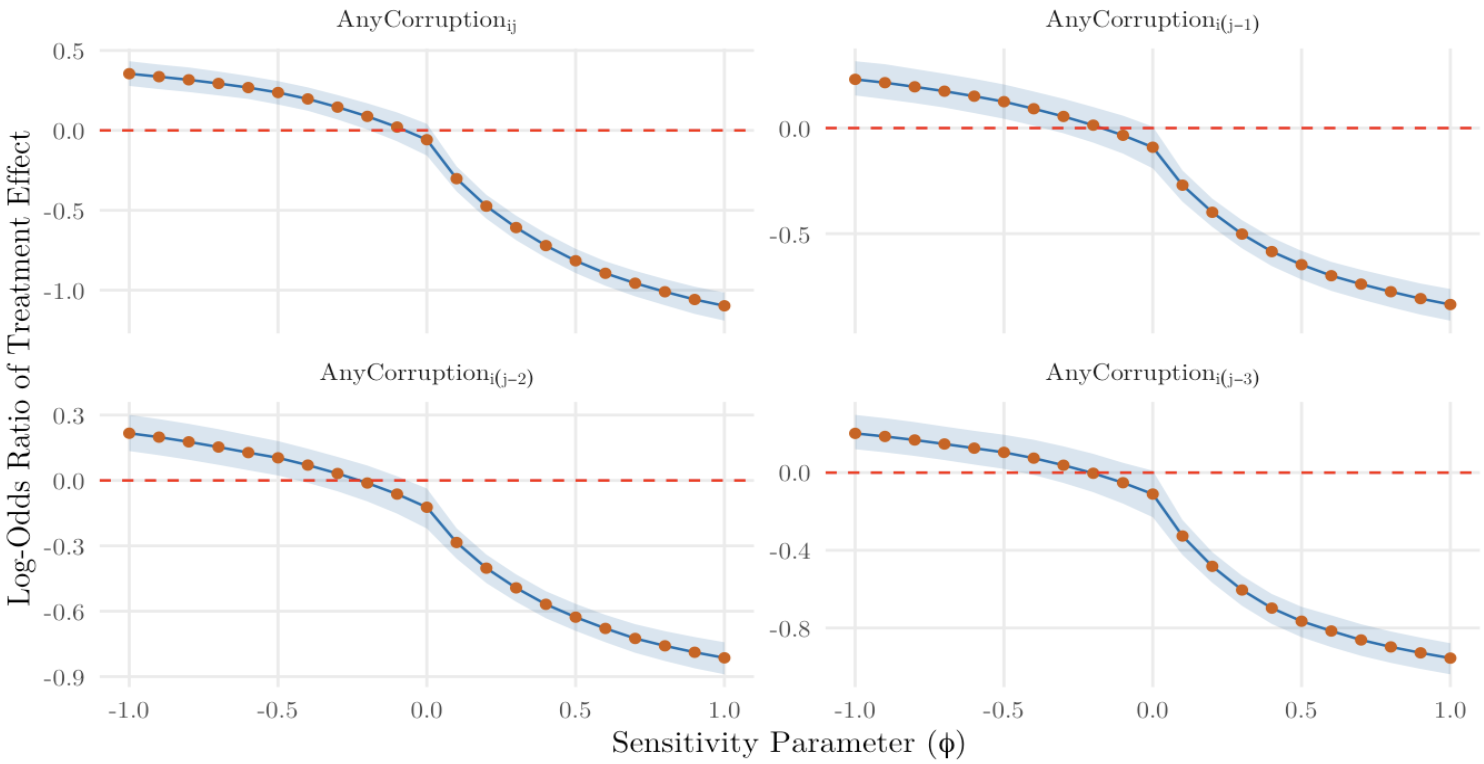}
    \label{fig:sensitivity Log-Odds Ratio of Treatment Effect vs. varphi- province fe, 3 lags}
    \end{subfigure}%
    \caption*{\footnotesize Note: The above results are a sensitivity check for the province-level fixed effects estimation (the Appendix's Table \ref{tab: binary MSM outcome model no time outcome, sequential treatments FE}, column (4)). The shaded blue band of Panel A reflects the 95\% confidence interval for the log-odds ratio of the treatment effect. Note also that pairs clustered bootstrap standard errors are used throughout these calculations. For Panel A, I confirm that the estimate at $\varphi = 0$ is equal to the estimate in the Appendix's Table \ref{tab: binary MSM outcome model no time outcome, sequential treatments FE}, column (4); this helps confirm a correct estimation.}
\end{figure}

As is apparent in the above graph, for $\varphi> 0$, the log-odds of the estimated treatment effect drop consistently, suggesting that the estimates derived under sequential ignorability, as reported in column (4), underestimate the true effect if unmeasured confounding exists. Recall that a positive $\varphi$ value corrects for the fact that indictments were over-assigned to high-corruption prefectures, and elucidating the direction of the bias, the sensitivity check suggests that the true deterrent effect of indictments on having any listings is larger in magnitude than originally estimated. In the initial outcome model (equation (\ref{eq: binary outcome no time})), the log-odds of the treatment effect became negative just before $\varphi =0$; this is still the case for the plot of $ACI_{ij}$, but for the three lags, $\varphi = -0.25$ is also negative. Thus, in the (unlikely) event that $\varphi$ is slightly less than 0 (i.e., $-0.25< \varphi <0$), the treatment effect will still be negative for the three lagged terms. The following plots with incremental effects on the y-axis (Figure \ref{fig:sensitivity Incremental Marginal Effect vs. varphi province fe, 3 lags}) helps to quantify the magnitude of the coefficient change in the presence of unmeasured confounding.

\vspace{3mm}
\begin{figure}[H]
    \centering
    \caption{\centering Sensitivity to $\varphi$ in Table \ref{tab: binary MSM outcome model no time outcome, sequential treatments FE}'s column (4) with province-level fixed effects in the IPTW: Incremental effects}
    \begin{subfigure}[t]{.75\linewidth}
        \centering
        \includegraphics[width=\linewidth]{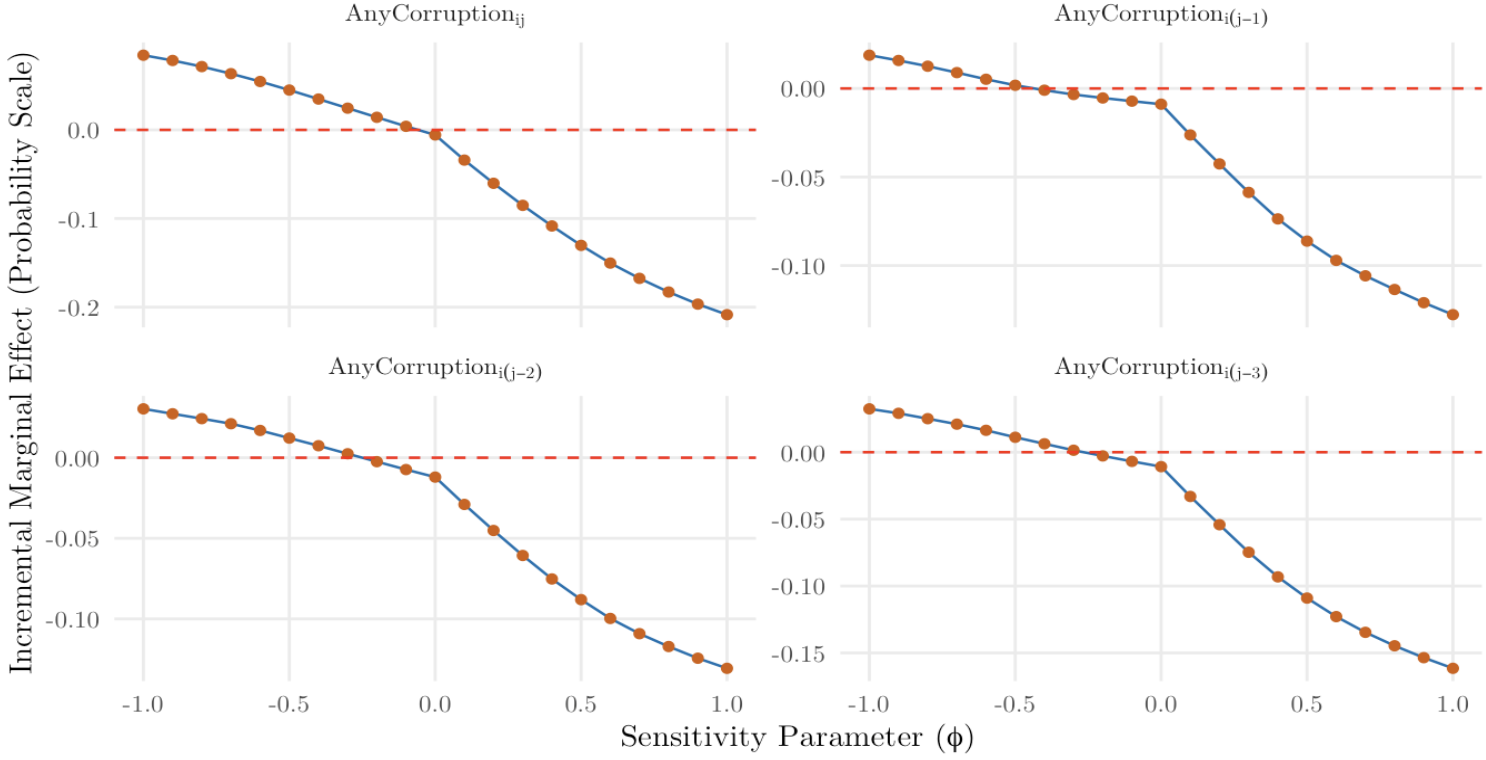}
    \end{subfigure}
    \caption*{\footnotesize Note: The above results are a sensitivity check for the estimation of Table \ref{tab: binary MSM outcome model no time outcome, sequential treatments FE}'s column (4) with province-level fixed effects in the IPTW model. Pairs clustered bootstrap standard errors are used throughout these calculations. }
    \label{fig:sensitivity Incremental Marginal Effect vs. varphi province fe, 3 lags}
\end{figure}

The apparent sharpness in the plots for $ACI_{i,j-1}, ACI_{i,j-2}$ and $ACI_{i,j-3}$ is due to points only being plotted for $\varphi$ values in increments of 0.25. Plotting at more frequent increments smooths out this apparent jump. As is apparent, even a small $\varphi$ value, such as 0.25, could have a significant effect on the probability of having any listings since the slope of the plot is quite steep near $\varphi = 0$, before stabilizing around $\varphi=2$.

I next conduct Ko, Hogan, and Mayer's sensitivity analysis for Table \ref{tab: binary MSM outcome model no time outcome, sequential treatments}'s column (5), which regresses $AnyListings$ on the cumulative sum of periods up to $j$ with nonzero corruption indictments.

\vspace{3mm}
\begin{figure}[H]
    \centering
    \caption{\centering Sensitivity to $\varphi$ for Table \ref{tab: binary MSM outcome model no time outcome, sequential treatments} column (5), with province-level fixed effects in the IPTW}
    \begin{subfigure}[t]{0.47\linewidth}
        \centering
        \includegraphics[width=\linewidth]{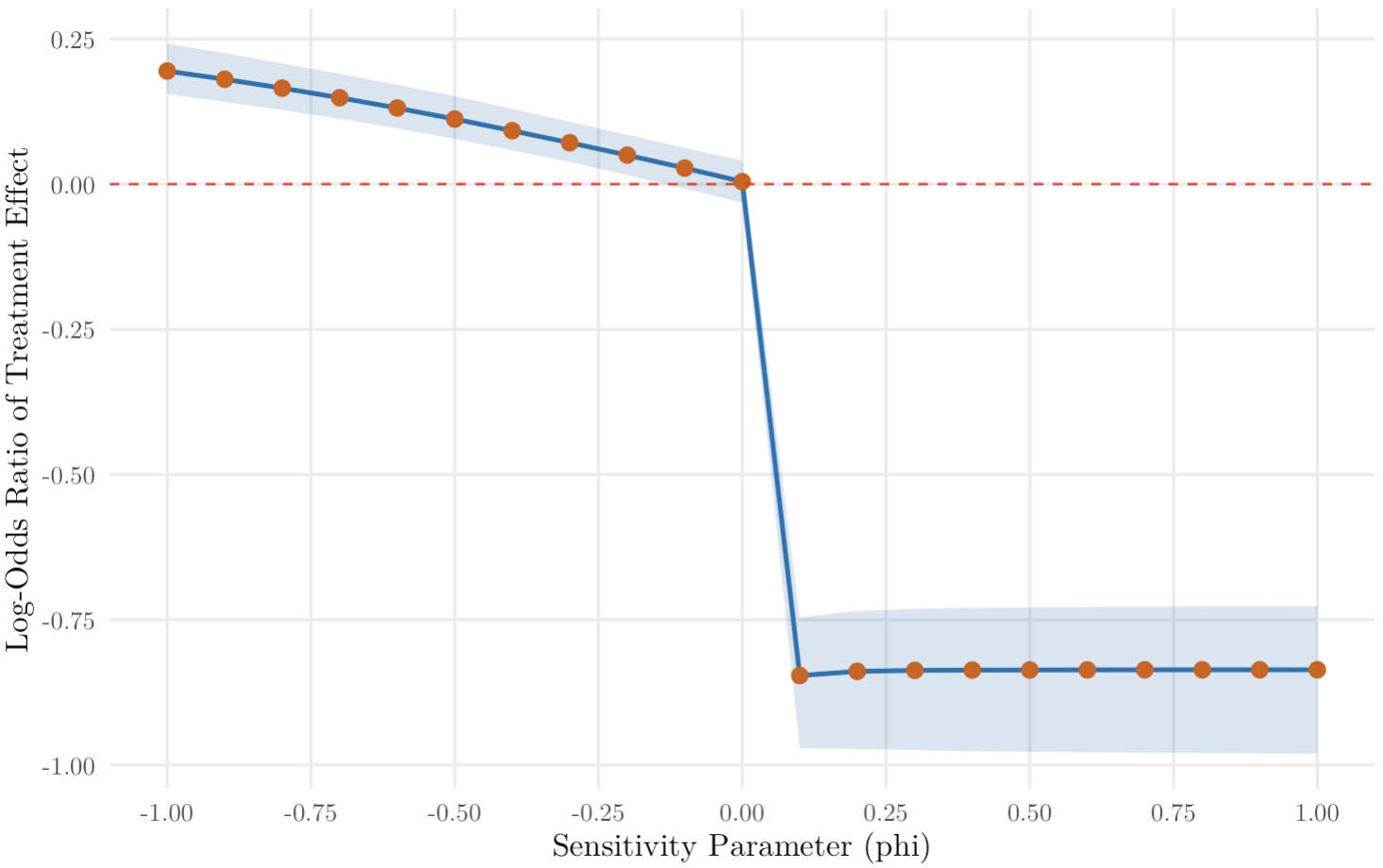}
        \caption{Log-odds ratio of treatment effect vs. $\varphi$}
    \label{fig:sensitivity Log-Odds Ratio of Treatment Effect vs. varphi- cumulative outcome prov}
    \end{subfigure}%
    \hspace{0.05\linewidth} 
    \begin{subfigure}[t]{.47\linewidth}
        \centering
        \includegraphics[width=\linewidth]{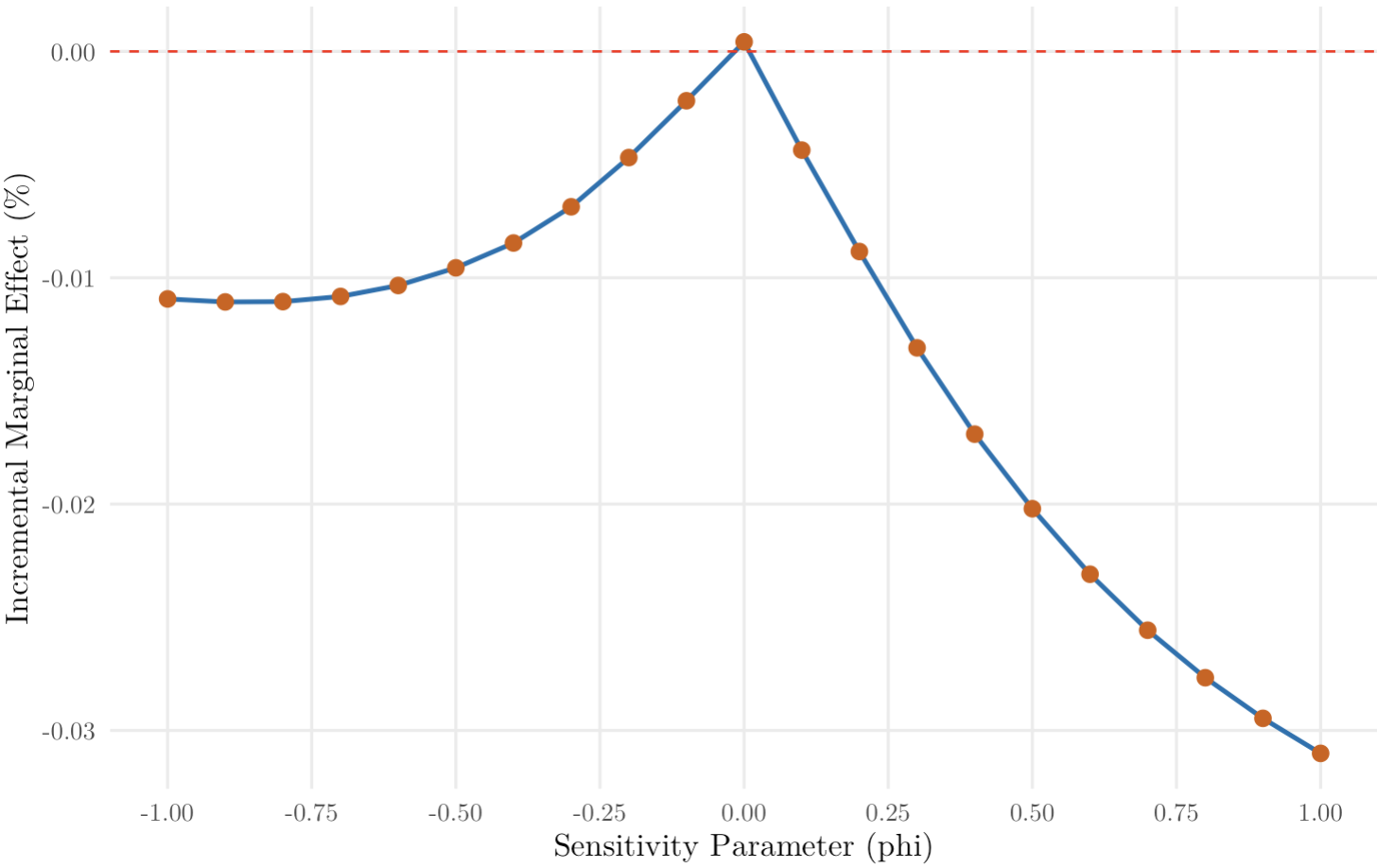}
        \caption{Incremental marginal effect vs. $\varphi$}
        \label{fig:sensitivity Incremental Marginal Effect vs. varphi- cumulative outcome prov}
    \end{subfigure}
    \caption*{\footnotesize The above results are a sensitivity check for Table \ref{tab: binary MSM outcome model no time outcome, sequential treatments}'s column (5) with province-level fixed effects in the IPTW. The shaded blue band of Panel A reflects the 95\% confidence interval for the log-odds ratio of the treatment effect. Points are estimated in increments of 0.1 for the interval [-1, 1]. Note also that pairs clustered bootstrap standard errors are used throughout these calculations. For Panel A, I confirm that the estimate at $\varphi = 0$ is equal to the estimate in the Appendix's Table \ref{tab: binary MSM outcome model no time outcome, sequential treatments FE}'s column (5); this helps confirm a correct estimation. The estimated incremental marginal effect at $\varphi = 0$ is also that reported for column (5).}
    \label{fig: sensivity to varphi- column 5 prov}
\end{figure}

These results closely mirror Figure \ref{fig: sensivity to varphi- column 5}; the plots are again much less smooth than those in Figures \ref{fig: sensivity to varphi}, \ref{fig: sensivity to varphi col 4}, and \ref{fig: sensivity to varphi- column (4)}. Like in Figure \ref{fig: sensivity to varphi- column 5}, the sharp drop in Panel A, which illustrates the log-odds ratio of the treatment effect, may reflect that the model is highly sensitive to the correction term near $\varphi = 0$ or that the outcome model is reacting nonlinearly to the corrected outcomes. Then, in Panel B with incremental marginal effects on the y-axis, as the marginal effects are a function of the coefficients and underlying predicted probabilities, it is apparent that any unmeasured confounders will cause a drop in the probability of having any listings. Notably, though, the scale of the incremental marginal effect is comparatively much smaller than that in Figures \ref{fig: sensivity to varphi}, \ref{fig: sensivity to varphi col 4}, or \ref{fig: sensivity to varphi- column (4)}\textemdash while the incremental marginal effects for the coefficients of these other plots is around 0.15-0.2 when $\varphi = 1$, it is only approximately 0.031 in Figure \ref{fig: sensivity to varphi- column 5}. The ``steep" decline of Panel B is therefore comparatively small in size. Thus, with the unusual shapes of these plots likely resulting from nonlinear interactions between the corrected outcome and the logistic model, as in Figure \ref{fig: sensivity to varphi- column 5}, they should not be interpreted as precise reflections of behavior but do suggest an extremely small change in the size of the coefficient (in terms of incremental effects) in the presence of unmeasured confounders. 

Finally, I implement \textcite{petersen_diagnosing_2012}'s version of the parametric bootstrap to diagnose possible positivity violations. For Table \ref{tab: binary MSM outcome model no time outcome, sequential treatments FE}'s column (4), I estimate the coefficients across 500 replicates and find that the estimated bias for $ACI_{ij}$ is -0.0221, which is smaller in magnitude than the standard error of 0.0528. The 95\% confidence interval is still inclusive of zero. I then perform this check on the other coefficients ($ACI_{i,j-1}$, $ACI_{i,j-2}$, $ACI_{i,j-3}$) and find similar results, with the bias estimate consistently below the standard error. For all coefficients except $ACI_{i,j-2}$, which has the confidence interval (-0.1991, -0.0027), the 95\% confidence interval includes zero. Thus, for all coefficients in column (4), the interpretation of the confidence interval does not substantively change, and the estimate of bias is well below the standard error. Then, for column (5), the estimated bias across 500 replicates is 0.0087. This number is again far below the standard error, and the 95\% confidence interval does not substantively change when bias is adjusted for, remaining almost symmetric around 0. Thus, there does not appear to be positivity violations that threaten the interpretation in either case. 

\pagebreak
\subsection{Sensitivity checks for MSM regressing corruption indictments on price} \label{Sensitivity checks for MSM regressing corruption indictments on price}

Sensitivity checks for the MSM regressing corruption indictments on price are performed below. Since the weights are the same as those used in the simple outcome model, I do not need to re-examine covariate balance, ESS, or propensity score overlap, for these all remain unchanged despite the different outcome models. I begin by performing the \textcite{ko_estimating_2003}-style implementation of \textcite{robins_association_1999}'s sensitivity check and then conduct \textcite{petersen_diagnosing_2012}'s parametric bootstrap to check for positivity violations. The sensitivity checks are focused on columns (1) and (3)\textemdash that is, the specifications with prefecture-level fixed effects in the IPTW with outcomes as the mean price per square meter (column (1)) and the median price per square meter (column (3)). 

For the Ko, Hogan, and Mayer check, the framework easily applies to cases with a continuous outcome, and the parameterization of $q_{ij}(t_j, t_j')$  described in Section \ref{sensitivity check sequential ignorability}\textemdash $q_{ij}(t_j, t_j') = \varphi\times(t_j - t_j')$ where $t_j' = 1- t_j$ and $\varphi$ gives the expected difference between $Y_{ij}(\overline{t}_j)$ given $T_{ij} = t_j$ versus  $T_{ij} = t_j'$\textemdash still holds. I examine the behavior of each coefficient of columns (1) and (3) in the presence of $\varphi$ values [-1, 1], as shown in Figures \ref{fig: sensitivity of Treatment Effect vs. varphi, mean price} and \ref{fig: sensitivity of Treatment Effect vs. varphi, median price}, respectively. Note that effect measured on the y-axis is already exponentiated and can be interpreted directly as a percent.

Recall that $\varphi = 0 $ corresponds to the assumption of no unmeasured confounders. If $\varphi > 0$, then on average, treatment is preferentially given to those units with higher $AvgPrice$ counterfactuals $\{Y_{ij}(\overline{t}_j)\}$ (i.e., corruption indictments are given to those prefectures where higher average prices would have occurred in the absence of indictments), even after controlling for past treatment and measured covariate history. Conversely, $\varphi < 0$ would imply that treatment was preferentially given to prefectures with lower $AvgPrice$ counterfactuals. As I expect that treatment is preferentially given to those units with higher (unmeasured) suspected corruption, I focus primarily on positive values of $\varphi$ in my sensitivity analysis, although I also include a range of negative $\varphi$ values for comparison. 

\vspace{3mm}
\begin{figure}[H]
    \centering
    \caption{Sensitivity to $\varphi$ for Table \ref{tab: Corruption indictments impacting sale price}'s columns (1) and (3)}
    \begin{subfigure}[t]{0.47\linewidth}
        \centering
        \includegraphics[width=\linewidth]{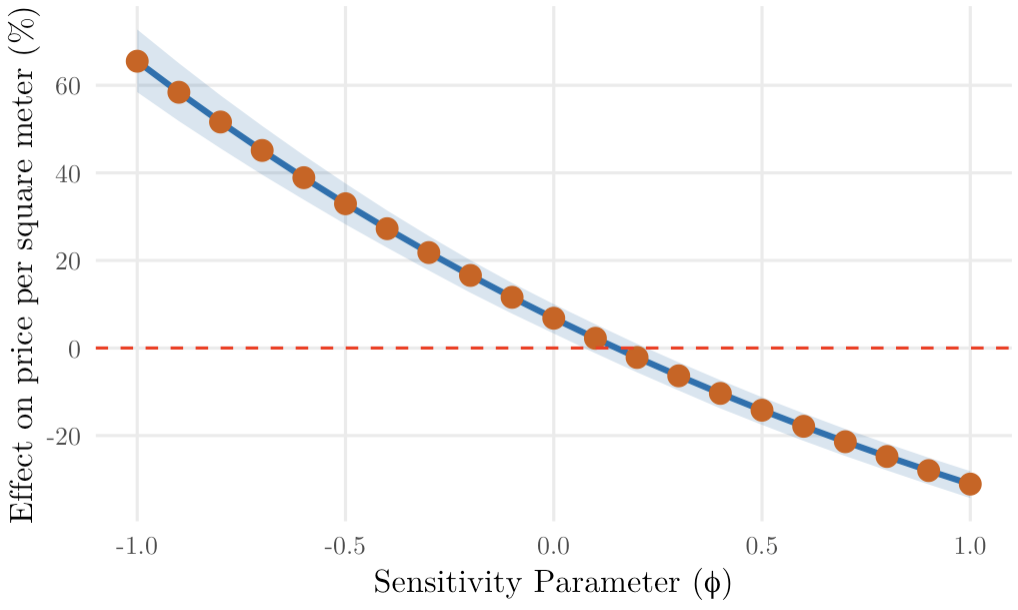}
        \caption{Mean price per square meter vs. $\varphi$}
    \label{fig: sensitivity of Treatment Effect vs. varphi, mean price}
    \end{subfigure}%
    \hspace{0.05\linewidth} 
    \begin{subfigure}[t]{.47\linewidth}
        \centering
        \includegraphics[width=\linewidth]{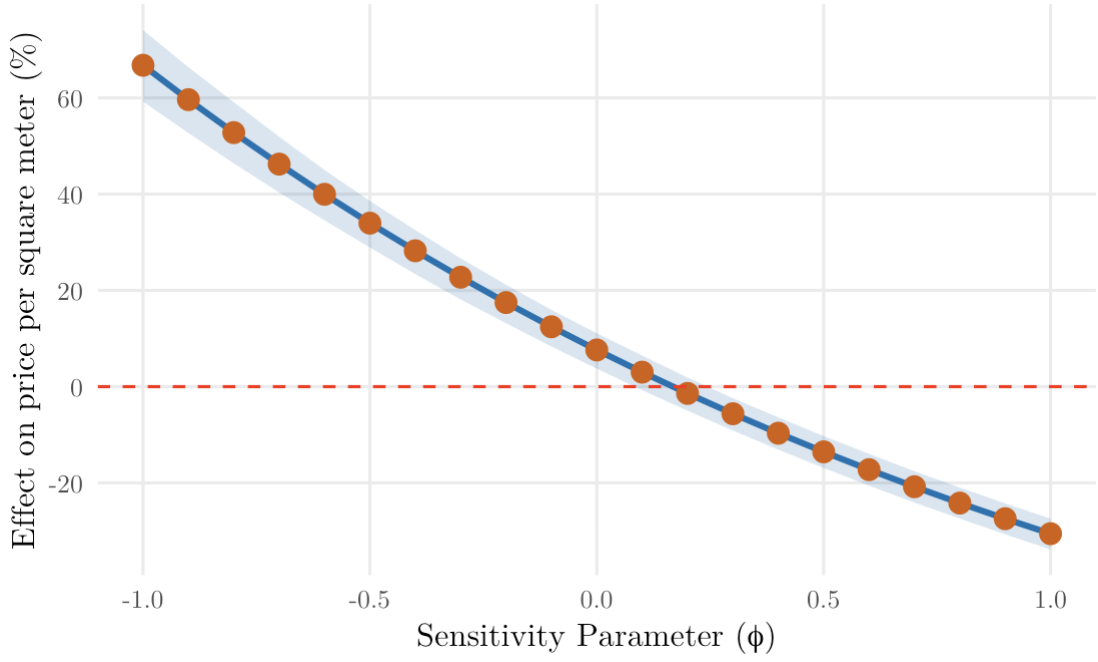}
        \caption{Median price per square meter vs. $\varphi$}
        \label{fig: sensitivity of Treatment Effect vs. varphi, median price}
    \end{subfigure}
    \caption*{\footnotesize Note: The above results are a sensitivity check for the prefecture-level fixed effects estimation (columns (1) and (3)). The shaded blue band reflects the 95\% confidence interval for the treatment effect. Note also that pairs clustered bootstrap standard errors are used throughout these calculations. For both panels, I confirm that the estimate at $\varphi = 0$ is equal to the estimate in Table \ref{tab: Corruption indictments impacting sale price}'s columns (1) and (3); this helps confirm a correct estimation.}
    \label{fig: sensivity to varphi, price}
\end{figure}

As is apparent in the above graph, the effect of the estimated treatment effect drops consistently, suggesting that the estimates derived under sequential ignorability as reported in columns (1) and (3) are overestimates when $\varphi> 0$\textemdash in the sense that they overestimate the true effect if treatment assignment is biased by selection. A positive $\varphi$ value would correct for the fact that indictments were over-assigned to high-price prefectures, and by accounting for this bias, it estimates that the true deterrent effect of indictments on having any listings is smaller than originally estimated, turning negative once $\varphi> 0.15$. 

Yet, while this test suggests that the direction of the results is sensitive to unmeasured confounding, the fixed effects in the IPTW absorb all prefecture-level time invariant confounding; any remaining concerns for unmeasured confounding thus come from time-varying confounders not directly controlled for, but there are likely not many of these in this prefecture-level case. Regardless, this sensitivity check is still useful in offering insight into what such confounding would imply for the results of Table \ref{tab: Corruption indictments impacting sale price}.

I next proceed with \textcite{petersen_diagnosing_2012}'s parametric bootstrap to diagnose possible positivity violations. For column (1), I estimate this coefficient across 500 replicates and find that the estimated bias is 0.0073. This falls below the (previously described) ``red flag" metrics of 1) the bias being of similar magnitude (or larger than) the estimator's standard error, which was 0.0154, per column (1), or 2) the bias changing the interpretation of the bias-corrected confidence interval. The recalculated 95\% confidence interval for column (1) when adjusted for this bias is now (0.0282, 0.0886), again remaining consistently positive. Thus, based on both conceptual groups and these checks, positivity seems to be broadly met for this specification. For column (3), the estimated bias is 0.0026, which again does not reach the ``red flag" level as it is below the estimated standard error. Likewise, the corrected confidence interval of (0.0400, 0.1050) again does not have a substantively different interpretation as it still remains positive. Thus, based on these results, there does not appear to be a positivity violation that would threaten causal estimates or interpretation.  

\end{document}